\documentclass[10pt]{article}

\usepackage{amsmath}
\usepackage{amssymb}
\usepackage{ifthen}
\usepackage{color}
\usepackage{colortbl}
\usepackage{rotating}
\usepackage{multirow}
\usepackage{url}
\usepackage{graphicx}
\usepackage{float}
\usepackage[title]{appendix}
\usepackage{authblk}

\newcolumntype{I}{!{\vrule width 1.5pt}}
\newlength\savedwidth
\newcommand\whline{\noalign{\global\savedwidth\arrayrulewidth
                            \global\arrayrulewidth 1.5pt}%
           \hline
           \noalign{\global\arrayrulewidth\savedwidth}}

\newboolean{IsWide}
\setboolean{IsWide}{true}

\newcommand{\FlaPartition}[2]{
\ifthenelse{\boolean{IsWide}}{{\bf partition } \hspace{-1em} #1 \hspace{-1em} #2}
{{\bf partition } \+ \\ #1 \+ \\ #2 \- \-}
}

\newcommand{\FlaRepartition}[2]{
\ifthenelse{\boolean{IsWide}}{{\bf repartition } \hspace{-1em} #1 \hspace{-1em} #2}
{{\bf repartition } \+ \\ #1 \+ \\ #2 \- \-}
}

\newcommand{\FlaContinue}[1]{
\ifthenelse{\boolean{IsWide}}{{\bf continue with } #1
}
{{\bf continue with } \+ \\ #1 \-
}
}

\newcommand{\blocksize}{1}

\newcommand{\repartitionings}{
\begin{minipage}[t]{3in}
\ \\
\ \\
\ \\
\end{minipage}
}

\newcommand{\repartitionsizes}{ \hspace{ 3.25in} }

\newcommand{\WSrepartitionBig}{
\begin{minipage}[t]{4.83in}
\ifthenelse{ \equal{\blocksize}{1} }{}
{%
\ifthenelse{ \equal{\blocksize}{blank} }{~}
{{\bf Determine block size $ \blocksize $}} \\
}
{\bf Repartition}
\begin{tabbing}
in \= in \= \+ \kill
\repartitionings \+ \\
{\bf where } \hspace*{-2ex} \repartitionsizes 
\end{tabbing}
\end{minipage}
}

\newcommand{\WSrepartitionNormal}{
\begin{minipage}[t]{4.22in}
\ifthenelse{ \equal{\blocksize}{1} }{}
{%
\ifthenelse{ \equal{\blocksize}{blank} }{~}
{{\bf Determine block size $ \blocksize $ }} \\
}
{\bf Repartition}
\begin{tabbing}
in \= in \= \+ \kill
\repartitionings \+ \\
{\bf where } \hspace*{-2ex} \repartitionsizes 
\end{tabbing}
\end{minipage}
}

\newcommand{\WSrepartitionMedium}{
\begin{minipage}[t]{3.60in}
\ifthenelse{ \equal{\blocksize}{1} }{}
{%
\ifthenelse{ \equal{\blocksize}{blank} }{~}
{{\bf Determine block size $ \blocksize $}} \\
}
{\bf Repartition}
\begin{tabbing}
in \= in \= \+ \kill
\repartitionings \+ \\
{\bf where } \hspace*{-2ex} \repartitionsizes 
\end{tabbing}
\end{minipage}
}

\newcommand{\WSrepartitionNarrow}{
\begin{minipage}[t]{3.00in}
\ifthenelse{ \equal{\blocksize}{1} }{}
{%
\ifthenelse{ \equal{\blocksize}{blank} }{~}
{{\bf Determine block size $ \blocksize $}} \\
}
{\bf Repartition}
\begin{tabbing}
in \= in \= \+ \kill
\repartitionings \+ \\
{\bf where } \hspace*{-2ex} \repartitionsizes 
\end{tabbing}
\end{minipage}
}

\newcommand{ \becomes }{:=}

\newcounter{mycounter}

\newcommand{\gemm}{{\sc gemm} }
\newcommand{\gemmns}{{\sc gemm}}

\newcommand{\ttsgemm}{{\tt sgemm()} }
\newcommand{\ttsgemmns}{{\tt sgemm()}}
\newcommand{\ttdgemm}{{\tt dgemm()} }
\newcommand{\ttdgemmns}{{\tt dgemm()}}
\newcommand{\ttcgemm}{{\tt cgemm()} }
\newcommand{\ttcgemmns}{{\tt cgemm()}}
\newcommand{\ttzgemm}{{\tt zgemm()} }
\newcommand{\ttzgemmns}{{\tt zgemm()}}

\newcommand{\ttstruct}{{\tt struct} }

\newcommand{\ttobjt}{{\tt obj\_t} }

\newcommand{\ttbligemm}{{\tt bli\_gemm()} }
\newcommand{\ttbligemmns}{{\tt bli\_gemm()}}

\newcommand{\updates}{\hspace{0.6mm}+\hspace{-1mm}=}
\newcommand{\real}[1]{#1^{r}}
\newcommand{\imag}[1]{#1^{i}}
\newcommand{\Rdom}{\mathbb R}
\newcommand{\Cdom}{\mathbb C}
\newcommand{\calRe}{{\cal Re}}
\newcommand{\calIm}{{\cal Im}}

\newcommand{\onem}{{\sc 1m} }

\newcommand{\onee}{{\sc 1e} }
\newcommand{\onerns}{{\sc 1r}}
\newcommand{\oner}{{\sc 1r} }

\newcommand{\fourm}{{\sc 4m} }

\newcommand{\zero}{{\sc 0} }
\newcommand{\onea}{{\sc 1a} }
\newcommand{\oneb}{{\sc 1b} }
\newcommand{\onec}{{\sc 1c} }
\newcommand{\twoab}{{\sc 2ab} }
\newcommand{\twobc}{{\sc 2bc} }
\newcommand{\twoac}{{\sc 2ac} }
\newcommand{\three}{{\sc 3} }
\newcommand{\zerons}{{\sc 0}}
\newcommand{\oneans}{{\sc 1a}}
\newcommand{\onebns}{{\sc 1b}}
\newcommand{\onecns}{{\sc 1c}}
\newcommand{\twoabns}{{\sc 2ab}}
\newcommand{\twobcns}{{\sc 2bc}}
\newcommand{\twoacns}{{\sc 2ac}}
\newcommand{\threens}{{\sc 3}}

\newcommand\tbstrut{\rule[-1.0ex]{0pt}{3.6ex}} 

\newcommand{\commentout}[1]{}

\newcommand{\NoShow}[1]{}

\NoShow{\markboth 
{%
F. G. Van Zee et al. 
}
{
Supporting mixed-domain mixed-precision matrix multiplication within the BLIS framework 
}
}
\title{
Supporting mixed-datatype matrix multiplication within the BLIS framework \\[0.2in]
	\large
	FLAME Working Note \#89
}

\newcommand*{\affaddr}[1]{#1} 
\newcommand*{\affmark}[1][*]{\textsuperscript{#1}}
\newcommand*{\email}[1]{\texttt{#1}}

\author{
	Field G. Van Zee\affmark[a,b],
	Devangi N. Parikh\affmark[a],
	Robert A. van de Geijn\affmark[a,b]\\
	\affaddr{\affmark[a]Institute for Computational Engineering and Sciences}\\
	\affaddr{\affmark[b]Department of Computer Science \\}
	The University of Texas at Austin, Austin, TX \\
	\email{{\tt \{field,dnp,rvdg\}@cs.utexas.edu}}
}

\NoShow{
\terms{Algorithms; Performance}


\acmformat{
Field G. Van Zee, Devangi N. Parikh, and Robert A. van de Geijn.
Supporting mixed-domain mixed-precision matrix multiplication within the BLIS framework.
}
}
\begin{document}
	\maketitle

	\begin{abstract}
		We approach the problem of implementing mixed-datatype support within the general matrix multiplication (\gemmns) operation of the BLIS framework, whereby each matrix operand $ A $, $ B $, and $ C $ may be stored as single- or double-precision real or complex values.
Another factor of complexity, whereby the computation is allowed to take place in a precision different from the storage precisions of either $ A $ or $ B $, is also included in the discussion.
We first break the problem into mostly orthogonal dimensions, considering the mixing of domains separately from mixing precisions.
Support for all combinations of matrix operands stored in either the real or complex domain is mapped out by enumerating the cases and describing an implementation approach for each.
Supporting all combinations of storage and computation precisions is handled by typecasting
the matrices at key stages of the computation---during packing and/or accumulation, as needed.
Several optional optimizations are also documented.
Performance results gathered on a 56-core Marvell ThunderX2 and a 52-core Intel Xeon Platinum demonstrate that high performance is mostly preserved, with modest slowdowns incurred from unavoidable typecast instructions.
The mixed-datatype implementation confirms that combinatoric intractability is avoided, with the framework relying on only two assembly microkernels to implement 128 datatype combinations.

	\end{abstract}

\section{Introduction}

%
%

The BLAS~\cite{BLAS3} defines the general matrix-matrix multiplication (\gemmns) operation to support any of the following computations:
\[
\begin{array}{l l l}
C := \alpha A   B   + \beta C, & C := \alpha A   B^T + \beta C, & C := \alpha A   B^H + \beta C, \\
C := \alpha A^T B   + \beta C, & C := \alpha A^T B^T + \beta C, & C := \alpha A^T B^H + \beta C, \\
C := \alpha A^H B   + \beta C, & C := \alpha A^H B^T + \beta C, & C := \alpha A^H B^H + \beta C.
\end{array}
\]
where $ C $ is $ m \times n $, the left-hand matrix product operand ($ A $, $ A^T $, or $ A^H $) is $ m \times k $, the right-hand matrix product operand ($ B $, $ B^T $, or $ B^H $) is $ k \times n $, and $ \alpha $ and $ \beta $ are scalars.

This matrix multiplication functionality is made available to software developers via the following application programming interfaces, or APIs:
\begin{center}
\vspace{2mm}
\tt sgemm( transa, transb, m, n, k, alpha, A, ldA, B, ldB, beta, C, ldC ) \\
\tt dgemm( transa, transb, m, n, k, alpha, A, ldA, B, ldB, beta, C, ldC ) \\
\tt cgemm( transa, transb, m, n, k, alpha, A, ldA, B, ldB, beta, C, ldC ) \\
\tt zgemm( transa, transb, m, n, k, alpha, A, ldA, B, ldB, beta, C, ldC ) \\
\vspace{2mm}
\end{center}
The first letter of the routine name uniquely encodes the datatype---that is, the domain and precision---of the matrix and scalar operands as well as the computation: single-precision real ({\tt s}); double-precision real ({\tt d}); single-precision complex ({\tt c}); and double-precision complex ({\tt z}).
The parameters {\tt transa} and {\tt transb} indicate if $ A $ and/or $ B $, respectively, should be computed upon as if they were transposed or conjugate-transposed.
The interfaces implicitly require that matrices be stored in column-major order.
Accordingly, the parameters {\tt ldA}, {\tt ldB}, and {\tt ldC} convey the so-called ``leading dimensions'' of the arrays {\tt A}, {\tt B}, and {\tt C}, respectively---that is, the number of elements that separate matrix element $ (i,j) $ from element $ (i,j+1) $ in memory.

While this interface has served the HPC community well, it has also become constraining.
For example, when computing tensor contractions (which often resemble matrix multiplications), one may need to refer to a sub-tensor that cannot be represented with column-major storage without making a temporary copy.
Similarly, some situations may call for conjugating (but not transposing) a matrix operand.
Indeed, such functionality is already supported by the BLIS framework, which exports BLAS-like operations and APIs~\cite{BLIS1}.
However, even BLIS only supports computation on operands with identical datatypes.
Consider the following:
\begin{itemize}
\item
There exist applications that may wish to update a complex matrix by the product of a complex matrix and a real matrix.
These include applications involving damped response~\cite{Kristensen09,Coriani12}, Green's functions methods~\cite{Nooijen}, Complex Absorbing Potential (CAP), and Complex Scaling (CS)~\cite{Jagau17}.
Applications in quantum chemistry may also benefit.
These mixed-domain instances of \gemm are currently improvised either by casting the operation in terms of {\tt cgemm} or {\tt zgemm}, in which case half of the floating-point operations are superfluous, or by performing two passes with {\tt sgemm} or {\tt dgemm}, which tends to be cumbersome and error-prone, requires extra workspace in which to make temporary copies of the real and imaginary parts of the complex matrix operands, and likely yields suboptimal performance.
\item
Similarly, there exist applications that could benefit from storing matrix operands in different precisions, and/or computing in a precision that is lower or higher than the storage precision of $ A $ and $ B $.
These include NWChem~\cite{NWChem,NWChem68} performing CCSD(T) computations~\cite{crawford_ccsdt,Stanton97}, and various applications in machine learning~\cite{fbgemm}.
Currently, this must be performed in an ad-hoc manner similar to the mixed-domain case, and with similar workspace and performance drawbacks.
\end{itemize}
Thus, there is likely a fair amount of pent-up demand for high-performance implementations to datatype-flexible BLAS-like APIs.

As alluded to above, the naive approach to supporting mixed-datatype functionality within the \gemm operation comes with obvious memory, performance, and productivity drawbacks:
typecasting matrix operands to a common domain and/or precision outside of the original implementation requires considerable workspace; the memory access patterns engendered by monolithic casting almost surely acts as a drag on performance;
and programming an ad-hoc solution in terms of the BLAS \gemm interfaces sometimes requires non-trivial skills.
Indeed, some would find providing the full combinatoric space of functionality daunting, and in response might attempt to survey the community and then only implement those cases for which interest was expressed.
Instead, our goal from the outset is to implement support for {\em all} cases, and to do so in a manner that delivers high or near-high performance.%
\footnote{
Understandably, some readers may question the utility of some mixed-datatype cases discussed in this article.
Skeptics may argue that one only needs to focus on the cases that ``make sense.''
We reason about the issue as follows. While we can identify certain cases that {\em are today} useful to {\em some} people or applications, we cannot say with certainty which cases {\em will never be} used by {\em any} person or application.
And because we cannot {\em a priori} identify the cases that will never be needed, we take the position that we must treat all cases as important enough to merit implementation.
}
Our approach, which builds on the BLIS framework, yields a comprehensive reference implementation with which consumers of this functionality can explore the benefits of mixed-domain and mixed-precision \gemm computation without being constrained by limitations in the interface, incomplete coverage within the implementation, or unnecessarily inefficient performance.

\subsection{Notation}

Our notation should be mostly self-evident to most readers of high-performance dense linear algebra literature.
We use uppercase Roman letters, such as $ A $, $ B $, and $ C $ to denote matrices, and lowercase Greek letters $ \alpha $ and $ \beta $ to represent scalars.

Real and complex domains are indicated by $ \Rdom $ and $ \Cdom $, respectively.
Occasionally, we refer to the real part of a matrix or matrix expression $ X $ with $ \calRe( X ) $ and to the imaginary part with $ \calIm( X ) $.
In other places, such as where this notation would be too cumbersome, we use superscripts, such as $ \real{\chi} $ and $ \imag{\chi} $ for the real and imaginary components, respectively, of a scalar $ \chi $.

When representing elements within a matrix, we use a subscript to encode the row and column indices.
For example, a scalar $ \alpha_{13} $ would reference the element located in the 2nd row and 4th column of a matrix $ A $.%
\footnote{
Our subscript notation starts counting from 0.
}


\section{Background}
\begin{figure}[tb!]
\begin{center}
\includegraphics[width=0.6\textwidth]{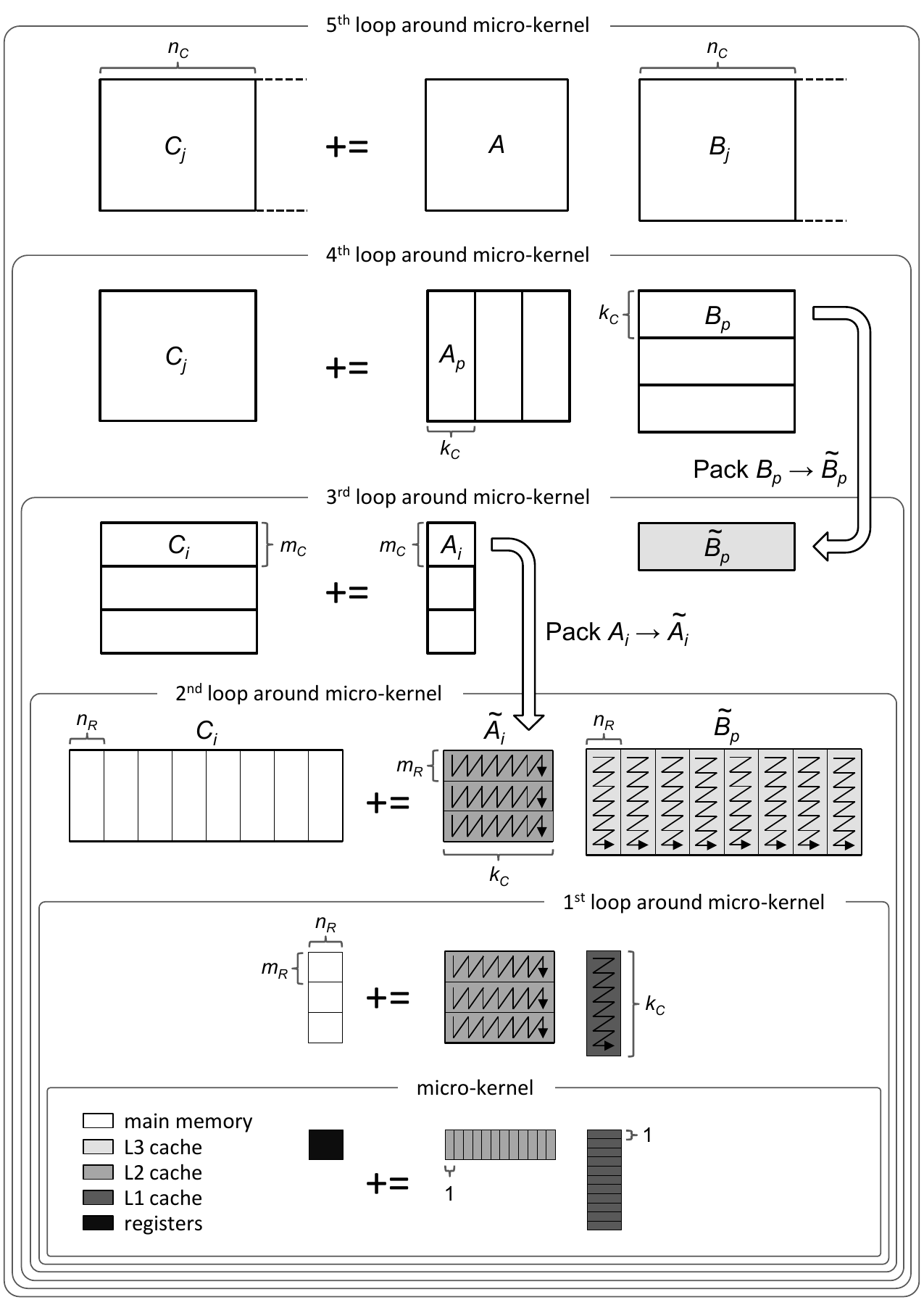}
\end{center}
\caption{
BLIS refactoring of the GotoBLAS algorithm as five loops around the microkernel~\cite{BLIS5}.
Used with permission.
}
\label{fig:BLIS}
\end{figure}

In this section, we review the approach to matrix multiplication taken within the BLIS framework as well as some related implementation details that will provide important context to discussions later in this article.

\subsection{Matrix Multiplication in BLIS}

The GotoBLAS algorithm~\cite{Goto1} for performing matrix multiplication underlies the highest-performing BLAS libraries for current general-purpose microprocessors.
The BLAS-like Library Instantiation Software (BLIS) framework, which implements \gemm and other matrix-matrix operations, refactors the GotoBLAS algorithm as pictured in Figure~\ref{fig:BLIS}.
BLIS isolates the code that needs to be optimized (in assembly code or with vector intrinsics) for different architectures in a {\em microkernel} that updates a very small submatrix, or microtile, of $ C $ with a sequence of rank-1 updates that are accumulated in registers~\cite{BLIS1}.
All other loops and supporting kernels are implemented portably in C99.
By contrast, Goto's implementation---also adopted by the OpenBLAS fork~\cite{OpenBLAS} of the GotoBLAS library---casts the computation into a larger assembly-coded kernel.
This larger unit of code, which corresponds to what BLIS refers to as the {\em macrokernel}, consists of the microkernel plus the logic that falls within the first two loops around the microkernel.

A key element of the GotoBLAS algorithm is that high-performance implementation of \gemm incorporates the packing of submatrices of $ B $ (into buffer $ \widetilde B $) and of $ A $ (into buffer $ \widetilde A $) to improve data locality during the execution of the microkernel.%
\footnote{
This reorganization of matrices $ A $ and $ B $ can also improve TLB performance~\cite{Goto}.
}
This feature has been used in the past to consolidate other functionality into the same framework:
implementation of other matrix-matrix operations (level-3 BLAS)~\cite{Goto2,BLIS1}, fusing sequences of matrix operations of importance to Machine Learning~\cite{Yu:2015:POK:2807591.2807601}, 
and implementation of practical Fast Matrix Multiplication (Strassen-like) algorithms~\cite{7967156,Huang:2016:SAR:3014904.3014983}.  
A final insight comes from Van Zee~\cite{BLIS6}, in which it is shown how complex matrix multiplication can be cast in terms of only microkernels designed for real domain \gemm without a significant performance penalty.

Given a target architecture, instantiating the traditional functionality of \gemm with BLIS requires only two microkernels, one each for single- and double-precision real domain computations, with the insight from~\cite{BLIS6} inducing the functionality typically provided by complex domain microkernels.
It also requires packing functions, which by default take the form of architecture-agnostic (C99) implementations provided by the framework, as well as architecture-specific cache and register blocking parameters.
BLIS exposes several other configure-time options, though they all default to values that typically need no further tweaking.

\subsection{Managing complexity in BLIS}

The combinatoric complexity of the \gemm operation in BLIS is mitigated in several ways.

\subsubsection{Storage}

BLIS tracks separate row and column strides for each matrix object.%
\footnote{
In BLIS, the row stride---like the leading dimension in row-major storage---expresses the number of elements that separate matrix element $ (i,j) $ and element $ (i+1,j) $ in memory.
The column stride---like the leading dimension in column-major storage---expresses the number of elements that separate elements $ (i,j) $ and $ (i,j+1) $.
}
Using these two stride parameters, BLIS supports three matrix storage formats in its end-user APIs: row-major storage (where the column stride is unit); column-major storage (where the row stride is unit); and general storage (where neither stride is unit).%
\footnote{
We often refer to row-major matrices as being row-stored, and column-major matrices as being column-stored. 
}
Each \gemm operand in BLIS may individually be stored in {\em any} of the three aforementioned storage formats.
If we consider all possible variations of general storage as a single format, this results in a total of $ 3^3 = 27 $ different storage combinations.
The \gemm operation supports these $ 27 $ storage combinations as follows: the packing function is written generically to allow it to read from any of the three storage formats when reading matrices $ A $ and $ B $ (during the packing of $ \widetilde A $ and $ \widetilde B $), and the microkernel is required to handle input/output of $ C $ in any of the three supported formats.

\subsubsection{Transposition}

BLIS easily accommodates transposition of matrices $ A $ or $ B $ by swapping the row and column strides (and their corresponding dimensions) just prior to packing.
This technique merely affects how the matrices are traversed rather than how they are stored; thus, we call this an ``induced'' (or logical) transposition, as no matrix elements are actually copied or moved.

\subsubsection{Conjugation}

In the case of complex matrices, optional conjugation%
\footnote{
In BLIS, all input matrix operands to \gemm and most other operations may be conjugated {\em without} transposition, which corresponds to a new {\tt trans} parameter value absent in the BLAS.
}
of $ A $ and $ B $ is handled during the packing into $ \widetilde A $ and $ \widetilde B $.

\subsubsection{Multithreaded parallelization}

The authors of~\cite{BLIS3} discuss how BLIS exposes many loops, each of which can be parallelized.
Crucially, the complexity over matrix storage datatypes (domain and precision), transposition/conjugation parameters ({\tt transA} and {\tt transb}), and matrix storage formats (row, column, generalized) are orthogonal to the issues pertaining to extracting multithreaded parallelism from \gemmns.
Therefore, the insights of~\cite{BLIS3} carry over to the parallelization when mixing domains and/or precisions.

\subsubsection{Optimizing input/output on $ C $}

High-performance microkernels accumulate their intermediate results using vector registers.
Thus, the microkernel author must decide whether to semantically assign the vector registers to contain contiguous rows or contiguous columns of the microtile submatrix.
We refer to this as the microkernel's {\em register orientation}.
Interestingly, the register orientation necessarily biases the microkernel towards loading and storing elements of $ C $ as either rows or columns, since performing IO on elements in the opposite orientation would require a sequence of costly permutation instructions.
BLIS tracks this intrinsic property, or IO preference, of the microkernel so that the framework can transform the matrix problem to the microkernel's liking.
For example, if our microkernel is row-preferential and the \gemm implementation is executed on a column-stored matrix $ C $, BLIS will employ a high-level transposition of the entire operation (to $ C^T \becomes B^T A^T $) so that, from the perspective of the microkernel, $ C^T $ {\em appears} to be stored in its preferred format---that is, a manner consistent with its vector register orientation.%
\footnote{
If, for whatever reason, this optimization is not employed, the microkernel in this example would use the general storage case to read and write to a column-stored $ C $, which would incur a small performance penalty due to an increased number of assembly instructions.
}
Thus, regardless of whether the microkernel is row- or column-preferential, the \gemm implementation will, generally-speaking, yield similar performance on row- and column-stored matrices $ C $.%
\footnote{
Typically, the general-storage case in the microkernel must be handled separately from the contiguous row- or column-storage case that is preferred.
This case usually incurs a performance penalty that is mostly unavoidable due to decreased spatial locality and an increased number of assembly instructions.
}

\subsection{For the busy reader}

We acknowledge that some readers may wish for a very short synopsis of the insights presented later in this article.
We sum up the techniques that allow implementing all cases
of mixed-domain and mixed-precision \gemm within the BLIS framework as follows:
The mixing of domains can be handled by enumerating the eight cases (six of them new), which largely reduce to either manipulating matrix metadata, and/or exposing the real and imaginary elements in a complex matrix in such a way that a real matrix multiplication may be performed to induce the desired result, in part motivated by insights from the \onem method~\cite{BLIS6}.
The mixing of precisions can be handled by typecasting between precisions, as needed, during the packing $ A $ and $ B $ (into $ \widetilde A $ and $ \widetilde B $), while the typecasting during the accumulation of the intermediate product $ A B $ may occur in special code wrappers that update $ C $ appropriately.


\section{API Considerations}

In the spirit of the BLAS API, a more complete interface that supports this richer, mixed-datatype environment could take the form of

\begin{figure}[h!]
\begin{center}
\begin{minipage}{5in}
\begin{verbatim}
gemm( m, n, k, alpha, transA, A, domainA, precA, rstrideA, cstrideA,
                      transB, B, domainB, precB, rstrideB, cstrideB,
                beta,         C, domainC, precC, rstrideC, cstrideC,
                                          precAB )
\end{verbatim}
\end{minipage}
\end{center}
\caption{
A hypothetical BLAS-like API for mixed-domain, mixed-precision \gemmns.
}
\label{fig:hypoapi}
\end{figure}

\noindent
where {\tt transX} indicates whether $ X $ should be computed upon as if it were (optionally) transposed, conjugate-transposed, or conjugated {\em without} transposition;
{\tt X} is the address where matrix $ X $ is stored; 
{\tt domainX} indicates whether $ X $ is stored as a matrix of real or complex elements;
{\tt precX} indicates the precision in which elements of $ X $ are stored (e.g., half-, single-, double-, or quad-precision);
{\tt rstrideX} and {\tt cstrideX} indicate the row and column strides, respectively, for storing%
\footnote{
Generally speaking, we consider these separate strides to support three storage formats---column-major, row-major, and so-called generalized storage (where neither the row stride nor column stride is unit).
} $ X $;
and
{\tt precAB} indicates the precision in which the matrix multiplication takes place (possibly implying promotion or demotion from the storage precision of either $ A $ or $ B $).
Note that column-major order and row-major order are the special cases where {\tt rstrideX} is unit and {\tt cstrideX} is unit, respectively.

The hypothetical API shown in Figure~\ref{fig:hypoapi} plainly exposes all of the dimensions of functionality along which the library developer must provide implementations.
However, we feel that such an interface is not very useful towards {\em inspiring} those implementations, as it subtly nudges the developer towards extending the use of separate interfaces for each parameter combination down the function stack, leading to solutions that are vertically siloed from each other, even if various subsets share many similarities.

BLIS preempted this problem by starting with an object-based foundation for encoding and expressing matrix (and vector) operands.
Each linear algebra entity (such as a matrix or vector) is encapsulated within a data structure, or more specifically, a custom \ttstruct type.
For example, BLIS currently exports the following object-based function prototype for invoking the \gemm operation:

\begin{figure}[h!]
\begin{center}
\begin{minipage}{5.27in}
\begin{verbatim}
void bli_gemm( obj_t* alpha, obj_t* a, obj_t* b, obj_t* beta, obj_t* c );
\end{verbatim}
\end{minipage}
\end{center}
\caption{
Function for computing \gemm provided by the object-based API in the BLIS framework.
}
\label{fig:gemm_cur}
\end{figure}

\noindent
The function \ttbligemm takes five arguments of type {\tt obj\_t*}, each of which corresponds to the address of a \ttstruct representing the floating-point operands traditionally passed into the \gemm operation.
The function exposes no other arguments, because all of the conventional parameters (such as {\tt transA} and {\tt transB}) may be interpreted as properties of one of the floating-point operands.

A simplified version of the \ttobjt type definition may be given as:

\begin{figure}[h!]
\begin{center}
\begin{minipage}{2.75in}
\begin{verbatim}
typedef struct obj_s
{
    dim_t     offset_m, offset_n;
    dim_t     dim_m, dim_n;
    inc_t     rstride, cstride;
    doff_t    diag_off;
    siz_t     elem_size;
    objbits_t info;
    char*     buffer;
    // Other fields as necessary...
} obj_t;
\end{verbatim}
\end{minipage}
\end{center}
\caption{
A simplified definition for the \ttstruct underlying the \ttobjt type used within BLIS.
}
\label{fig:obj_t}
\end{figure}

\noindent
Here, {\tt dim\_t}, {\tt inc\_t}, {\tt doff\_t}, {\tt siz\_t}, and {\tt objbits\_t} represent various integer types defined within BLIS for representing dimensions, strides and increments, diagonal offsets, byte sizes, and object property bitfields, respectively.
The idea behind the \ttstruct example in Figure~\ref{fig:obj_t} is that matrices may be represented by a collection of properties, or metadata, and that these properties may be set---for example, when the object is initialized and its underlying data buffer is allocated---and then subsequently queried or modified using a collection of object-based accessor functions.
Encapsulating matrix properties within objects helps hide details that need not be exposed at certain levels of the implementation.

The key observation to make now is that the {\tt domainX} and {\tt precX} arguments shown in Figure~\ref{fig:hypoapi} can be completely hidden within the object API of BLIS.
Indeed, the current definition of \ttobjt within the framework {\em already} includes domain and precision bits within the {\tt info} bitfield.
We only need to add an additional parameter, or designate additional bits within the {\tt info} property, to support the computation precision (labeled in Figure~\ref{fig:hypoapi} as {\tt precAB}).
Thus, it is possible to add mixed-datatype support to \gemm without any modification to the function interface to \ttbligemmns.

A more thorough walkthrough of BLIS's object API is well beyond the scope of this article.%
\footnote{
Curious readers may find tutorial-like example codes included alongside the BLIS source code, which is primarily distributed via GitHub~\cite{blisweb}.
Markdown documentation is also made available, and may be conveniently rendered via GitHub with modern web browsers.
}
The main takeaway from this discussion is that the original author of BLIS designed the framework around an object-based core with the keen understanding that additional APIs of arbitrary format, including (but not limited to) those in the style of the BLAS, could always be layered on top of this more general abstraction.
Consequently, any such APIs built above and beyond the underlying object layer are only incidental to the framework;
they merely constitute syntactic re-expressions of some subset of the functionality made possible by the object foundation.%
\footnote{
The BLAS API provided by BLIS serves as a classic example of this kind of layering, as it builds on the object API to arrive an an interface that exactly mimics the BLAS, even if doing so precludes access to functionality and features that would otherwise be available.
}

\section{Supporting Mixed Domain Computation}
\label{sec:mixeddom}
We consider the storage domain (real or complex) of the matrix to be orthogonal to the storage precision (single, double, etc).
In this section, we consider how to handle mixing matrix operands of different domains.
For now, the reader should assume that the storage precision is held constant across all matrix operands, and therefore can be ignored.
We also ignore scalars $ \alpha $ and $ \beta $ for the time being, which simplifies the general matrix multiplication operation to $ C := A B + C $.

\subsection{The 1m method}
\label{sec:1m}

The author of~\cite{BLIS6} recently presented a novel method of computing complex matrix multiplication without relying upon kernels that explicitly perform complex arithmetic at the scalar level, as is typically the case in high-performance BLAS libraries.
Instead, the so-called \onem method relies {\em only} upon matrix primitives (kernels) that compute real matrix multiplication.
And unlike the older and more easily understood \fourm method~\cite{BLIS5}, \onem replaces each complex matrix multiplication with only a {\em single} real matrix multiplication.

The key to \onem is a pair of special packing formats, which Van Zee denotes \onee and \onerns.
The author illustrates the role of these two packing formats using the following example of complex matrix multiplication $ C \updates A B $ where $ m = 3 $, $ n = 4$, and $ k = 2 $.
\renewcommand{\arraystretch}{1.2}
\begin{align}
\left( 
\begin{array}{cccc}
\real{\gamma_{00}} & \real{\gamma_{01}} & \real{\gamma_{02}} & \real{\gamma_{03}} \\
\imag{\gamma_{00}} & \imag{\gamma_{01}} & \imag{\gamma_{02}} & \imag{\gamma_{03}} \\
\real{\gamma_{10}} & \real{\gamma_{11}} & \real{\gamma_{12}} & \real{\gamma_{13}} \\
\imag{\gamma_{10}} & \imag{\gamma_{11}} & \imag{\gamma_{12}} & \imag{\gamma_{13}} \\
\real{\gamma_{20}} & \real{\gamma_{21}} & \real{\gamma_{22}} & \real{\gamma_{23}} \\
\imag{\gamma_{20}} & \imag{\gamma_{21}} & \imag{\gamma_{22}} & \imag{\gamma_{23}} \\
\end{array} 
\right)
&
\updates
\left(
\begin{array}{crcr}
\real{\alpha_{00}} & -\imag{\alpha_{00}} & \real{\alpha_{01}} & -\imag{\alpha_{01}} \\
\imag{\alpha_{00}} &  \real{\alpha_{00}} & \imag{\alpha_{01}} &  \real{\alpha_{01}} \\
\real{\alpha_{10}} & -\imag{\alpha_{10}} & \real{\alpha_{11}} & -\imag{\alpha_{11}} \\
\imag{\alpha_{10}} &  \real{\alpha_{10}} & \imag{\alpha_{11}} &  \real{\alpha_{11}} \\
\real{\alpha_{20}} & -\imag{\alpha_{20}} & \real{\alpha_{21}} & -\imag{\alpha_{21}} \\
\imag{\alpha_{20}} &  \real{\alpha_{20}} & \imag{\alpha_{21}} &  \real{\alpha_{21}} \\
\end{array} 
\right)
\left(
\begin{array}{cccc}
\real{\beta_{00}} & \real{\beta_{01}} & \real{\beta_{02}} & \real{\beta_{03}} \\
\imag{\beta_{00}} & \imag{\beta_{01}} & \imag{\beta_{02}} & \imag{\beta_{03}} \\
\real{\beta_{10}} & \real{\beta_{11}} & \real{\beta_{12}} & \real{\beta_{13}} \\
\imag{\beta_{10}} & \imag{\beta_{11}} & \imag{\beta_{12}} & \imag{\beta_{13}} \\
\end{array} 
\right)
\label{eq:1mc}
\end{align}
In this example, it is assumed that matrix $ C $ is column-stored, which prescribes that the \onee format be applied to $ A $ and the \oner format be applied to $ B $.
This can be confirmed by inspection: applying a real matrix multiplication to the left- and right-hand matrix product operands would not correctly compute the complex matrix multiplication if $ C $ were row-stored because the intermediate elements of $ A B $ would update the wrong elements of $ C $.
However, \onem provides a cure to this situation. Namely, symmetry in the method allows for a row-oriented variant where $ C $ is row-stored.
\renewcommand{\arraystretch}{1.2}
\begin{align}
\left( 
\begin{array}{cccccc}
\real{\gamma_{00}} & \imag{\gamma_{00}} & \real{\gamma_{01}} & \imag{\gamma_{01}} & \real{\gamma_{02}} & \imag{\gamma_{02}} \\
\real{\gamma_{10}} & \imag{\gamma_{10}} & \real{\gamma_{11}} & \imag{\gamma_{11}} & \real{\gamma_{12}} & \imag{\gamma_{12}} \\
\real{\gamma_{20}} & \imag{\gamma_{20}} & \real{\gamma_{21}} & \imag{\gamma_{21}} & \real{\gamma_{22}} & \imag{\gamma_{22}} \\
\real{\gamma_{30}} & \imag{\gamma_{30}} & \real{\gamma_{31}} & \imag{\gamma_{31}} & \real{\gamma_{32}} & \imag{\gamma_{32}} \\
\end{array} 
\right)
&
\updates
\left(
\begin{array}{cccc}
\real{\alpha_{00}} & \imag{\alpha_{00}} & \real{\alpha_{01}} & \imag{\alpha_{01}} \\
\real{\alpha_{10}} & \imag{\alpha_{10}} & \real{\alpha_{11}} & \imag{\alpha_{11}} \\
\real{\alpha_{20}} & \imag{\alpha_{20}} & \real{\alpha_{21}} & \imag{\alpha_{21}} \\
\real{\alpha_{30}} & \imag{\alpha_{30}} & \real{\alpha_{31}} & \imag{\alpha_{31}} \\
\end{array} 
\right)
\left(
\begin{array}{rcrcrc}
 \real{\beta_{00}} & \imag{\beta_{00}} & \real{\beta_{01}} & \imag{\beta_{01}} & \real{\beta_{02}} & \imag{\beta_{02}} \\
-\imag{\beta_{00}} & \real{\beta_{00}} &-\imag{\beta_{01}} & \real{\beta_{01}} &-\imag{\beta_{02}} & \real{\beta_{02}} \\
 \real{\beta_{10}} & \imag{\beta_{10}} & \real{\beta_{11}} & \imag{\beta_{11}} & \real{\beta_{12}} & \imag{\beta_{12}} \\
-\imag{\beta_{10}} & \real{\beta_{10}} &-\imag{\beta_{11}} & \real{\beta_{11}} &-\imag{\beta_{12}} & \real{\beta_{12}} \\
\end{array} 
\right)
\label{eq:1mr}
\end{align}
In this case, the application of the packing formats is reversed, such that \onee is applied to $ B $ and \oner is applied to $ A $.
Van Zee later points out that it is not actually the storage of $ C $ that determines whether Eq.~\ref{eq:1mc} or~\ref{eq:1mr} is employed, but rather the SIMD register orientation of the underlying real domain microkernel---which determines the input/output preference of the microtile and thus the natural method of performing SIMD reads and write instructions on $ C $.

While the \onem method was originally articulated as a way to implement complex domain matrix multiplication on operands of identical datatypes, we will soon show that not only can it be extended to mixed-precision complex domain computation, but it also indirectly supports a particular combination of mixed-domain operands.

\subsection{Enumerating the cases}
\label{sec:cases}

Consider for simplicity $ C := A B + C $, ignoring the scaling factors $ \alpha $ and $ \beta $.
Each of $ \{ A, B, C \} $ can be stored in either the real or complex domains, leading to $ 2^3 = 8 $ different combinations.
Figure~\ref{fig:mixeddom} enumerates and names each possible case.
We will now discuss each case, how it is interpreted, and how it is implemented within the BLIS framework.

\newcommand{\descwidth}{3.97in}
\begin{figure}[t!]
\hspace{2mm}
\begin{center}
\begin{tabular}{| c | c | c | c I p{\descwidth} |} \hline
Case & $ C $ & $ A $ & $ B $ &
\begin{minipage}[c][6mm][c]{1.0in}
Description/Approach
\end{minipage}
\\ \whline
\zero &
$ \Rdom $ & $ \Rdom $ & $ \Rdom $ &
\begin{minipage}[c][9mm][c]{\descwidth}
Supported by the original framework. Performs $ 2mnk $ flops.
\end{minipage}
\\ \hline
\oneb &
$ \Rdom $ & $ \Rdom $ & $ \Cdom $ & 
\begin{minipage}[c][9mm][c]{\descwidth}
Interpreted as $ C := A \calRe( B ) + C $. Ignore $ \calIm(B) $ and compute as Case 0. Performs $ 2mnk $ flops.
\end{minipage}
\\ \hline
\onea &
$ \Rdom $ & $ \Cdom $ & $ \Rdom $ & 
\begin{minipage}[c][9mm][c]{\descwidth}
Interpreted as $ C := \calRe( A ) B + C $. Ignore $ \calIm(A) $ and compute as Case 0. Performs $ 2mnk $ flops.
\end{minipage}
\\ \hline
\twoab &
$ \Rdom $ & $ \Cdom $ & $ \Cdom $ & 
\begin{minipage}[c][9mm][c]{\descwidth}
Interpreted as $ C := \calRe( A B ) + C $. Use {\sc 1r} packing format from {\sc 1m} method to compute only $ \calRe( A B ) $. Performs $ 4mnk $ flops. 
\end{minipage}
\\ \hline
\onec &
$ \Cdom $ & $ \Rdom $ & $ \Rdom $ & 
\begin{minipage}[c][9mm][c]{\descwidth}
Compute $ A B $ and accumulate into $ \calRe(C) $. Performs $ 2mnk $ flops.
\end{minipage}
\\ \hline
\twobc &
$ \Cdom $ & $ \Rdom $ & $ \Cdom $ & 
\begin{minipage}[c][17mm][c]{\descwidth}
Compute as if $ A \in \Cdom $, but avoid all computations with $ \calIm(A) $. Represent $ C $ and $ B $ with real and imaginary elements indistinguishable within a $ m \times 2n $ and $ k \times 2n $ real matrices, respectively. Requires a row-preferential microkernel. Performs $ 4mnk $ flops.
\end{minipage}
\\ \hline
\twoac &
$ \Cdom $ & $ \Cdom $ & $ \Rdom $ & 
\begin{minipage}[c][21mm][c]{\descwidth}
Compute as if $ B \in \Cdom $, but avoid all computations with $ \calIm(B) $. Represent $ C $ and $ A $ with real and imaginary elements indistinguishable within $ 2m \times n $ and $ 2m \times k $ real matrices, respectively. Requires a column-preferential microkernel. Performs $ 4mnk $ flops.
\end{minipage}
\\ \hline
\three &
$ \Cdom $ & $ \Cdom $ & $ \Cdom $ &
\begin{minipage}[c][13mm][c]{\descwidth}
Supported by the original framework (via the the {\sc 1m} method and/or assembly-coded complex microkernels). Performs $ 8mnk $ flops.
\end{minipage}
\\ \hline
\end{tabular}
\end{center}
\caption{
A table summarizing the eight possible cases of mixed-domain computation within the \gemm operation.
The first column identifies a name for each case, with the number identifying the number of complex matrix operands and the letters identifying the matrices that are complex.
The second, third, and fourth columns explicitly identify the domains of each matrix operand.
The last column describes the interpretation of each case within BLIS along a brief comment on how the case is implemented (where applicable) and a (minimum) flop count when implemented optimally.
}
\label{fig:mixeddom}
\end{figure}

\subsubsection{Cases \zerons, \threens}

The trivial case where all matrices are stored in the real domain, which we refer to as Case 0, is already supported by the framework via algorithms based on real domain microkernels.
Similarly, Case 3, which applies when all matrices are stored in the complex domain, is also already supported.
Support for Case 3 is provided in BLIS via conventional algorithms based on complex domain microkernels as well as via the \onem method, which is particularly useful when complex microkernels are not available.
Cases 0 and 3 incur $ 2mnk $ and $ 8mnk $ flops, respectively.

\subsubsection{Cases \oneans, \onebns}

Case \onea captures situations where $ C $ and $ B $ are real while $ A $ is complex.
We interpret such an operation as $ C := \calRe( A ) B + C $.
Implementing this case in BLIS is rather straightforward: we ignore the imaginary part of $ A $ and compute as if all matrices were real.
Because BLIS tracks both row and column strides for each matrix operand, ignoring the imaginary elements amounts to a temporary change to the dimension and stride metadata contained within the object representing $ A $.
Case \oneb involves a complex matrix $ B $ and real matrices $ C $ and $ A $, but is otherwise handled similarly.
Since these case are ultimately implemented in terms of Case 0, they both performs $ 2mnk $ flops.

\subsubsection{Case \twoabns}

Case \twoab is applicable when $ A $ and $ B $ are complex while the matrix to which they accumulate, $ C $, is real.
We interpret this somewhat curious scenario as a matrix product that takes place in the complex domain, but one for which the imaginary result is discarded: $ C := \calRe( A B ) + C $.
Since $ \calIm(AB) $ is not needed, only $ 4mnk $ flops need to be performed.
Thus, this case provides an opportunity for computational savings when properly implemented.
BLIS implements \twoab by borrowing the \oner packing format used by the \onem method.%
\footnote{
This is the indirect support alluded to at the conclusion of Section~\ref{sec:1m}.
}
Specifically, BLIS packs {\em both} matrices $ A $ and $ B $ using the \oner format while simultaneously conjugating $ B $. This has the effect of allowing a subsequent real matrix multiplication over the packed matrices to correctly compute only the real half of the complex matrix multiplication update.

\subsubsection{Case \onecns}

The opposite of \twoabns---Case \onecns---refers to settings in which matrix $ C $ is complex while $ A $ and $ B $ are real.
Since the matrix product takes place entirely in the real domain, the natural interpretation is that $ A B $ updates only $ \calRe( C ) $, and
$ \calIm( C ) $ is left untouched.
Generally speaking, BLIS implements \onec using a strategy similar to the one used with \onea and \onebns.
That is, a temporary change to the object metadata describing matrix $ C $ allows us to isolate $ \calRe( C ) $, which once again reduces the problem to Case \zerons.
Accordingly, this case requires only $ 2mnk $ flops.

\subsubsection{Cases \twoacns, \twobcns}

Consider Case \twoacns, in which matrices $ C $ and $ A $ are complex and matrix $ B $ is real.
We interpret this situation as performing a complex matrix product $ A B $ to update both real and imaginary parts of $ C $.
However, all computation involving the imaginary part of $ B $, which is implicitly zero, may be ignored.
This means that the computation requires only $ 4mnk $ flops.
Now, if $ C $ and $ A $ were guaranteed to be column-stored, BLIS could handle this case with a simple change of metadata that recasts those complex matrices as real, with the real and imaginary elements treated equally and indistinguishably.
However, BLIS also allows row storage (and general storage) for all matrices, and thus the solution is not quite so simple.
Instead, BLIS handles \twoac as follows.

If the original problem fits into Case \twoac and the microkernel is column-preferential, $ A $ is packed as a real matrix (with imaginary elements stored traditionally, in element-wise interleaved fashion) and $ B $ is packed normally.
A real domain macrokernel (Case \zerons) is then executed, which will properly update $ C $.
However, if the microkernel is row-preferential, the operation is logically transposed and Case \twobc is executed instead (whereby matrices $ C $ and $ B $ are complex and $ A $ is real).
Thus, the effective case employed, \twoac or \twobcns, depends on the register orientation of the microkernel, {\em not} the storage of $ C $, and does so for the same reason that the \onem method, discussed previously in Section~\ref{sec:1m}, depends on the same property.

After the aforementioned logical transposition is applied (or not), the microkernel input/output preference may differ from the storage of matrix $ C $.
If this is the case, then BLIS calls a {\em virtual} microkernel%
\footnote{
In BLIS, virtual microkernels share the same type signature of conventional (``native'') microkernels and ultimately compute the same operation.
The only difference is that virtual microkernels typically implement the microkernel operation in the form of additional logic before (and sometimes after) the call to the native microkernel.
Sometimes, as with the \onem method, the native microkernel being called is the real-domain equivalent relative to the virtual microkernel's type signature (e.g., a virtual \ttzgemm microkernel implemented in terms of the native \ttdgemm microkernel).
Thus, the term is somewhat general-purpose and additional context is needed to identify its specific nature.
}, instead of calling the microkernel directly, allowing for logic that will use a very small amount of temporary storage---equivalent to one microtile---in order to facilitate the proper use of the microkernel (whether it be row- or column-preferential) and then copy and/or accumulate the temporary microtile result back to the appropriate location within the output matrix $ C $.

\subsection{Computation domain}

Unlike the computation precision, which is discussed in the next section, the computation domain is implied according to the case-specific interpretations covered in Section~\ref{sec:cases} and summarized in Figure~\ref{fig:mixeddom}.
Alternate interpretations exist, however.
For example, the computation of Case \twoab could be interpreted as
taking place in the real domain, which would result in $ \calIm( A ) $ and $ \calIm( B ) $ being ignored.
Similar interpretations---all of which would change the update to $ C $---could be applied to Cases \twoacns, \twobcns, or even \threens.
However, we do not immediately see significant utility in exposing these cases.%
\footnote{
If an intrepid user wished to access such functionality, it could easily be done currently with BLIS by manipulating the matrix object metadata accordingly.
Of course, if users show interest in this functionality, we will reconsider official support within the framework.
}
Thus, our implementation in BLIS does not presently allow the caller to explicitly specify the computation domain.


\section{Supporting Mixed Precision Computation}
\label{sec:mixedprec}

Now that variation among the storage domains has been fully explored, we turn our attention to the storage precision of the matrices $ A $, $ B $, and $ C $.
Once again, we set aside the scalars $ \alpha $ and $ \beta $, focusing only on variation among matrix operands.
We also limit the initial discussion to variation within the set of precisions that includes only single and double precision.

\subsection{Supporting all cases}

Each of $ \{ A, B, C \} $ can be stored in single or double precision.
Furthermore, we define a separate {\em computation} precision to identify the precision in which the matrix product takes place.
Combining the storage precision of each matrix with the computation precision, we find that there exist $ 2^4 = 16 $ different cases.

At first glance, it may seem worthwhile to enumerate all mixed-precision cases as we did for mixed-domain computation in Section~\ref{sec:mixeddom}.
However, there is a more concise and systematic way of describing how to support all cases, one that happens to coincide closely with how mixed-precision support was ultimately implemented in BLIS. 

While not part of the runtime logic for implementing mixed-precision computation, we must first modify the packing facility so that the source and destination precisions may differ.
For example, we must be able to pack from a single-precision real matrix to a double-precision real matrix, or vice versa.
Once the mixed-precision functionality is in place for the packing operation, the runtime logic may be encoded in three broad steps, as follows.

\subsubsection{Identify the computation precision}
\label{sec:prec1}

First, we must identify the computation precision.
In BLIS, we provide a field for the computation precision within the metadata of any matrix object.
However, semantically, we deem only the field within the object for matrix $ C $ to be relevant.
Thus, upon calling the \gemm operation, we query the computation precision from $ C $.
Note that in BLIS, when objects are created, the computation precision is initialized to be equal to the object's storage precision.
Consequently, if it is not explicitly set by the caller prior to invoking \gemmns, the computation precision will automatically default to the storage precision of $ C $.

\subsubsection{Construct the target datatypes for $ A $ and $ B $}
\label{sec:prec2}

Next, we embed {\em target} datatypes within the metadata for objects representing matrices $ A $ and $ B $.
BLIS defines the target datatype for $ A $ and $ B $ as being the storage datatype (domain and precision) of the matrix during its packed state---in other words, the datatype to which the matrix must be typecast before it can be computed upon.%
\footnote{
Note that we do not need to track the target datatype for $ C $ since the storage datatype of $ C $ does not change in the course of the mixed-datatype \gemm operation.
}
The target datatype for matrix $ A $ is constructed by combining the storage domain of $ A $ with the operation's computation precision, with a similar process resulting in the target datatype for matrix $ B $.
The target datatype for each matrix is then embedded within the metadata of the corresponding object.

\subsubsection{Determine whether typecasting is needed on accumulation}
\label{sec:prec3}

If the computation precision differs from the storage precision of $ C $, then the intermediate result from computing the product $ A B $ must be typecast before it is accumulated into $ C $.
This may be implemented outside the microkernel by implementing additional macrokernels that write the microkernel result to a temporary microtile (allocated on the function stack) before typecasting and accumulating the temporary values back to $ C $.
Alternatively, this logic may hidden within a virtual microkernel.
BLIS opts for the former solution, which somewhat reduces function call overhead at the cost of a somewhat higher
object (binary) code footprint.

If the computation precision is identical to the storage precision of $ C $, then $ A B $ does not require any typecasting before being accumulated.
This corresponds to use of the traditional macrokernel.

\subsection{Using the \onem method for Case 3}

As alluded to in the closing of Section~\ref{sec:1m}, the \onem method can be extended to encompass all combinations of mixed-precision operands---that is, all precision combinations that fall within mixed-domain Case \threens.
This amounts primarily to (1) adding the ability to pack to the \onee and \onerns%
\footnote{
The ability to typecast while packing to the \oner format is also required by mixed-precision instances of Case \twoabns.
}
formats when the target datatype differs from the storage datatype
and (2) adding the ability to scale by $ \alpha $ (when $ \imag\alpha \ne 0 $) during mixed-precision packing of $ A $ and $ B $.
With these changes in place, the \onem virtual microkernel may be used as-is since its functioning is undisturbed by the typecasting logic encoded in the additional macrokernels mentioned previously in Section~\ref{sec:prec3}.%
\footnote{
In the course of our work, the BLIS testsuite was updated to allow testing of its \onem method implementation with mixed-precision operands.
}

\subsection{Summary}

By addressing the three areas described in Sections~\ref{sec:prec1} through~\ref{sec:prec3}, and adding support for typecasting within the packing function, we can handle all 16 cases of mixing single and double precisions.
Similarly, relatively minor changes to BLIS's implementation of the \onem method enable all mixed-precision instances of Case \three to be handled even if conventional, assembly-coded complex microkernels are unavailable.

Interestingly, the hypothetical impact of adding support for an additional floating-point precision, such as half-precision, would manifest almost exclusively in the form of additional support to the packing function.%
\footnote{
A real domain microkernel that performs computations in the new precision would also be needed.
}
The mixed-precision runtime logic, described above, would then trivially extend to support the two additional datatypes (one each for real and complex domains).


\section{Optimizations}
\label{sec:opts}

After retrofitting the mixed-domain/mixed-precision functionality into the \gemm implementation, it is possible to apply various optimizations to certain cases.
In this section, we briefly explore some of these optimizations.

\subsection{Avoid virtual microkernel overhead with two microkernels} 
\label{sec:opt1}

If and when BLIS adopts a regime whereby each hardware architecture is supported simultaneously by {\em two} microkernels per datatype, one with a row preference and one with a column preference, then the virtual microkernel logic becomes unnecessary, and both \twoac and \twobc may be fully implemented without additional overhead.

\subsection{Avoid non-contiguous/non-SIMD access on $ C $}
\label{sec:opt2}

Some mixed-domain cases---at least as described in Section~\ref{sec:cases}---result in accessing complex matrix $ C $ by individual real and imaginary elements.
The best example of this is Case \onecns, in which $ \calRe( C ) $ is updated by the product of real matrices $ A $ and $ B $.
However, in order to isolate $ \calRe( C ) $, the metadata describing matrix $ C $ must be tweaked in such a way that the matrix then has non-unit stride in both dimensions.
While BLIS microkernels may update such matrices, doing so on current hardware comes with an unavoidable performance penalty that does not manifest when accessing contiguous elements via SIMD load and store instructions.
Thus, it may be advantageous to internally allocate a temporary $ m \times n $ matrix $ C_{temp} $ in which to compute $ A B $, after which the result is copied back to $ C $.
As long as $ C_{temp} $ is created (a) in the real domain and (b) as either row- or column-stored, the microkernel will be able to update $ C_{temp} $ efficiently with SIMD instructions.

This optimization tends to be most worthwhile when $ k > k_C $, as it would imply that the computation of $ A B $ unfolds as multiple rank-$ k_C $ updates of $ C $ (that is, multiple iterations of the 4th loop around the microkernel), with the non-contiguous load/store penalty otherwise being incurred for each update.

\subsection{Reduce typecasting costs and round-off error accumulation}
\label{sec:opt3}

When the storage precision of $ C $ differs from the computation precision, the \gemm implementation executes typecasting instructions emitted by the compiler in order to properly convert from one floating-point datatype to another---single- to double-precision or double- to single-precision.
These instructions can be costly, especially when $ k > k_C $, since each element of $ C $ is typecast once per rank-$ k_C $ update.
In addition to the performance cost of these typecasting instructions, they can also incur a numerical cost.
Specifically, when the storage precision of $ C $ is lower than the computation precision, repeated round-off error can occur during accumulation of the intermediate matrix product, which is once again exacerbated for large values of $ k $.
But as with the repeated cost incurred from non-contiguous access described in Section~\ref{sec:opt2}, these costs can be avoided by allocating a temporary $ m \times n $ matrix $ C_{temp} $.
The key difference is that, in this case, $ C_{temp} $ is created with its storage precision equal to the computation precision, which will avoid intermediate typecasting (and thus reduce round-off error), leaving only typecasts on input ($ C_{temp} \becomes C $) and on output ($ C \becomes C_{temp} $).
The total number of typecasts needed may be further reduced to those on output provided that $ C_{temp} $ is initialized to zero and the final $ A B $ product be accumulated, rather than copied, back to $ C $. 

\subsection{Avoid virtual microkernel overhead with $ C_{temp} $}
\label{sec:opt4}

Notice that Cases \twoac and \twobcns, as described in Section~\ref{sec:cases}, may require use of a virtual microkernel if a row-preferential microkernel (needed by \twobcns) must be used on a column-stored (or general stored) matrix, or if a column-preferential microkernel (needed by \twoacns) must be used on a row-stored (or general stored) matrix.
The overhead of the virtual microkernel, while small, may still be noticeable and is incurred for each rank-$ k_C $ update.
The dual-microkernel strategy described in Section~\ref{sec:opt1} solves this issue.
However, if only a single microkernel (per datatype) is available, a temporary $ m \times n $ matrix $ C_{temp} $ may be allocated, with the important distinction being that $ C_{temp} $ is created with storage (by columns or rows) to match the preference of the available microkernel.
This allows the implementation to avoid the virtual microkernel altogether during intermediate accumulation into $ C_{temp} $.

\subsection{Summary}

The optimizations described above are optional.
At the time of this writing, BLIS implements all except the dual-microkernel strategy described in Section~\ref{sec:opt1}.
BLIS also allows the the user to optionally disable all uses of $ C_{temp} $ at configure-time, which avoids the extra workspace that would otherwise be needed by the optimizations discussed in Sections~\ref{sec:opt2} through~\ref{sec:opt4}.


\section{Handling scalars}

Before concluding our discussion of how to implement and support mixed-datatype \gemmns, we turn our attention to scalars $ \alpha $ and $ \beta $, which have been omitted from our discussion thus far.

\subsection{Mixed precision}

If the precision of $ \alpha $ differs from the computation precision, a decision must be made as to how to proceed.
Numerous possible policies exist for handling such situations.
Three examples follow:
\begin{enumerate}
\item Typecast $ \alpha $ to match the computation precision.
\item Typecast the computation precision to match that of $ \alpha $.
\item Unconditionally promote the lower precision value to the precision of the other (higher precision) value. 
\end{enumerate}
Our mixed-datatype extension to BLIS currently opts for option (1).

A similar decision must be made for handling the precision of $ \beta $.
The choices here are even more numerous, because while we consider the precision of $ \beta $ and the storage precision of $ C $ to be the main inputs to the runtime logic, one could also argue for considering the disagreement with the computation precision that governs the computation of $ A B $.
A few possible policies are:
\begin{enumerate}
\item Typecast $ \beta $ to match the storage precision of $ C $.
\item Perform the scaling $ \beta C $ in the higher precision of the two values, typecasting back to the storage precision of $ C $, as necessary.
\item Typecast both $ \beta $ and $ C $ to the computation precision so that all suboperations within $ \beta C + A B $ can occur in the same precision before being typecast back to the storage precision of $ C $.
\end{enumerate}
Once again, our solution in BLIS opts for the relatively simple solution alluded to in (1), with $ \alpha A B $ being typecast to the storage precision on accumulation to $ C $.

\subsection{Mixed domain}

Real values of $ \beta $ are easily handled for all eight cases enumerated in Section~\ref{sec:cases}.

For Cases \zerons, \oneans, \onebns, and \onecns, complex $ \beta $ may be projected into the real domain (which discards $ \calIm(\beta) $ entirely) since, with $ C \in \Rdom $, $ \calIm(\beta) $ cannot change the final result.
Similarly, complex values of $ \beta $ are handled as expected in the four cases where $ C \in \Cdom $.
Specifically, Case \three already handles complex $ \beta $ while Cases \twoabns, \twoacns, and \twobc support $ \calIm(\beta) \ne 0 $ via extra logic in the virtual microkernel.

Real and complex values of $ \alpha $ are already handled in Cases \zero and \threens, respectively.
Real $ \alpha $ are also already handled by Case \three since $ \Rdom \subset \Cdom $.
In Case \zerons, a complex $ \alpha $ can be projected to the real domain since $ \calIm(\alpha) $ would not change the computation, even if we had storage in which to save the final result.

For Cases \oneans, \onebns, \onecns, \twoab, \twoacns, and \twobcns, a non-zero imaginary component in $ \alpha $ could presumably change the final computation under a literal interpretation of the computation---that is, one in which all five operands' domains are taken at face value.
However, implementing this logic is non-trival.
For example, consider our approach to handling Case \oneans, as discussed in Section~\ref{sec:cases}.
In this case, we perform the computation according to Case \zerons, as if the imaginary components of $ A $ were zero.
That case's handling is completely consistent with its mathematics, since in that scenario $ A $ is the only operand with non-zero imaginary values, and thus they would have no impact on the final result.
However, if both $ A $ {\em and} $ \alpha $ are complex, then the imaginary components could combine to change the real component of the scalar-matrix product $ \alpha A $.
Adjusting for this new possible use case would require a different approach in the implementation, perhaps using temporary workspace to store a copy of $ A $ while it is scaled by $ \alpha $, after which the imaginary components may be ignored (assuming they were even computed to begin with).%
\footnote{
Alternatively (and more preferably), logic that packs $ \calRe( \alpha A ) $ where $ \alpha, A \in \Cdom $ may be encoded within a special packing function.
}
However, while it is clear that going through such motions would maintain deeper fidelity to the literal mathematics expressed in the mixed-domain scenario, it's not clear to us that this additional functionality would be vital for most applications.
As we continue to solicit feedback from the community, we will pay close attention to whether users expect or request support for non-zero imaginary values of $ \alpha $ in Cases \oneans, \onebns, \onecns, \twoab, \twoacns, and \twobcns.
For now, our mixed-datatype solution supports only real values of $ \alpha $ for those six cases, and prints an error message if the scalar is given with a non-zero imaginary component.


\section{Performance}
\label{sec:perf}

In this section, we discuss performance results for our mixed-datatype implementations on two servers with modern hardware architectures.

\subsection{Platform and implementation details}

\subsubsection{Marvell ThunderX2}

The first system upon which we measured performance is a single compute node consisting of two 28-core Marvell ThunderX2 CN9975 processors.%
\footnote{
While four-way symmetric multithreading (SMT) was available on
this hardware, the feature was disabled at boot-time so
that the operating system detects only one logical
core per physical core and schedules threads accordingly.
}
Each core, running at a clock rate of 2.2 GHz, provides a single-core
peak performance of 17.6 gigaflops (GFLOPS) in double precision and 35.2 GFLOPS in single precision.
Each of the two sockets has a 32MB L3 cache that is shared among its local cores, and each core has a private 256KB L2 cache and 32KB L1 (data) cache.
The installed operating system was Ubuntu 16.04 running the Linux 4.15.0 kernel.
Source code was compiled by the GNU C compiler ({\tt gcc}) version 7.3.0.%
\footnote{
The following optimization flags were used during compilation on the ThunderX2:
{\tt -O3 -ftree-vectorize -mtune=cortex-a57}.
In addition to those flags, the following flags were also used
when compiling assembly kernels:
{\tt -march=armv8-a+fp+simd -mcpu=cortex-a57}.
}

\subsubsection{Intel Xeon Platinum}

The second system is a single node consisting of two 26-core Intel Xeon Platinum 8167M processors.%
\footnote{
Two-way SMT (which Intel refers to as ``Hyperthreading'') was available. However, we employed processor affinity settings that limited the operating system to utilizing only one logical core per physical core.
}
Each core ran at a clock rate of 2.0 GHz, providing single-core
peak performance of 64 gigaflops (GFLOPS) in double precision and 128 GFLOPS in single precision.
Each of the two sockets has a 35.75MB L3 cache that is shared among its local cores, and each core has a private 1MB L2 cache and 32KB L1 (data) cache.
The installed operating system was Ubuntu 18.04.1 running the Linux 4.15.0 kernel.
Source code was compiled by the GNU C compiler ({\tt gcc}) version 7.3.0.%
\footnote{
The following optimization flags were used during compilation on the Xeon Platinum:
{\tt -O3}.
In addition to those flags, the following flags were also used
when compiling assembly kernels:
{\tt -mavx512f -mavx512dq -mavx512bw -mavx512vl -mfpmath=sse -march=skylake-avx512}.
}

\subsubsection{Implementations}

On both the ThunderX2 and the Xeon Platinum, the version of BLIS used was based on an inter-version release that preceded 0.5.1.%
\footnote{
This version of BLIS may be uniquely identified, with high probability,
by the first 10 digits of its
{\tt git} ``commit'' (SHA1 hash) number: {\tt cbdb0566bf}.
} In both cases, BLIS was configured with OpenMP-based multithreading.
Architecture-specific configuration, which determines settings such as kernel sets and cache blocksizes, was performed automatically via the {\tt auto} target to the {\tt configure} script.

We showcase two (or, in some cases, three) implementations, with a third (or fourth) provided for reference:
\begin{itemize}
\item
{\bf Internal without extra memory.} This refers to the BLIS implementation described in Sections~\ref{sec:mixeddom} and \ref{sec:mixedprec} in which all logic for supporting the mixing of domains and precisions occurs opaquely inside of BLIS.
This implementation does not, however, employ the use of a temporary matrix $ C_{temp} $ discussed in Sections~\ref{sec:opt2}--\ref{sec:opt4}.
\item
{\bf Internal with extra memory.}
This implementation is identical to the previous implementation, except that $ C_{temp} $ is made available, thus incurring extra workspace requirements in certain situations.
It is worth pointing out that this implementation does not differ from its extra-memory-avoiding counterpart for all 128 mixed-datatype cases.
Thus, we will only present these results when one of the conditions laid out in Sections~\ref{sec:opt2}--\ref{sec:opt4} is applicable.
Also, note that we do not employ any cache blocking or parallelism outside of the underlying call to \gemmns, such as when copying/accumulating $ C_{temp} $ to/from $ C $.
\item
{\bf Ad-hoc.} This refers to an implementation that is formulated \emph{outside} of BLIS, using temporary workspace and matrix copies wherever needed.
The purpose behind such an implementation is to show the best a knowledgeable computational scientist could reasonably expect to achieve using only a BLAS library.
Thus, while we link this solution to BLIS, we do so via the framework's BLAS compatibility layer.
Furthermore, we do not employ any cache blocking or parallelism outside of the underlying call to \gemmns.
\item
{\bf Reference.} This refers to the \ttsgemmns, \ttdgemmns, \ttcgemmns, and \ttzgemm provided by BLIS, whichever is appropriate, as determined by the mixed-datatype case presented by each graph.
These Reference curves are provided as a visual ``high-water mark'' to show how much performance is ceded by the Internal and Ad-hoc mixed-datatype implementations.
As of this writing, we had not yet developed native complex domain \gemm microkernels, and therefore the \ttcgemm and \ttzgemm curves represent BLIS implementations based on the \onem method.
\end{itemize}

\subsection{Results}
\label{sec:results}

\subsubsection{Scope and Conventions}

Performance data for sequential, multithreaded within one socket (28 threads), and multithreaded across two sockets (56 threads) was gathered on the ThunderX2.
Similarly, sequential, single-socket (26 threads), and dual-socket (52 threads) data was gathered on the Xeon Platinum.
This produced a rather large set of data, which we formatted into 768 individual graphs.
As a practical matter, we have relegated this complete set of performance results to Online Appendix~\ref{sec:appendix}.
Here, in the main body of the article, we limit our presentation to a slice of data that we feel is broadly representative of the full set.

For any given graph, the $ x $-axis denotes the problem size (where $ m = n = k $), the $ y $-axis shows observed floating-point performance in units of GFLOPS per
core, and the theoretical peak performance of the hardware coincides with the top of the graph.
Problem sizes for sequential instances of \gemm were run from 40 to 2000 in increments of 40 while multithreaded executions were run from 120 to 6000 in increments of 120.%
\footnote{
Some readers may wish that we had run our multithreaded experiments to a somewhat larger maximum problem sizes, perhaps 8,000 or 10,000.
We sympathize with these readers.
However, the results presented in this article, including Online Appendix~\ref{sec:appendix}, required a total of 302,400 invocations of \gemm performed over a period of several days.
Limiting the maximum problem size was necessary so that the experiments would finish in a reasonable amount of time.
}
The data points in all performance graphs report the best
of three trials.

Individual graphs are labeled according to the mixed-datatype case of its Internal and Ad-hoc implementations.
The datatypes are encoded as $ cabx $, where the characters $ c $, $ a $, and $ b $ encode the storage datatypes of $ C $, $ A $, and $ B $, respectively, while $ x $ encodes the computation precision.
For example, a case labeled ``zcsd'' would refer to mixed-domain Case \twoacns, where matrices $ A $ and $ B $ are stored in single-precision (complex and real, respectively), matrix $ C $ is double-precision complex, and the computation occurs in double-precision arithmetic.

All experiments reflect the use of randomized, column-stored
matrices with \gemm scalars $ \alpha = 1 $ and $ \beta = 1 $.

%
%

\newcommand{\graphhspace}{-3.0mm}
\newcommand{\graphwidth}{5.35in}
\newcommand{\trimleft}{3.7cm}
\newcommand{\trimlower}{2.8cm}
\newcommand{\trimright}{3.7cm}
\newcommand{\trimupper}{2.5cm}

\newcommand{\sentencezero}[1]{Sequential (top) and multithreaded with 28 threads (bottom) performance of ``Internal'' and ``Ad-hoc'' implementations of \gemm for select datatype combinations on a Marvell ThunderX2 CN9975 processor. }
\newcommand{\sentenceone}[2]{The 12 graphs on the left side and right sides report computation in single- and double-precision, respectively. }
\newcommand{\sentencetwo}{The theoretical peak performance coincides with the top of each graph. }

\begin{figure}[tp!]
\begin{center}
\begin{tabular}{l}
\hspace{\graphhspace}
\includegraphics[width=\graphwidth,trim={\trimleft, \trimlower, \trimright, \trimupper},clip]{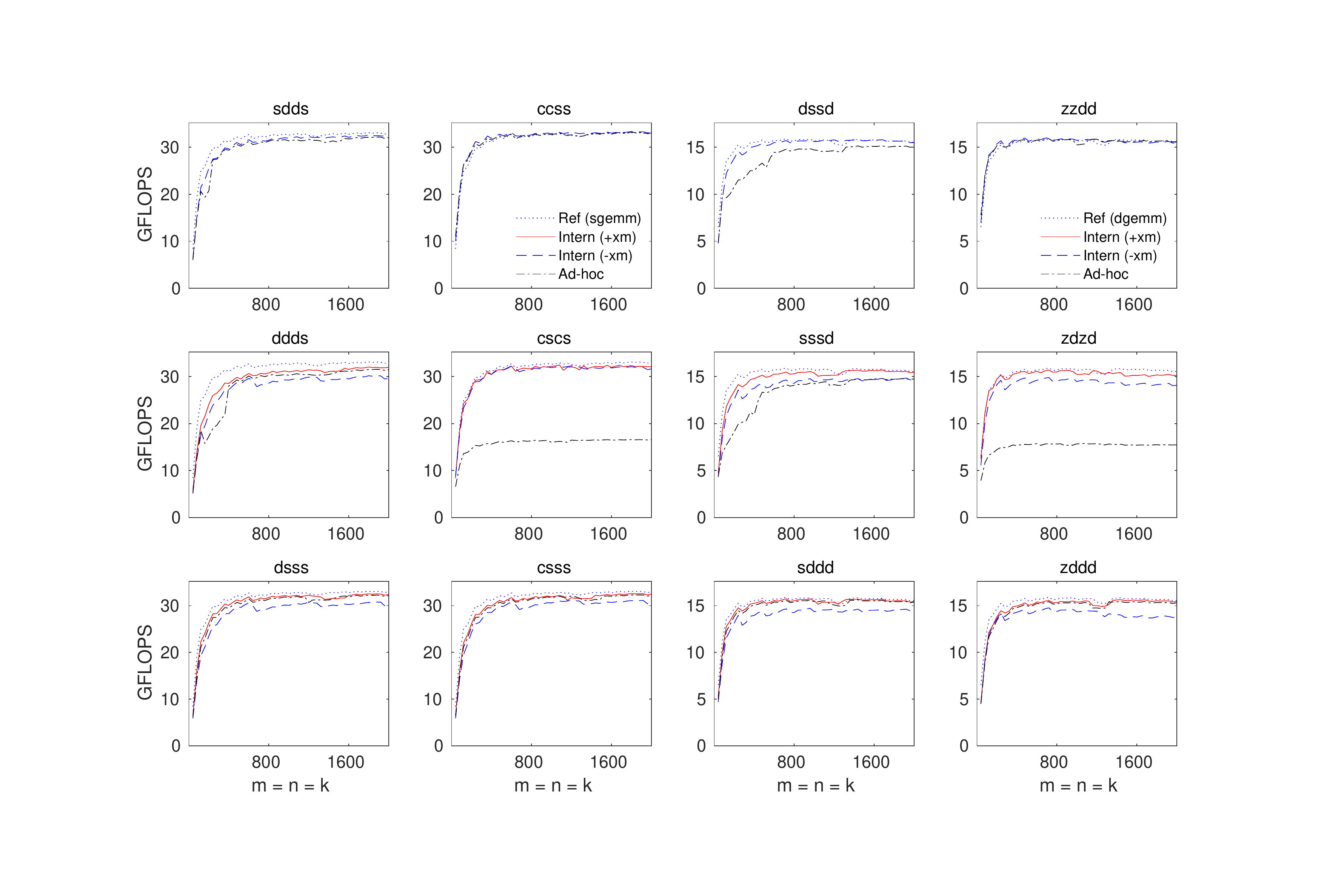} \\ \whline
\hspace{\graphhspace}
\includegraphics[width=\graphwidth,trim={\trimleft, \trimlower, \trimright, \trimupper},clip]{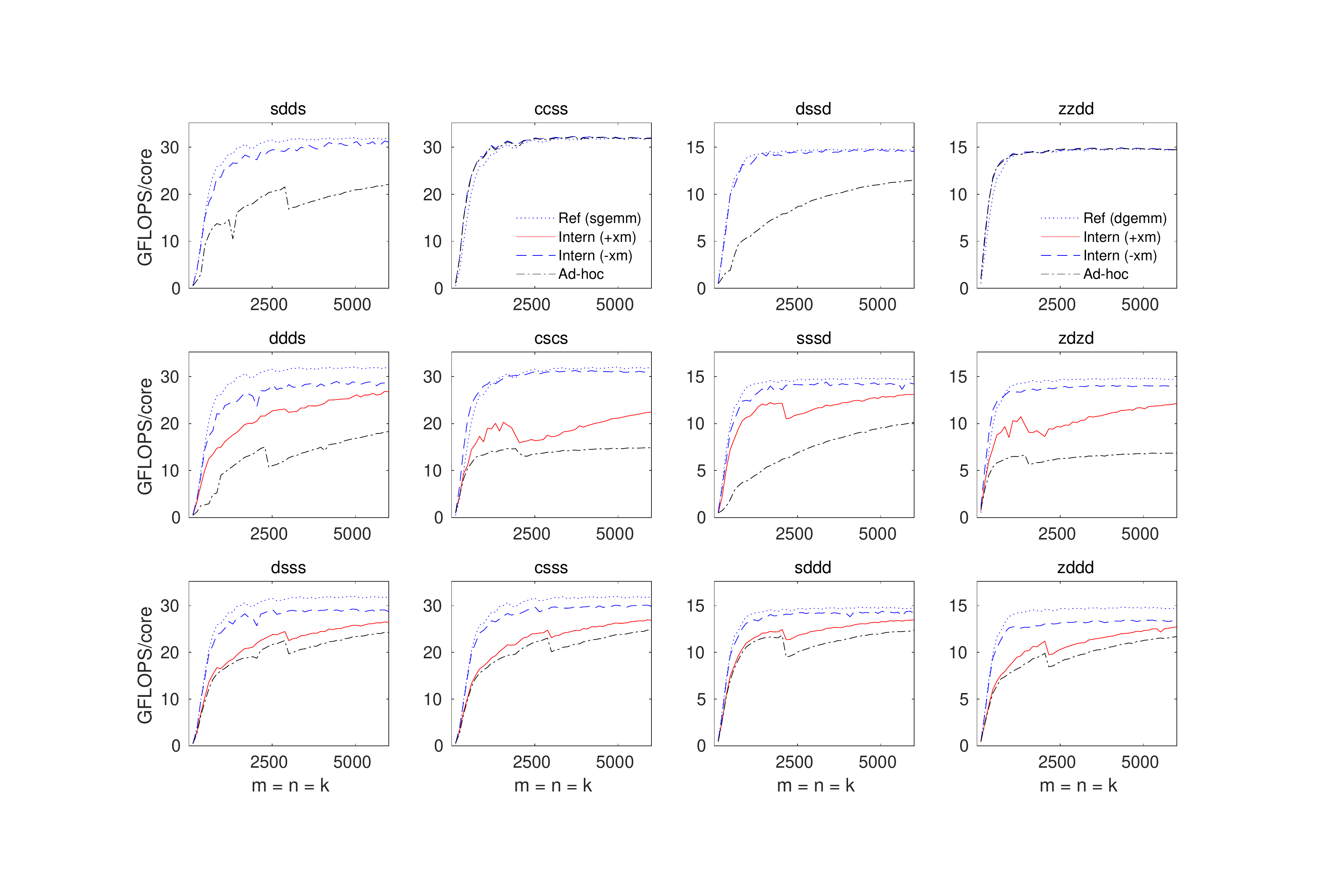}
\end{tabular}
\end{center}
\caption{
\sentencezero{\zerons}
\sentenceone{\ttsgemmns}{\ttdgemmns} \sentencetwo
}
\label{fig:perf_select_tx2}
\end{figure}

%
%

\renewcommand{\graphhspace}{-3.0mm}
\renewcommand{\graphwidth}{5.35in}
\renewcommand{\trimleft}{3.3cm}
\renewcommand{\trimlower}{2.8cm}
\renewcommand{\trimright}{3.7cm}
\renewcommand{\trimupper}{2.5cm}

\renewcommand{\sentencezero}[3]{Sequential (top) and multithreaded with 26 threads (bottom) performance of ``Internal'' and ``Ad-hoc'' implementations of \gemm for select datatype combinations on a Intel Xeon Platinum 8167M processor. }
\renewcommand{\sentenceone}[2]{The 12 graphs on the left side and right sides report computation in single- and double-precision, respectively. }
\renewcommand{\sentencetwo}{The theoretical peak performance coincides with the top of each graph. }

\begin{figure}[tp!]
\begin{center}
\begin{tabular}{l}
\hspace{\graphhspace}
\includegraphics[width=\graphwidth,trim={\trimleft, \trimlower, \trimright, \trimupper},clip]{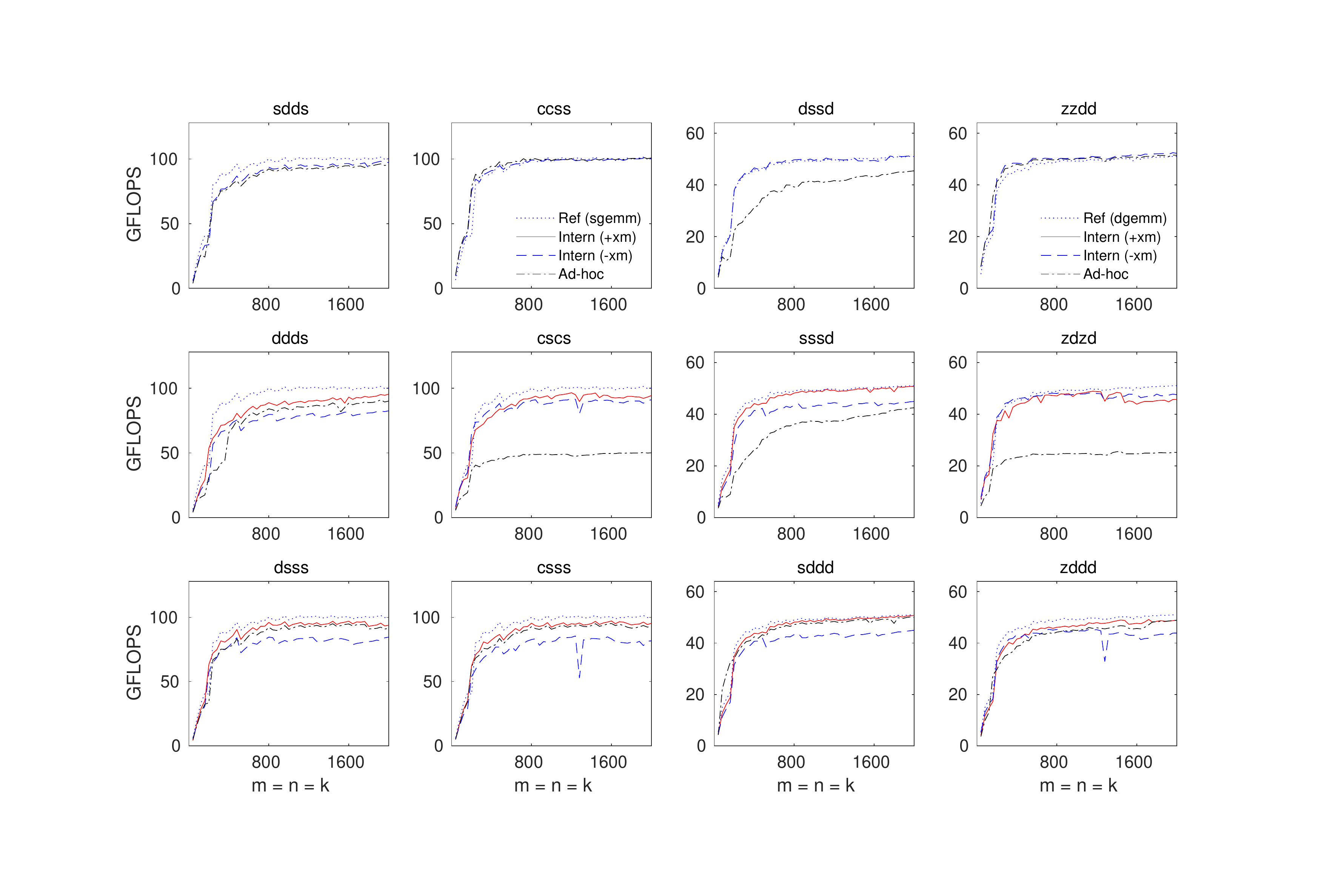} \\ \whline
\hspace{\graphhspace}
\includegraphics[width=\graphwidth,trim={\trimleft, \trimlower, \trimright, \trimupper},clip]{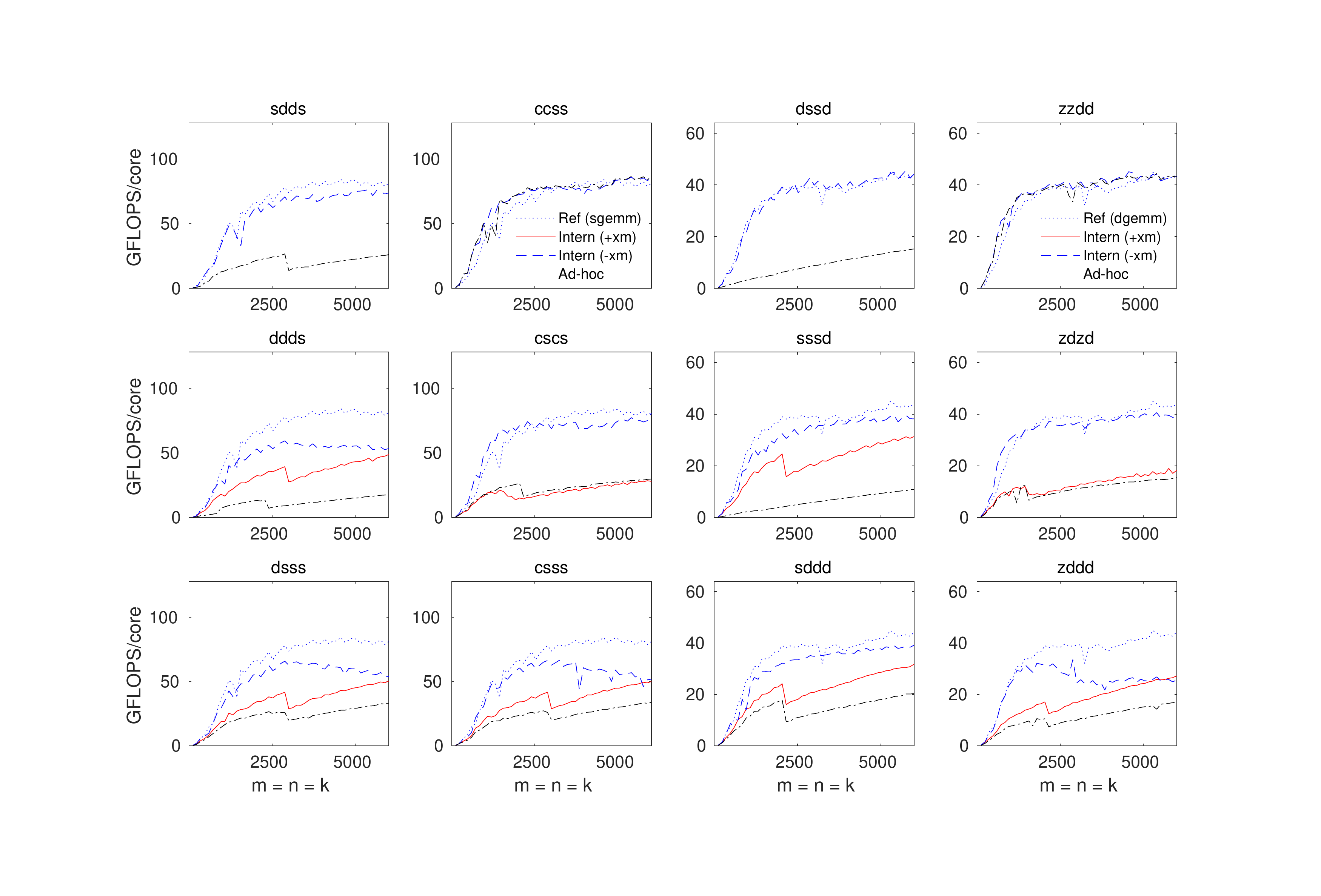}
\end{tabular}
\end{center}
\caption{
\sentencezero{\onecns}{\twoacns}{\twobcns}
\sentenceone{\ttsgemmns}{\ttdgemmns} \sentencetwo
}
\label{fig:perf_select_skx}
\end{figure}

\subsubsection{Exposition}

Figure~\ref{fig:perf_select_tx2} (top) reports sequential performance on the Marvell ThunderX2.
The six graphs on the left half of Figure~\ref{fig:perf_select_tx2} (top) report performance for six select mixed-datatype cases that we felt are interesting: sdds, ddds, dsss, ccss, cscs, and csss.
All of these mixed-datatype cases perform their computation in single-precision.
On the right half, we display graphs that correspond to the precision-toggled analogues of the graphs on the left---dssd, sssd, sddd, zzdd, zdzd, and zddd---all of which perform their computation in double-precision.
Both Internal implementations (with and without extra memory) are shown in four of the six graphs in each group, with the remaining two graphs displaying performance only for the implementation where extra memory is disabled, since the optimization is not applicable for those cases.

The legends are shown once for each group of six graphs.
Within the legend, the curves labeled ``Intern (+xm)'' and ``Intern (-xm)'' refer to the Internal implementations with and without extra memory, respectively.
Additionally, the label for the Reference implementation is augmented, in parentheses, with the conventional \gemm routine that serves as the reference curve within that group of six graphs.

Organized identically as those in the top, the graphs in Figure~\ref{fig:perf_select_tx2} (bottom) report multithreaded performance on one socket (28 threads) of the ThunderX2.

Finally, Figure~\ref{fig:perf_select_skx} (top) and (bottom) report sequential and single-socket (26 threads) performance, respectively, on the Intel Xeon Platinum using the same mixed-datatype cases and organization as shown in Figure~\ref{fig:perf_select_tx2}.

\subsubsection{Observations and Analysis}

Let us turn first to the sequential performance results from the ThunderX2. 

The first thing we notice is that, in five of the six mixed-datatype cases, the Ad-hoc approach is capable of performing quite well relative to the Internal implementations.
The sixth case, which falls within mixed-domain Case \twobcns, suffers because, unlike with \twoacns, we are unable to cast the problem in terms of \ttsgemm or \ttdgemm by manipulating matrix metadata, and therefore must resort to using \ttcgemm and \ttzgemmns.
And because $ \imag{A} = 0 $, this approach necessarily wastes half of the floating-point computations.

Next, we notice that employing $ C_{temp} $ often affords a modest but noticeable increase in performance relative to forgoing extra memory.
This is expected for all of the reasons described in Sections~\ref{sec:opt2}--\ref{sec:opt4}.

Turning to the multithreaded performance on ThunderX2, we notice that the effect of using extra memory in the Internal implementation is not only reversed, but also magnified.
Here, the additional costs incurred within the virtual microkernel are overwhelmed by the cost of the accumulation of $ C_{temp} $ back to $ C $ that must be performed after the \gemm operation, which is not parallelized.%
\footnote{
This copy/accumulation operation, while not parallelized for any of the implementations tested for this article, could in principle be parallelized.
However, speedup for that component of the \gemm would likely be limited as the many threads quickly saturate the available memory bandwidth.
}
The performance of the Ad-hoc implementation also suffers, for similar reasons to that of the Internal implementation using $ C_{temp} $.
The effect is even worse for Ad-hoc, however, because that implementation must make whole copies of matrices up-front, and does so sequentially, before executing the underlying \gemm operation.
By contrast, the Internal implementations benefit from typecasting $ A $ and $ B $ during packing, which is already parallelized.

Overall, the extra-memory-avoiding Internal implementation performs quite well relative to its \ttsgemm and \ttdgemm benchmarks.

Turning to the Intel Xeon Platinum results in Figure~\ref{fig:perf_select_skx}, we find the data largely tells the same story.
Here, the multithreaded performance degradation caused by employing extra memory is even more severe, and the Ad-hoc performance is similarly attenuated.
Once again, in both sequential and multithreaded cases, one of the Internal implementations matches or exceeds (sometimes by a large margin) that of the Ad-hoc approach.
However, for some datatype cases, even the memory-avoiding Internal implementation lags noticeably behind its \ttsgemm or \ttdgemm benchmark.
The cause of this is not immediately clear, but may be related to memory bandwidth becoming strained in cases where the the virtual microkernel is relied upon to update the output matrix in two steps, using a temporary microtile as intermediate storage.

These results strongly suggest that, in general, BLIS should employ the use of Internal implementation with $ C_{temp} $ for sequential invocations of \gemmns, but avoid $ C_{temp} $ in the case of many-threaded execution.
As a function of the number of threads, the crossover point between the two Internal implementations will likely depend on the amount of parallelism that can be extracted within the accumulation of $ C_{temp} $ back to $ C $ before memory bandwidth is saturated.
We leave this topic for future exploration.


\section{Measuring the impact on code size}

Prior to reading the article, a casual reader might have been skeptical of the practicality of our solution. However, Sections~\ref{sec:mixeddom} and~\ref{sec:mixedprec} decompose the problem into mostly orthogonal use cases, giving hope that the ultimate impact on library code size is much more manageable.

Indeed, by multiple measures, the BLIS library grew only modestly after introducing mixed-datatype support.

\begin{figure}[h!]
\hspace{2mm}
\begin{center}
\begin{tabular}{| l I r | r I r | r |} \hline
\begin{minipage}[c][8mm][c]{0.8in}
Framework source code
\end{minipage}
&
Total lines & Change &
\begin{minipage}[c][9mm][c]{0.7in}
Total size \\ (Kilobytes)
\end{minipage}
& Change \\ \whline
Before \tbstrut &
148,862 &
        &
  4,706 &
        \\ \hline
After \tbstrut &
154,962 &
 +6,100 &
  4,892 &
   +186 \\ \hline
\end{tabular}
\end{center}
\caption{
The total number of lines and the total size (in kilobytes) of source code in the BLIS framework (excluding the build system, kernels, and the testsuite) before ({\tt git} commit {\tt 667d3929}) and after ({\tt git} commit {\tt 5fec95b9}) support was added for mixed-datatype computation via the \gemm operation.
}
\label{fig:src_code_frame}
\end{figure}

\begin{figure}[h!]
\hspace{2mm}
\begin{center}
\begin{tabular}{| l I r | r I r | r |} \hline
\begin{minipage}[c][9mm][c]{0.8in}
Testsuite source code
\end{minipage}
&
Total lines & Change &
\begin{minipage}[c][9mm][c]{0.7in}
Total size \\ (Kilobytes)
\end{minipage}
& Change \\ \whline
Before \tbstrut &
 22,891 &
        &
    680 &
        \\ \hline
After \tbstrut &
 24,356 &
 +1,465 &
    722 &
    +42 \\ \hline
\end{tabular}
\end{center}
\caption{
The total number of lines and the total size (in kilobytes) of source code in the BLIS testsuite before ({\tt 667d3929}) and after ({\tt 5fec95b9}) support was added for mixed-datatype computation via the \gemm operation.
}
\label{fig:src_code_testsuite}
\end{figure}

\begin{figure}[h!]
\hspace{2mm}
\begin{center}
\begin{tabular}{| l I r | r I r | r I r | r |} \hline
\begin{minipage}[c][9mm][c]{1.05in}
Object code size \\
(Kilobytes)
\end{minipage}
&
\begin{minipage}[c][9mm][c]{0.45in}
Static library
\end{minipage}
& Change &
\begin{minipage}[c][9mm][c]{0.45in}
Shared library
\end{minipage}
& Change &
\begin{minipage}[c][13mm][c]{0.62in}
Statically-linked testsuite
\end{minipage}
& Change \\ \whline
Before \tbstrut &
  3,141 &
        &
  2,286 &
        &
  1,632 &
        \\ \hline
After (disabled) \tbstrut &
  3,253 &
   +112 &
  2,382 &
    +94 &
  1,720 &
    +88 \\ \hline
After (enabled) \tbstrut &
  3,366 &
   +225 &
  2,486 &
   +200 &
  1,820 &
   +188 \\ \hline
\end{tabular}
\end{center}
\caption{
The total object code size for three build products: BLIS built as a static library; BLIS built as a shared library; and the BLIS testsuite linked against the aforementioned static library. Object code sizes are given using the: BLIS library just prior to mixed-datatype support ({\tt 667d3929}); with mixed-datatype support present ({\tt 5fec95b9}) but disabled at configure-time; and with mixed-datatype support present ({\tt 5fec95b9}) and enabled at configure-time.
}
\label{fig:object_code}
\end{figure}

The second column in Figure~\ref{fig:src_code_frame} shows the total number of lines%
\footnote{
In this section, we uniformly report {\em total} lines of code, including blank lines and comments.
}
of code present in the BLIS framework proper---which excludes other components such as the build system, kernels, and the testsuite---before and after
mixed-datatype support was added to the \gemm operation.%
\footnote{
The ``before'' and ``after'' snapshots of BLIS are uniquely identified with high probability by the first eight digits of the {\tt git} commit (SHA1 hash) numbers.
Commit {\tt 667d3929} identifies the code just before mixed-datatype support was added, while {\tt 5fec95b9} identifies the first commit in which mixed-datatype support is present.
This latter commit includes virtually all changes discussed in this article with the exception of the mixed-precision support for the \onem method, which was added in a later commit ({\tt 375eb30b}).
}
The fourth column shows the total size in kilobytes of the source code.
The change between the ``before'' and ``after'' values for total lines and total size are shown in the third and fifth columns, respectively.
Mixed-datatype support for the \gemm operation adds approximately 4\% each to the total number of lines and total bytes of source code.

Figure~\ref{fig:src_code_testsuite} lists similar metrics for the BLIS testsuite, which is capable of testing the vast majority of BLIS's computational operations.
Here, the support for testing all combinations of mixed-datatype execution, with any combination of matrix storage storage or transposition and/or combinations, increases the source code footprint by approximately 6\% in both lines and total size.

Finally, Figure~\ref{fig:object_code} shows the object (binary) code size for three build products: BLIS built as a static library; BLIS built as a shared library; and the BLIS testsuite linked against the static library.%
\footnote{
These object codes were built using GNU {\tt gcc} 5.4.0 on an Intel Xeon E3-1271 v3 (Haswell) workstation.
}
This figure shows three rows of values: before mixed-datatype support for \gemm was added ({\tt 667d3929}); after mixed-datatype support was added ({\tt 5fec95b9}) where the feature was disabled at configure-time; and after mixed-datatype support was added ({\tt 5fec95b9}) where the feature was enabled at configure-time.
(In the the latter two cases, the ``Change'' columns represent the change from the ``before'' state.)
With mixed-datatype support present and enabled, the size of the static library increases by only 255KB, or 8\% of the original library size.
In the case of the shared library, the increase is just under 9\%.
And a statically-linked instance of the BLIS testsuite increases by about 11\%.

Thus, no matter the metric, the increase in code footprint is quite modest relative to the scope of functionality added.


\section{Final Thoughts}

We conclude this article by sharing with the reader insights and observations that we have drawn to-date about BLIS, BLAS, and the adoption of new software functionality by the broader HPC community.

\subsection{Case studies}


In late 1990's, various community participants convened multiple meetings of the BLAS Technical (BLAST) Forum to discuss extensions to the original BLAS~\cite{BLAST}.
Some of these extensions targeted extended and mixed-precision functionality and were eventually implemented and branded as XBLAS~\cite{lawn149,xblasweb}.
In the end, the mixed-precision extensions were not widely adopted.
We speculate that this was in part due to the fact that the reference implementation falls short of achieving high performance.
By contrast, when the reference codes of the original BLAS~\cite{BLAS3,blasweb} were introduced, compilers could easily translate them to binary implementations that would yield high performance on the vector supercomputers that were prominent at the time.
Though computer architectures have evolved since then, we suspect that the initial availability of high-performance reference BLAS implementations helped give rise to a network of institutional experts and expertise, laying the foundation for today's well-established constellation of BLAS solutions.

A classic example of the release of a new standard hand-in-hand with a high quality implementation was the Message-Passing Interface (MPI)~\cite{MPI1,MPI2}.  
From the start, an implementation that later become known as MPICH provided a high-performance reference implementation~\cite{MPICH}.
Another example is TBLIS, a library and framework for performing efficient contractions and related operations on tensors~\cite{tblis_sisc}.
TBLIS borrows some of the insights of BLIS and then goes further to construct a general-purpose tensor library from scratch.
This new software architecture avoids the drawbacks of ad-hoc, BLAS-based solutions that must first reorder tensors into column-major storage simply for the purpose of calling \ttdgemmns.
The result is a comprehensive set of tensor functionality that exports flexible APIs (in C89 and C++11) while also facilitating higher performance for both sequential and multithreaded applications.
In contrast to the BLAST extensions, we feel that MPICH and TBLIS serve as important examples of software projects that each made an impact in their community by coupling new APIs and functionality with a complete high-performance reference implementation.

\subsection{Managing complexity}

Supporting the goal of providing a high-performance reference implementation with new APIs, while a laudable step in the right direction, is ultimately insufficient if the software is not designed to be practically implementable and maintainable. 
Suppose we attempted to solve the problem of mixed-datatype \gemm by implementing a solution in the style of the original reference BLAS.
Such an approach tends to lead to implementations that are duplicative and vertically siloed from one another, with support for each datatype, storage case, and transposition/conjugation scenario resulting in a fully independent block of code.
On top of supporting four (potentially differing) storage datatypes across all three matrix operands, a fully independent computation precision, and an fourth ``conjugation only'' transposition parameter value, BLIS also supports three different storage formats (row, column, and general storage) for each matrix operand.
This would lead to 31,104 separate implementations.
%
%
%
If two additional precisions are supported---half-precision and quad-precision, for example---this number grows to 497,664.

Our takeaway from this analysis, and our past experiences, is that achieving a complete high-performance reference implementation for mixed-datatype \gemm requires careful management of complexity in the implementation.
Complexity must be managed not only within interfaces (e.g., via object-based APIs) but also internally by allowing feature ``decision points'' (e.g., typecasting during packing) to work together in sequence, rather than in duplication, to enable the desired combinatoric space of functionality while collapsing its corresponding axial dimensions within the source code.
Otherwise, the solution quickly becomes unwieldy, if not hopelessly intractable, for its maintainers.


\subsection{Thesis}

It is our conjecture that interfaces to new functionality, such as the mixed-datatype \gemm presented in this article, should be accompanied by a corresponding high-performance reference implementation, and that the key to making such a goal attainable is managing combinatoric software complexity.
Aside from serving several obvious purposes---a research proof-of-concept, a reference for other developers, a working solution upon which end-users can rely---the reference implementation, once instantiated, confers another, less tangible benefit.
Namely, it creates the initial conditions in which a community can form and a tangible product around which its members can organize, collaborate, and advance towards its shared objectives.
Therefore, this approach ultimately benefits all stakeholders.

\section{Software availability}

The software referenced in this article may be found at the BLIS project page~\cite{blisweb} on GitHub along with documentation, examples, links to discussion forums, and other related resources.

\section{Acknowledgments}
We kindly thank Jeff R. Hammond and Devin A. Matthews for referring us to appropriate citations for various applications and use cases that stand to benefit from implementing mixed-domain and mixed-precision matrix multiplication functionality.
We also acknowledge Matthews for participating in and facilitating early discussions on the topic of mixed-domain and mixed-precision computation.
Also, we thank Marvell and Oracle for providing access to the Marvell ThunderX2 CN9975 and Intel Xeon Platinum 8167M servers, respectively, on which performance data for this article was gathered.
Finally, we thank members of the Science of High-Performance Computing (SHPC) group for their contributions throughout the research towards and drafting of this article.

This research was partially sponsored by grants from
Oracle, Huawei, and
the National Science Foundation
(Awards ACI-1550493 and ACI-1714091).
{\em Any opinions, findings and conclusions or recommendations expressed in this material are those of the author(s) and do not necessarily reflect the views of the National Science Foundation (NSF).}

\NoShow{
	\begin{acks}
We kindly thank Jeff R. Hammond and Devin A. Matthews for referring us to appropriate citations for various applications and use cases that stand to benefit from implementing mixed-domain and mixed-precision matrix multiplication functionality.
We also acknowledge Matthews for participating in and facilitating early discussions on the topic of mixed-domain and mixed-precision computation.
Also, we thank Marvell and Oracle for providing access to the Marvell ThunderX2 CN9975 and Intel Xeon Platinum 8167M servers, respectively, on which performance data for this article was gathered.
Finally, we thank members of the Science of High-Performance Computing (SHPC) group for their contributions throughout the research towards and drafting of this article.
\end{acks}
}

\bibliographystyle{plain}
\bibliography{biblio}


\pagebreak
\begin{appendices}
\section{Complete performance results}
\label{sec:appendix}

This section contains complete performance results using the same hardware, implementations, and test parameters discussed in Section~\ref{sec:perf}.
We report 128 performance graphs for each combination of sequential, single-socket, and dual-socket execution on both the Marvell ThunderX2 and Intel Xeon Platinum, resulting in a total of $ 128 \times 3 \times 2 = 768 $ graphs.

Performance graphs are organized into one set for each mixed-domain case, with each set containing the 16 possible precision combinations within that case. 
The mixed-domain sets of graphs appear in pairs (top and bottom) across Figures~\ref{fig:perf_tx2_t1_0} to~\ref{fig:perf_skx_t52_3}.
Within a figure, graphs in the left and center-left columns report performance using a computation precision of single precision while those in the right and center-right columns show performance using a computation precision of double precision.
Furthermore, the graphs are organized such that, for any given graph in the single-precision computation subgroup, the graph located two spots to its right corresponds to the experiments where all precisions are toggled (from single to double or vice versa).

Within the graphs for any given mixed-domain set, the legends are omitted from all except the top-right graph within each computation precision subgroup---that is, the top graph in the center-left and right columns.
As with the graphs presented in Section~\ref{sec:perf}, the ``Reference'' curves are listed as ``Ref ({\tt ?gemm})'', where the {\tt ?} indicates one of {\tt \{s,d,c,z\}}.
This added detail serves to remind the reader which datatype-specific variant of BLIS's conventional (datatype-homogeneous) \gemm is the most appropriate curve against which to judge the ``Internal'' and ``Ad-hoc'' mixed-datatype implementations.
In general, the graphs in the left and right halves of Figures~\ref{fig:perf_tx2_t1_0}--\ref{fig:perf_skx_t52_3} (top and bottom) use \ttsgemm and \ttdgemm as reference curves, respectively, except for the mixed-domain Case \three results in the bottom halves of Figures~\ref{fig:perf_tx2_t1_0}, \ref{fig:perf_tx2_t28_0}, \ref{fig:perf_tx2_t56_0}, \ref{fig:perf_skx_t1_0}, \ref{fig:perf_skx_t26_0}, and \ref{fig:perf_skx_t52_0}, which compare against \ttcgemm and \ttzgemmns in the left and right halves, respectively.

We provide the following table to aid the reader in finding the performance graphs associated with any given mixed-domain case, for each hardware and threading configuration.

\vspace{3mm}
\begin{center}
\begin{tabular}{| l | c I c | c | c | c |} \hline
Hardware & Threads
&
Cases \zerons, \three &
\oneans, \oneb &
\twoabns, \onec &
\twobcns, \twoac \\ \whline
\multirow{3}{*}{ThunderX2} & 1 &
Fig.~\ref{fig:perf_tx2_t1_0} &
Fig.~\ref{fig:perf_tx2_t1_1} &
Fig.~\ref{fig:perf_tx2_t1_2} &
Fig.~\ref{fig:perf_tx2_t1_3} \\ 
 & 28 &
Fig.~\ref{fig:perf_tx2_t28_0} &
Fig.~\ref{fig:perf_tx2_t28_1} &
Fig.~\ref{fig:perf_tx2_t28_2} &
Fig.~\ref{fig:perf_tx2_t28_3} \\ 
 & 56 &
Fig.~\ref{fig:perf_tx2_t56_0} &
Fig.~\ref{fig:perf_tx2_t56_1} &
Fig.~\ref{fig:perf_tx2_t56_2} &
Fig.~\ref{fig:perf_tx2_t56_3} \\ \hline
\multirow{3}{*}{Xeon Platinum} & 1 &
Fig.~\ref{fig:perf_skx_t1_0} &
Fig.~\ref{fig:perf_skx_t1_1} &
Fig.~\ref{fig:perf_skx_t1_2} &
Fig.~\ref{fig:perf_skx_t1_3} \\ 
 & 26 &
Fig.~\ref{fig:perf_skx_t26_0} &
Fig.~\ref{fig:perf_skx_t26_1} &
Fig.~\ref{fig:perf_skx_t26_2} &
Fig.~\ref{fig:perf_skx_t26_3} \\ 
 & 52 &
Fig.~\ref{fig:perf_skx_t52_0} &
Fig.~\ref{fig:perf_skx_t52_1} &
Fig.~\ref{fig:perf_skx_t52_2} &
Fig.~\ref{fig:perf_skx_t52_3} \\ \hline
\end{tabular}
\end{center}

\newcommand{\sentencezeroa}[2]{}
\newcommand{\sentenceonea}{}
\newcommand{\sentencetwoa}{}


\renewcommand{\graphhspace}{-3.0mm}
\renewcommand{\graphwidth}{4.4in}
\renewcommand{\trimleft}{3.35cm}
\renewcommand{\trimlower}{2.7cm}
\renewcommand{\trimright}{3.7cm}
\renewcommand{\trimupper}{2.2cm}

\renewcommand{\sentencezeroa}[2]{Sequential performance of ``Internal'' and ``Ad-hoc'' implementations of \gemm for all precision combinations within mixed-domain Cases #1 (top) and #2 (bottom) on a Marvell ThunderX2 CN9975 processor. }
\renewcommand{\sentenceonea}{The 16 graphs on the left side and right sides report computation in single- and double-precision, respectively. }
\renewcommand{\sentencetwoa}{The theoretical peak performance coincides with the top of each graph. }

%
%
\begin{figure}[hp!]
\begin{center}
\begin{tabular}{l}
\hspace{\graphhspace}
\includegraphics[width=\graphwidth,trim={\trimleft, \trimlower, \trimright, \trimupper},clip]{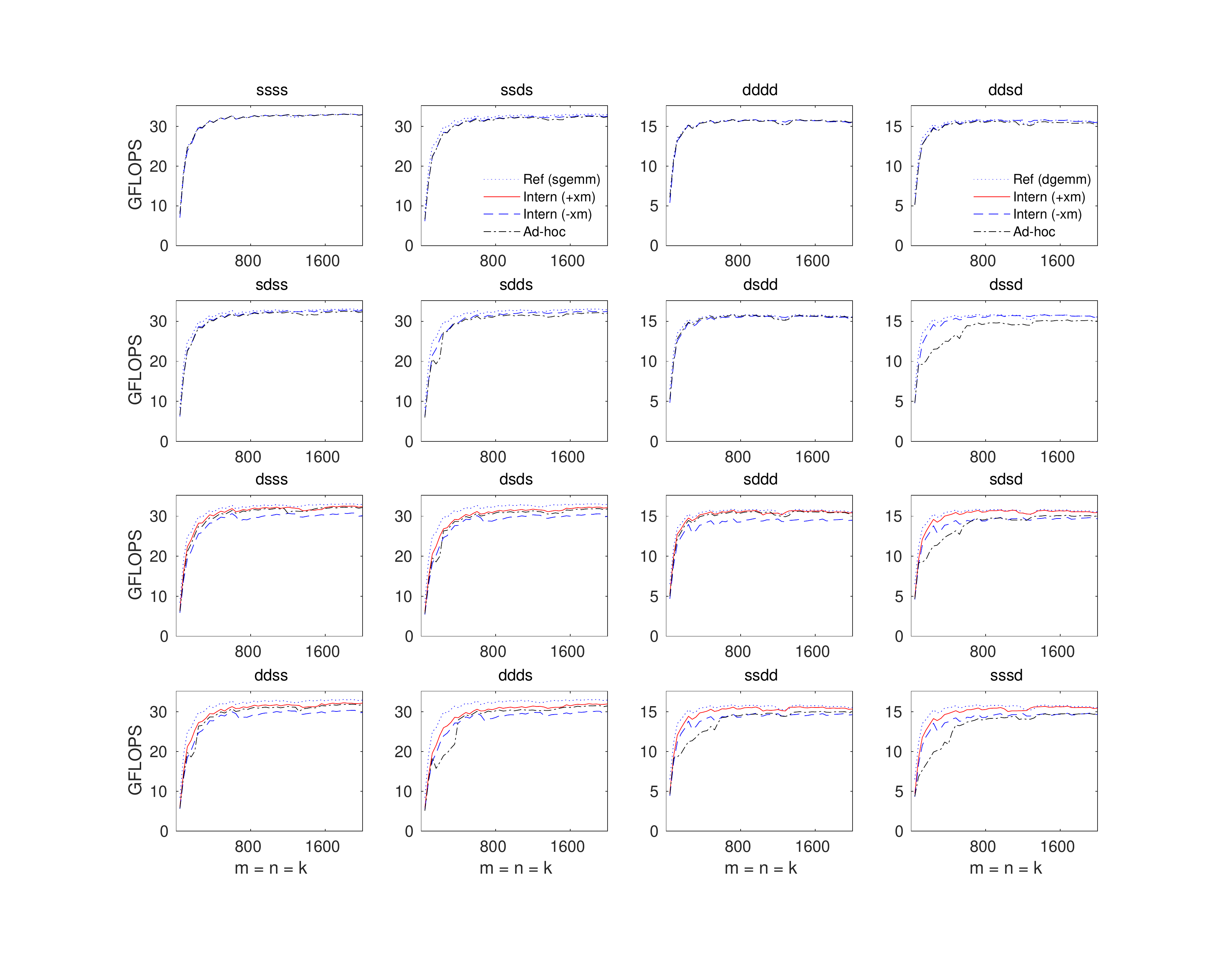} \\ \whline
\hspace{\graphhspace}
\includegraphics[width=\graphwidth,trim={\trimleft, \trimlower, \trimright, \trimupper},clip]{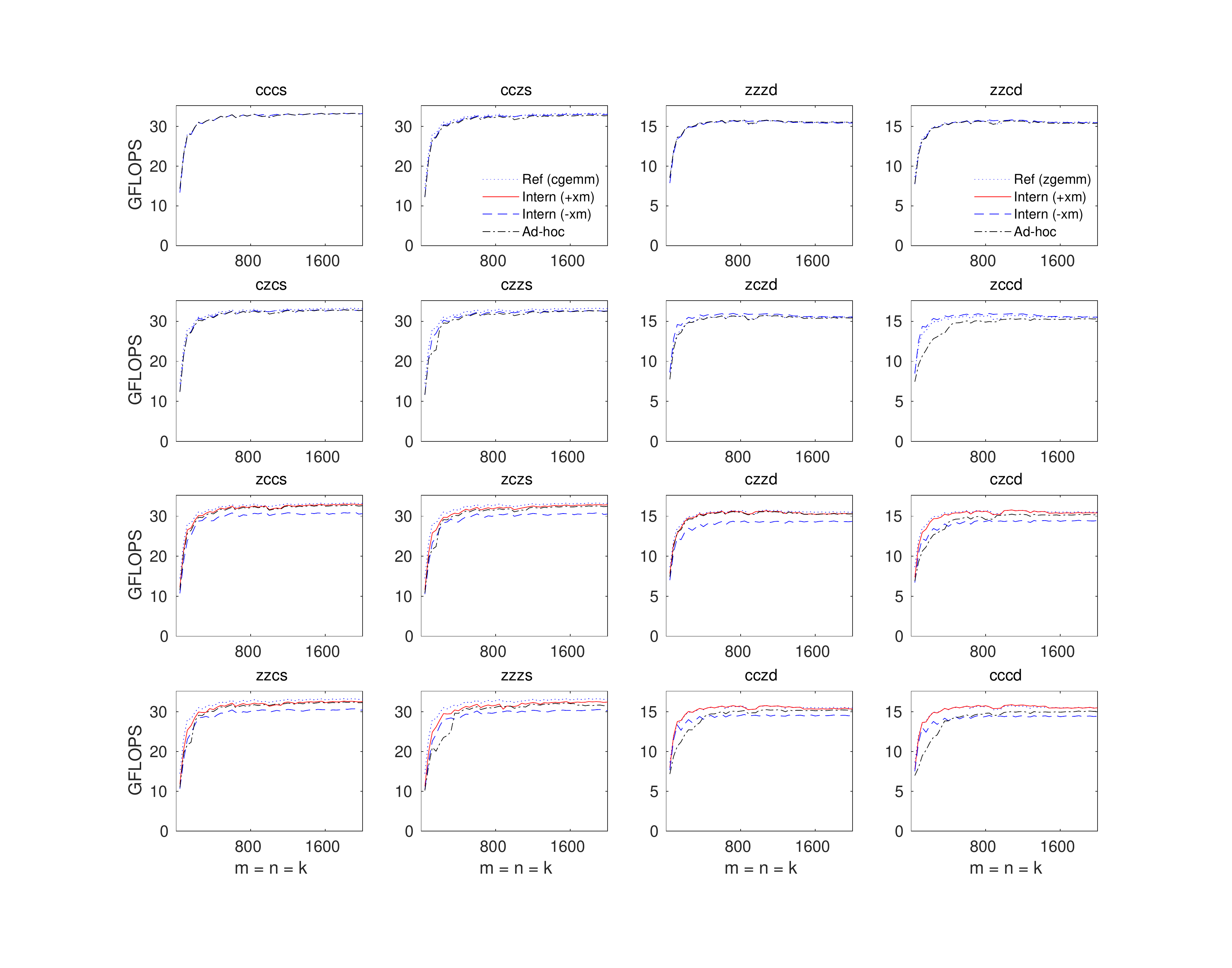}
\end{tabular}
\end{center}
\caption{
\sentencezeroa{\zerons}{\threens}\sentenceonea\sentencetwoa
}
\label{fig:perf_tx2_t1_0}
\end{figure}

%
%
\begin{figure}[hp!]
\begin{center}
\begin{tabular}{l}
\hspace{\graphhspace}
\includegraphics[width=\graphwidth,trim={\trimleft, \trimlower, \trimright, \trimupper},clip]{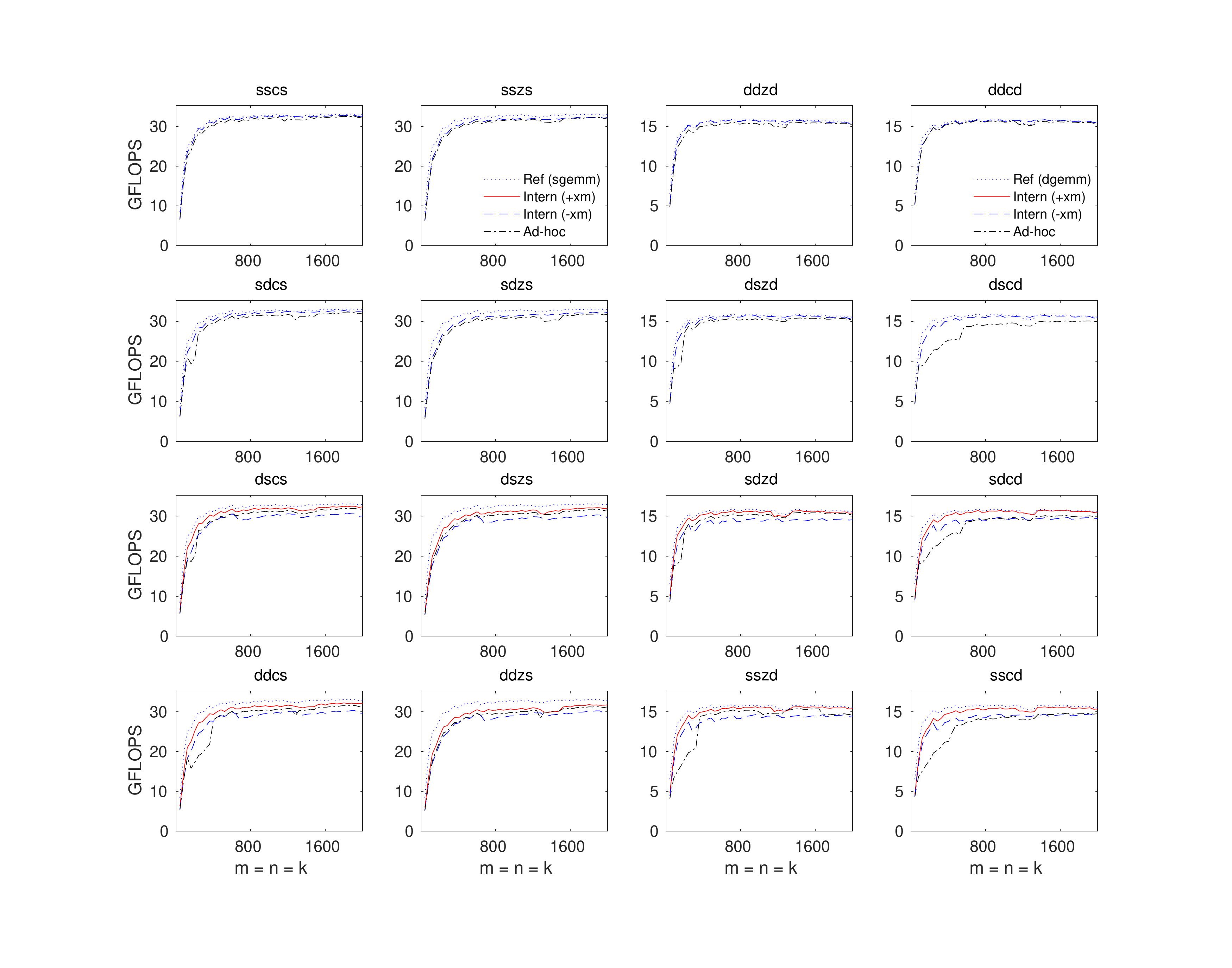} \\ \whline
\hspace{\graphhspace}
\includegraphics[width=\graphwidth,trim={\trimleft, \trimlower, \trimright, \trimupper},clip]{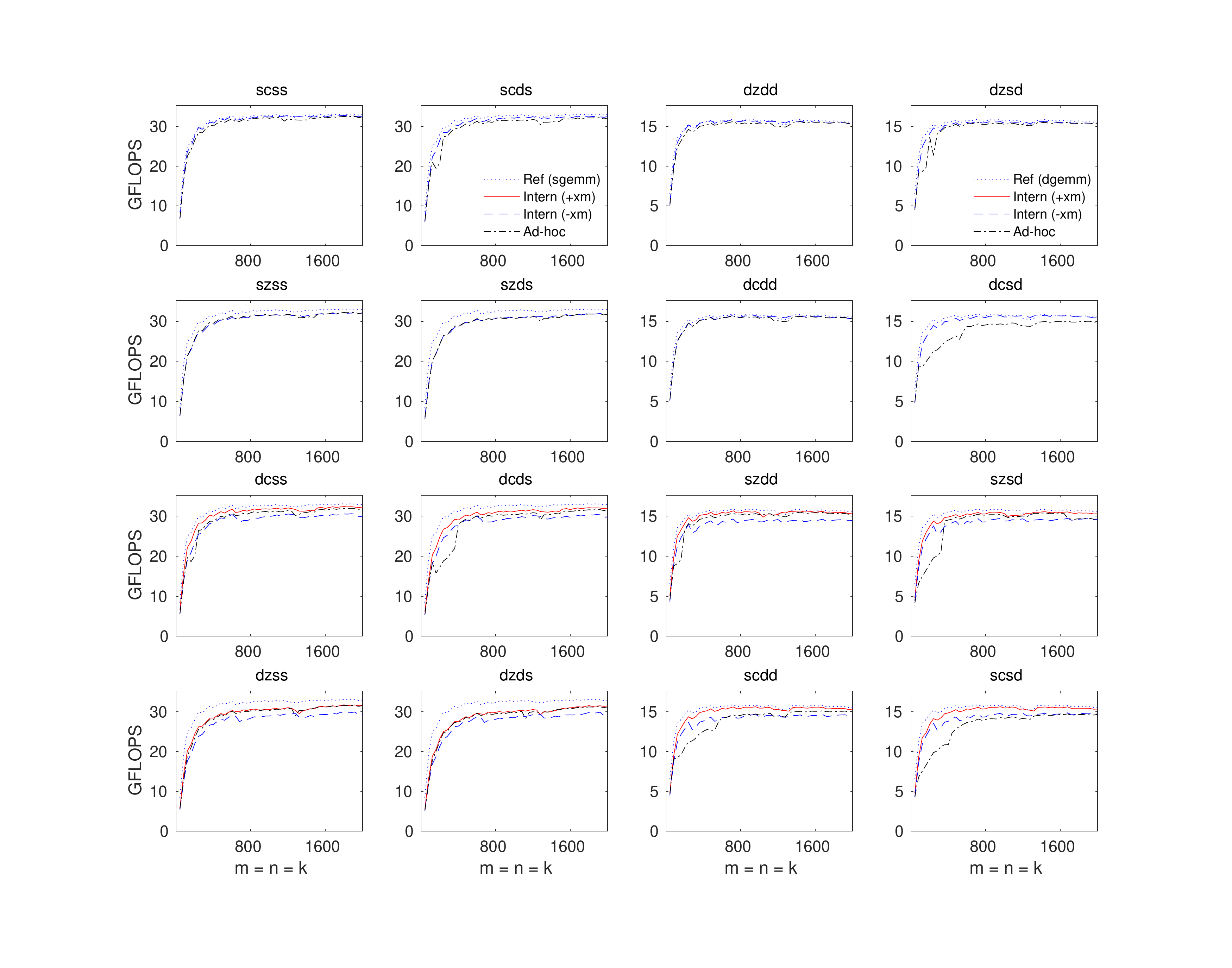}
\end{tabular}
\end{center}
\caption{
\sentencezeroa{\oneans}{\onebns}\sentenceonea\sentencetwoa
}
\label{fig:perf_tx2_t1_1}
\end{figure}

%
%
\begin{figure}[hp!]
\begin{center}
\begin{tabular}{l}
\hspace{\graphhspace}
\includegraphics[width=\graphwidth,trim={\trimleft, \trimlower, \trimright, \trimupper},clip]{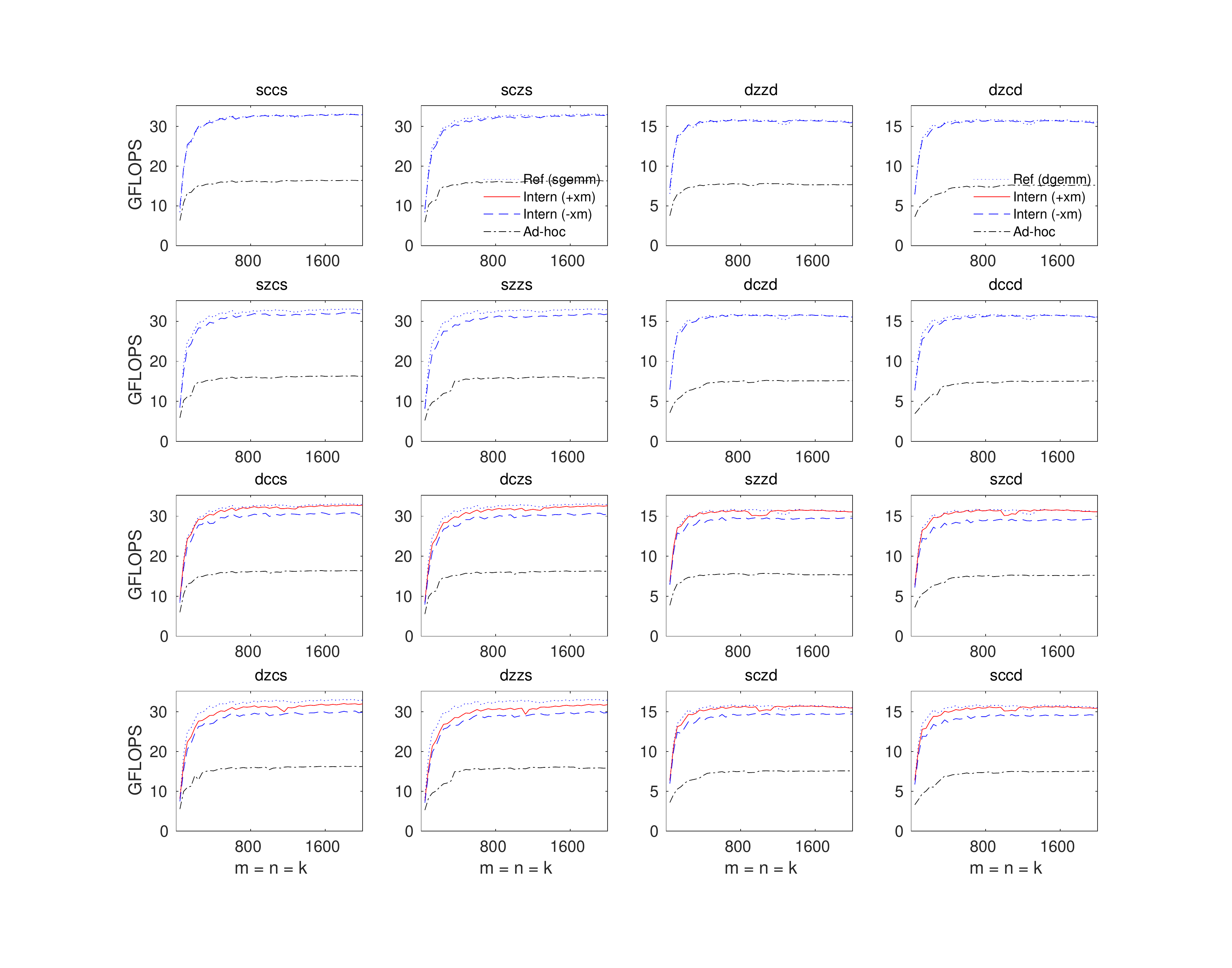} \\ \whline
\hspace{\graphhspace}
\includegraphics[width=\graphwidth,trim={\trimleft, \trimlower, \trimright, \trimupper},clip]{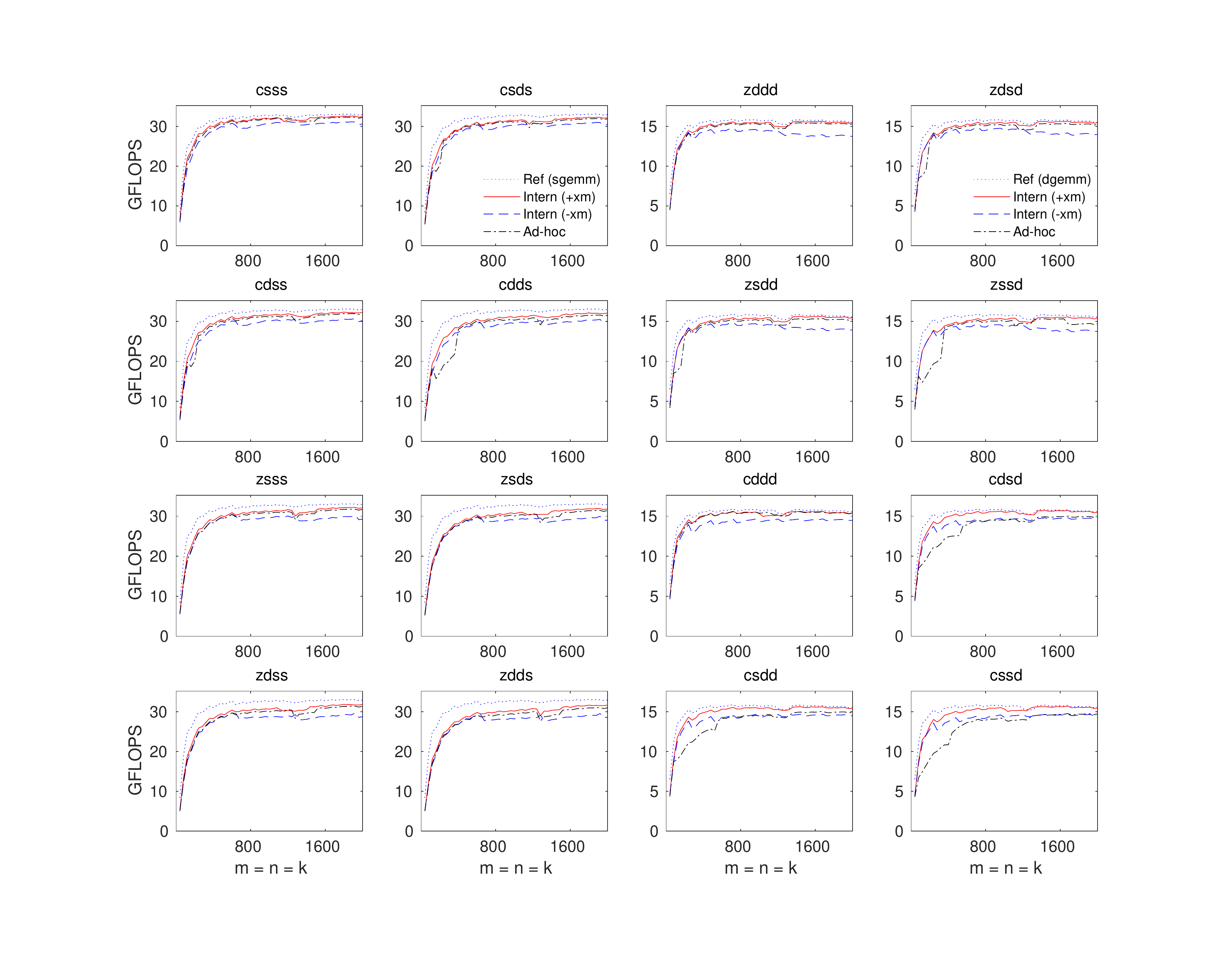}
\end{tabular}
\end{center}
\caption{
\sentencezeroa{\twoabns}{\onecns}\sentenceonea\sentencetwoa
}
\label{fig:perf_tx2_t1_2}
\end{figure}

%
%
\begin{figure}[hp!]
\begin{center}
\begin{tabular}{l}
\hspace{\graphhspace}
\includegraphics[width=\graphwidth,trim={\trimleft, \trimlower, \trimright, \trimupper},clip]{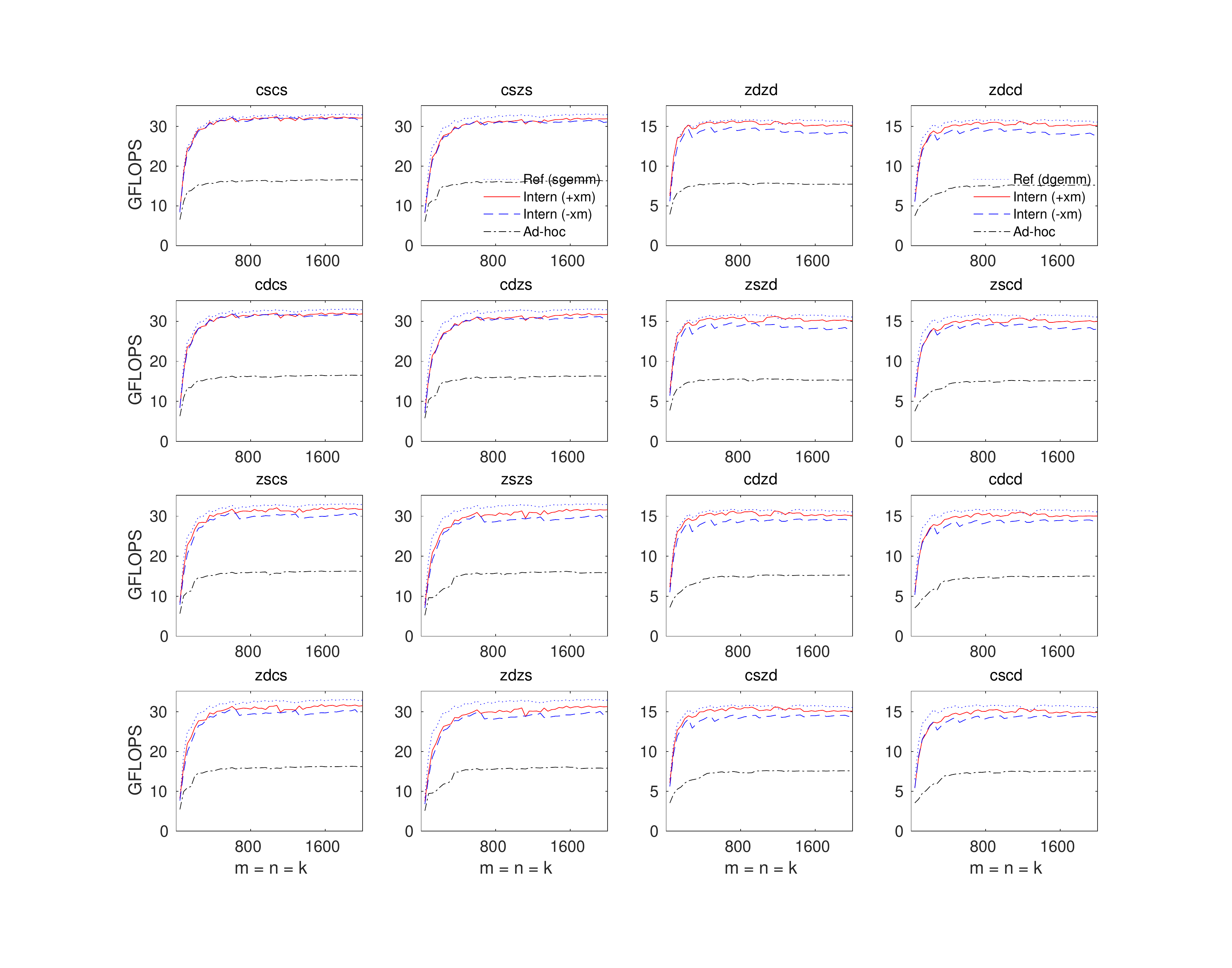} \\ \whline
\hspace{\graphhspace}
\includegraphics[width=\graphwidth,trim={\trimleft, \trimlower, \trimright, \trimupper},clip]{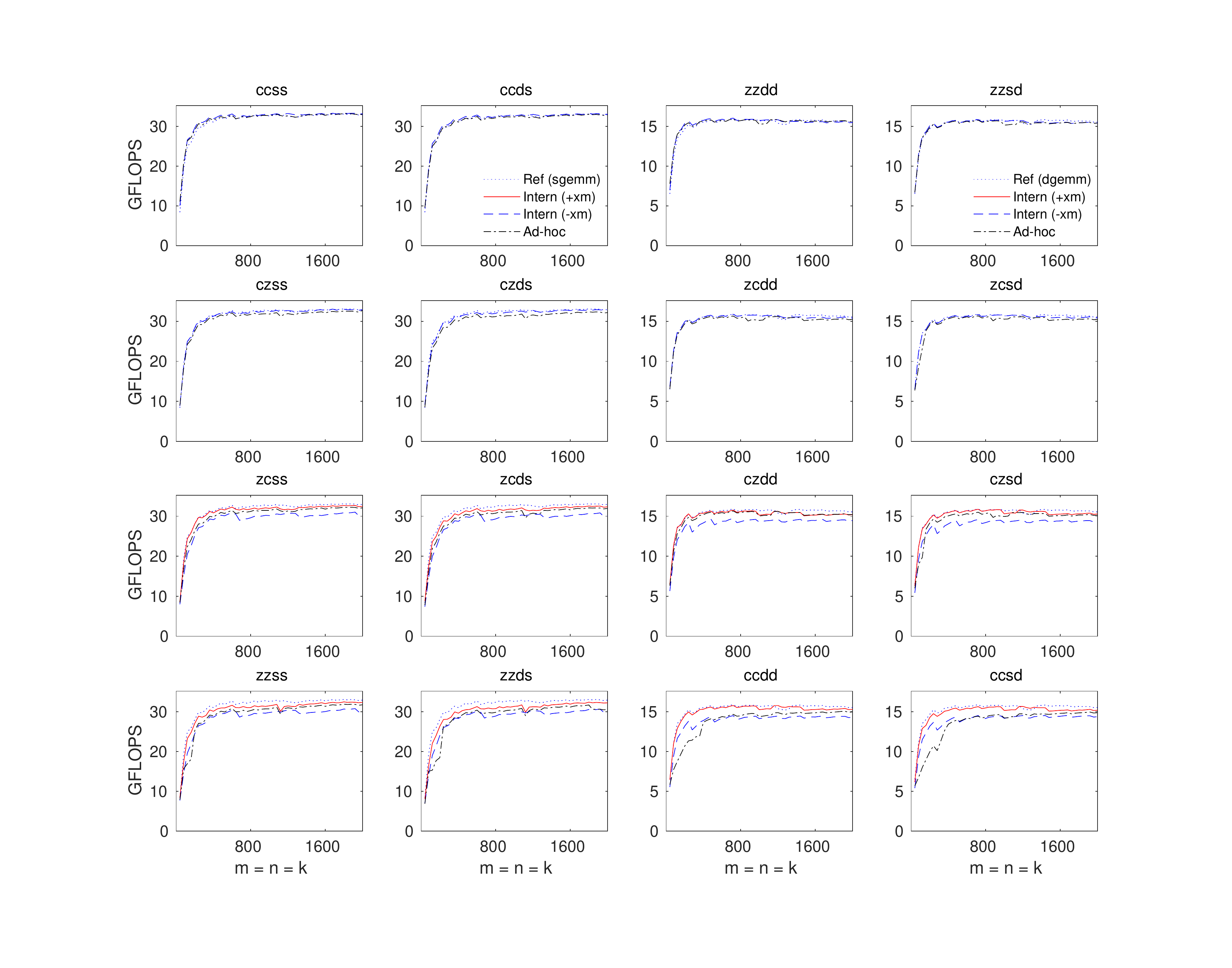}
\end{tabular}
\end{center}
\caption{
\sentencezeroa{\twobcns}{\twoacns}\sentenceonea\sentencetwoa
}
\label{fig:perf_tx2_t1_3}
\end{figure}

\renewcommand{\graphhspace}{-3.0mm}
\renewcommand{\graphwidth}{4.4in}
\renewcommand{\trimleft}{3.35cm}
\renewcommand{\trimlower}{2.7cm}
\renewcommand{\trimright}{3.7cm}
\renewcommand{\trimupper}{2.2cm}

\renewcommand{\sentencezeroa}[2]{Multithreaded (28 threads) performance of ``Internal'' and ``Ad-hoc'' implementations of \gemm for all precision combinations within mixed-domain Cases #1 (top) and #2 (bottom) on a Marvell ThunderX2 CN9975 processor. }
\renewcommand{\sentenceonea}{The 16 graphs on the left side and right sides report computation in single- and double-precision, respectively. }
\renewcommand{\sentencetwoa}{The theoretical peak performance coincides with the top of each graph. }

%
%
\begin{figure}[hp!]
\begin{center}
\begin{tabular}{l}
\hspace{\graphhspace}
\includegraphics[width=\graphwidth,trim={\trimleft, \trimlower, \trimright, \trimupper},clip]{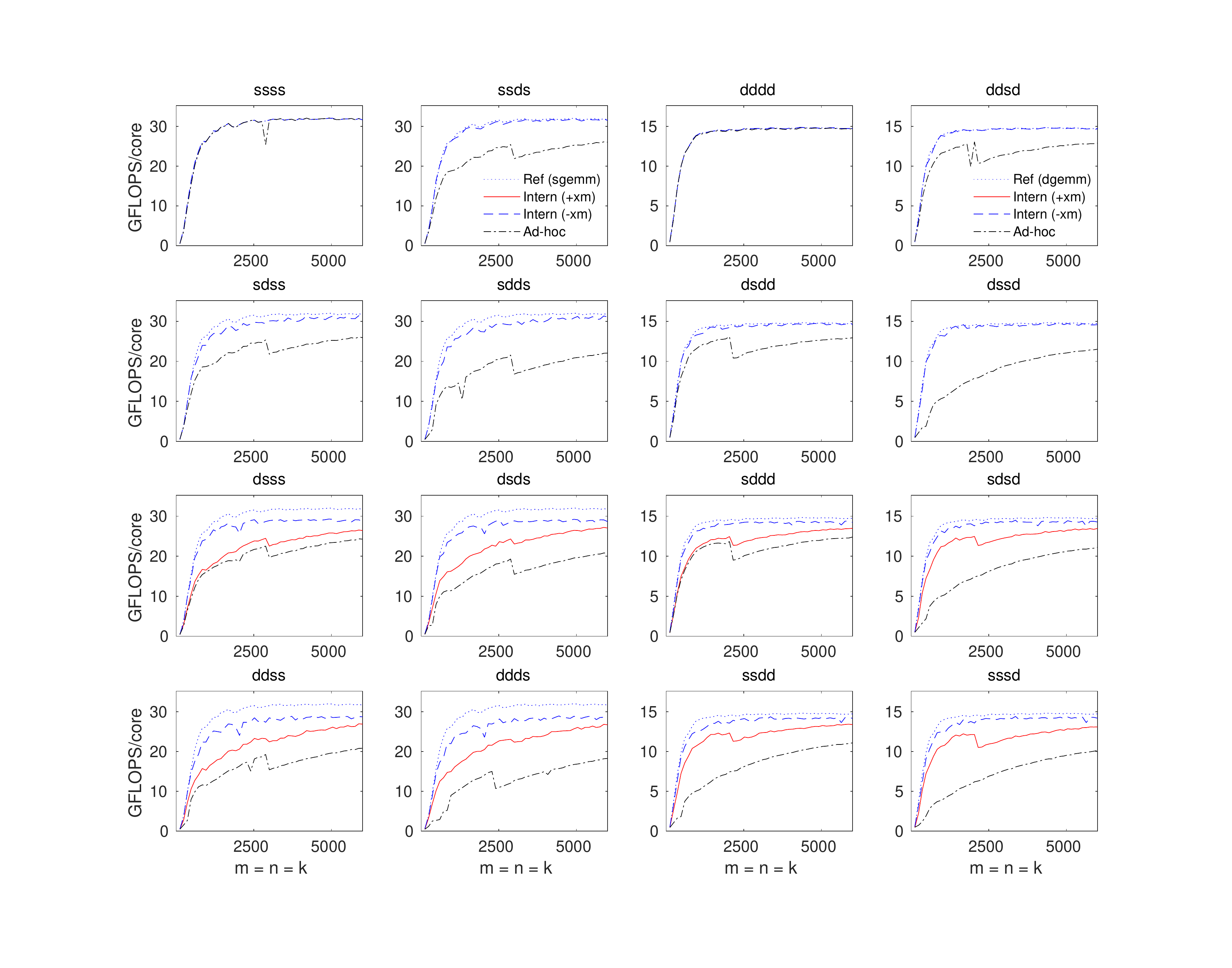} \\ \whline
\hspace{\graphhspace}
\includegraphics[width=\graphwidth,trim={\trimleft, \trimlower, \trimright, \trimupper},clip]{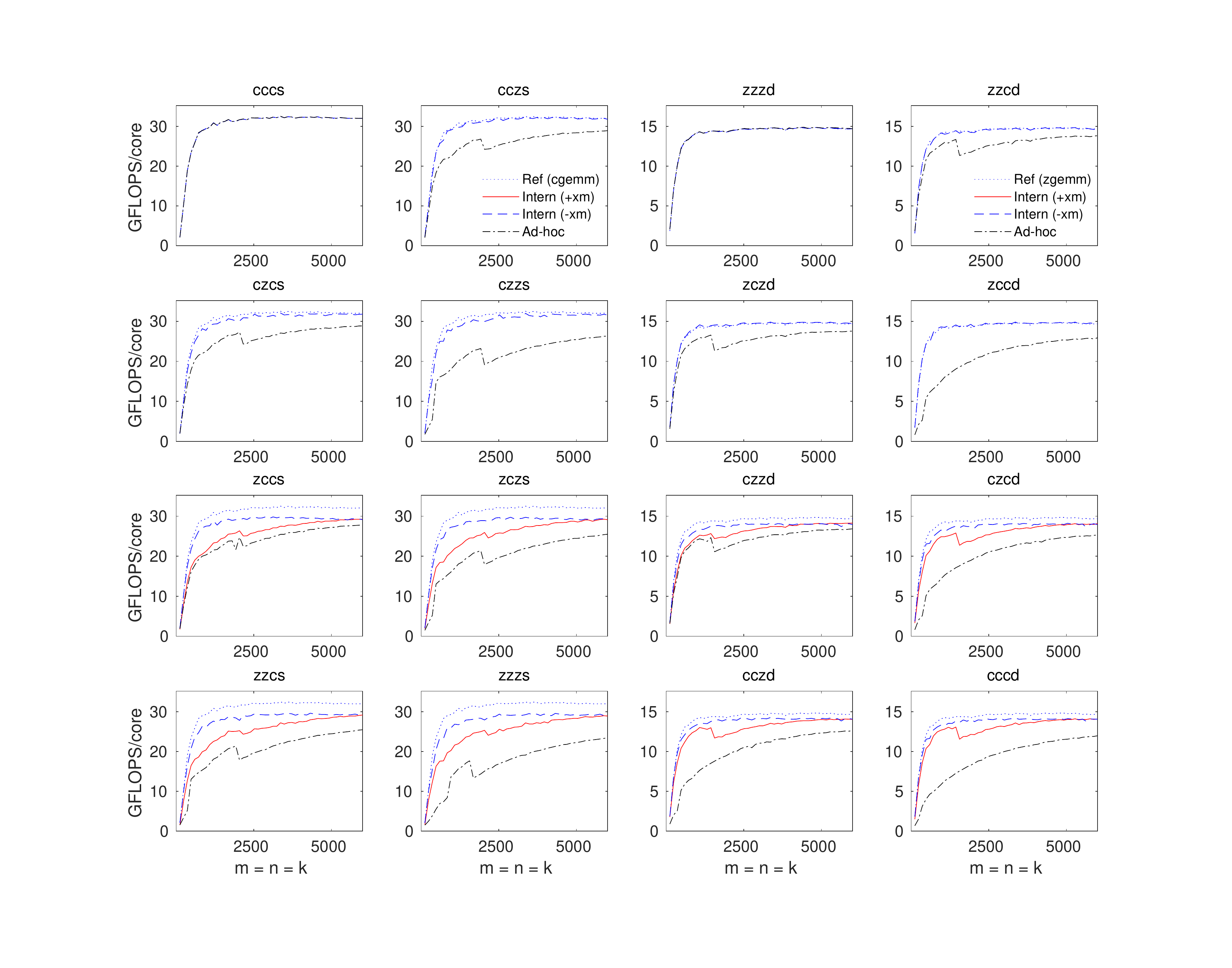}
\end{tabular}
\end{center}
\caption{
\sentencezeroa{\zerons}{\threens}\sentenceonea\sentencetwoa
}
\label{fig:perf_tx2_t28_0}
\end{figure}

%
%
\begin{figure}[hp!]
\begin{center}
\begin{tabular}{l}
\hspace{\graphhspace}
\includegraphics[width=\graphwidth,trim={\trimleft, \trimlower, \trimright, \trimupper},clip]{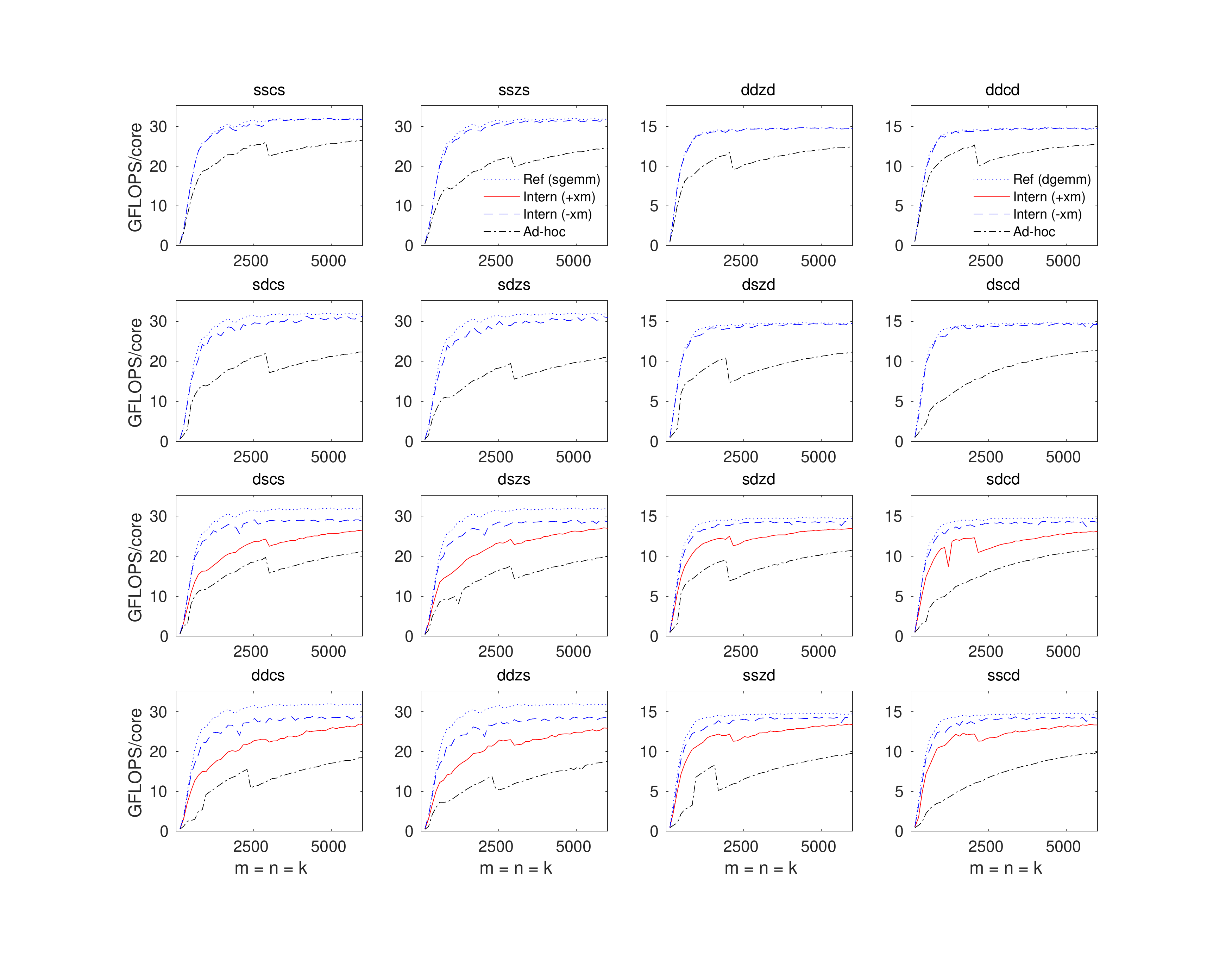} \\ \whline
\hspace{\graphhspace}
\includegraphics[width=\graphwidth,trim={\trimleft, \trimlower, \trimright, \trimupper},clip]{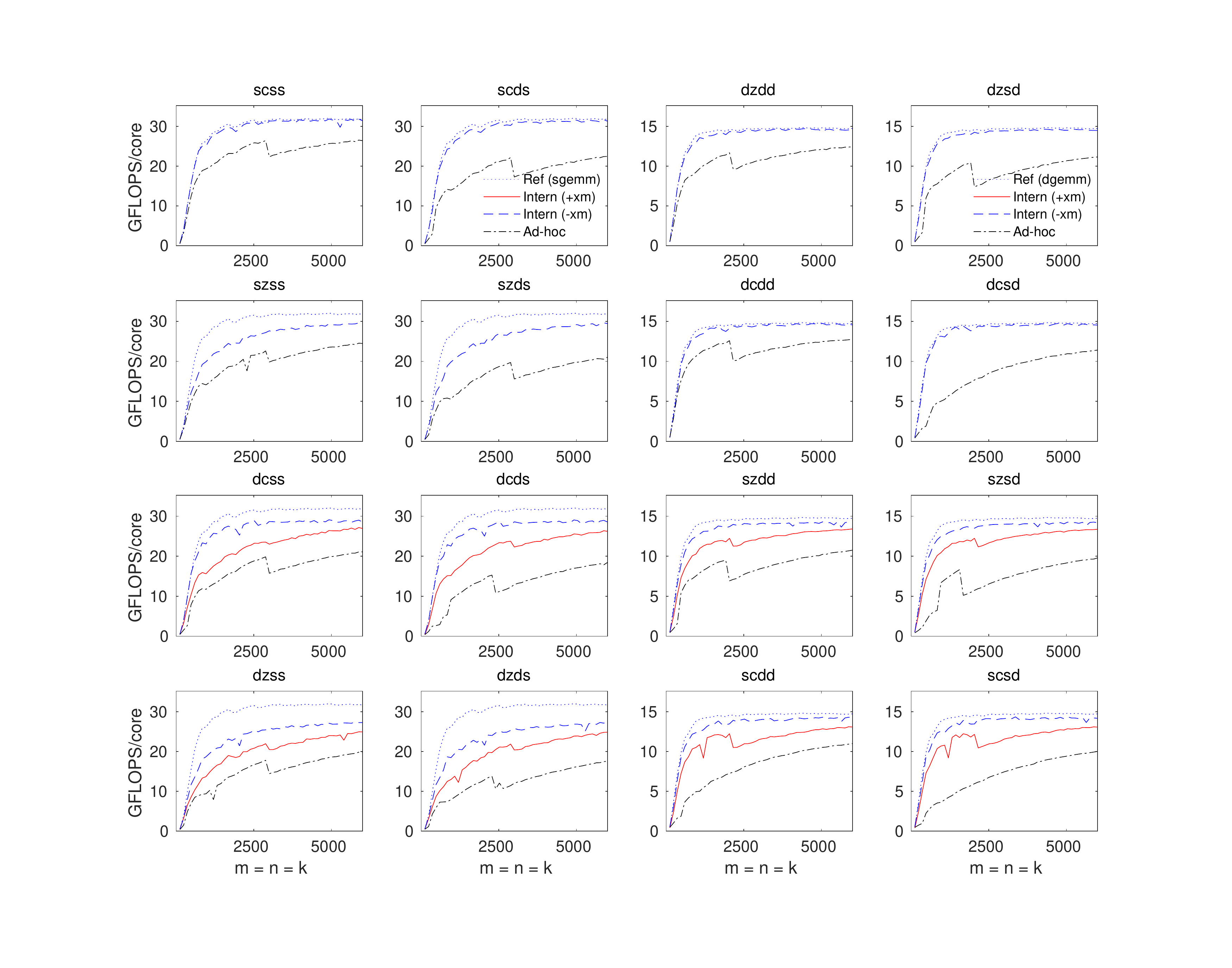}
\end{tabular}
\end{center}
\caption{
\sentencezeroa{\oneans}{\onebns}\sentenceonea\sentencetwoa
}
\label{fig:perf_tx2_t28_1}
\end{figure}

%
%
\begin{figure}[hp!]
\begin{center}
\begin{tabular}{l}
\hspace{\graphhspace}
\includegraphics[width=\graphwidth,trim={\trimleft, \trimlower, \trimright, \trimupper},clip]{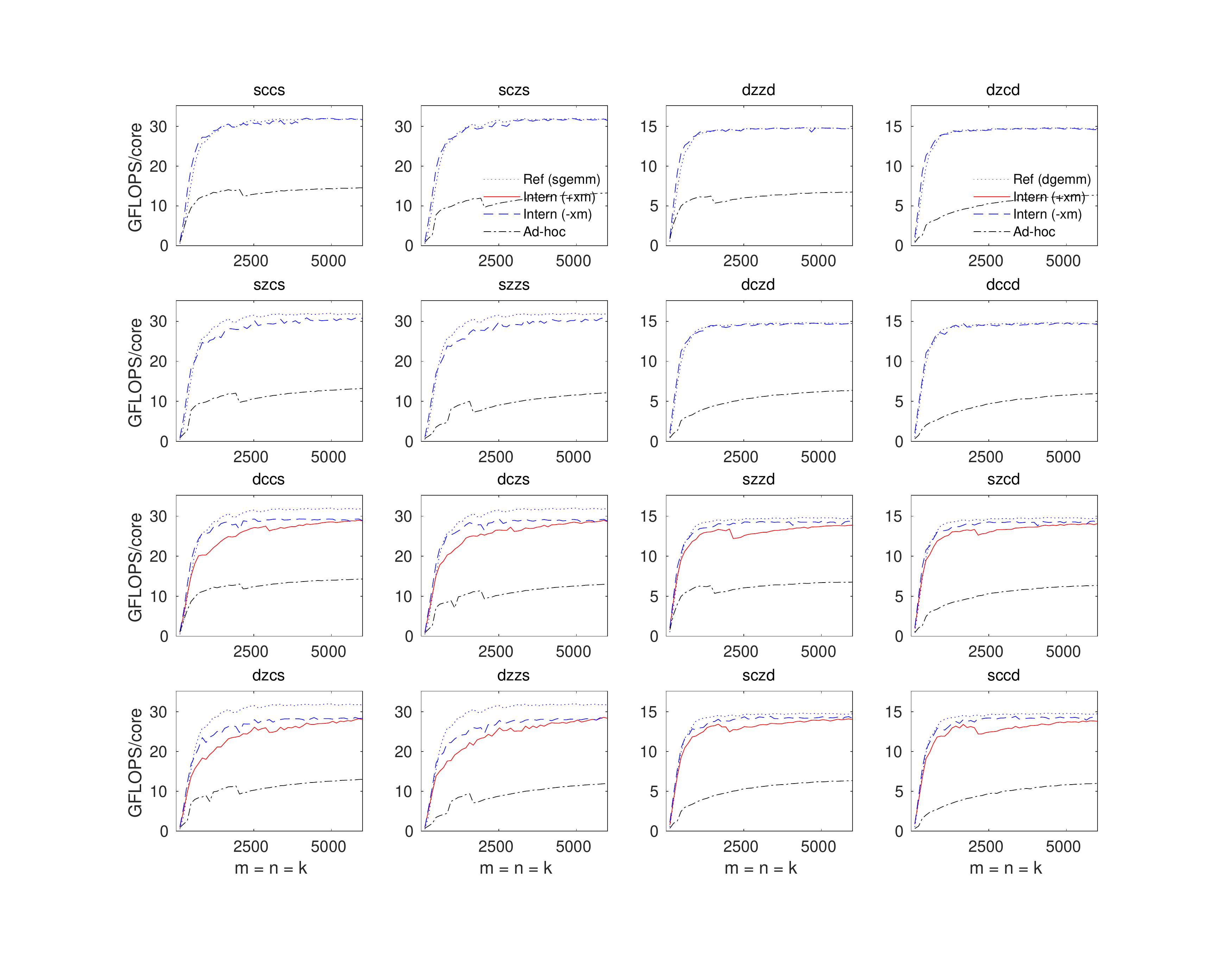} \\ \whline
\hspace{\graphhspace}
\includegraphics[width=\graphwidth,trim={\trimleft, \trimlower, \trimright, \trimupper},clip]{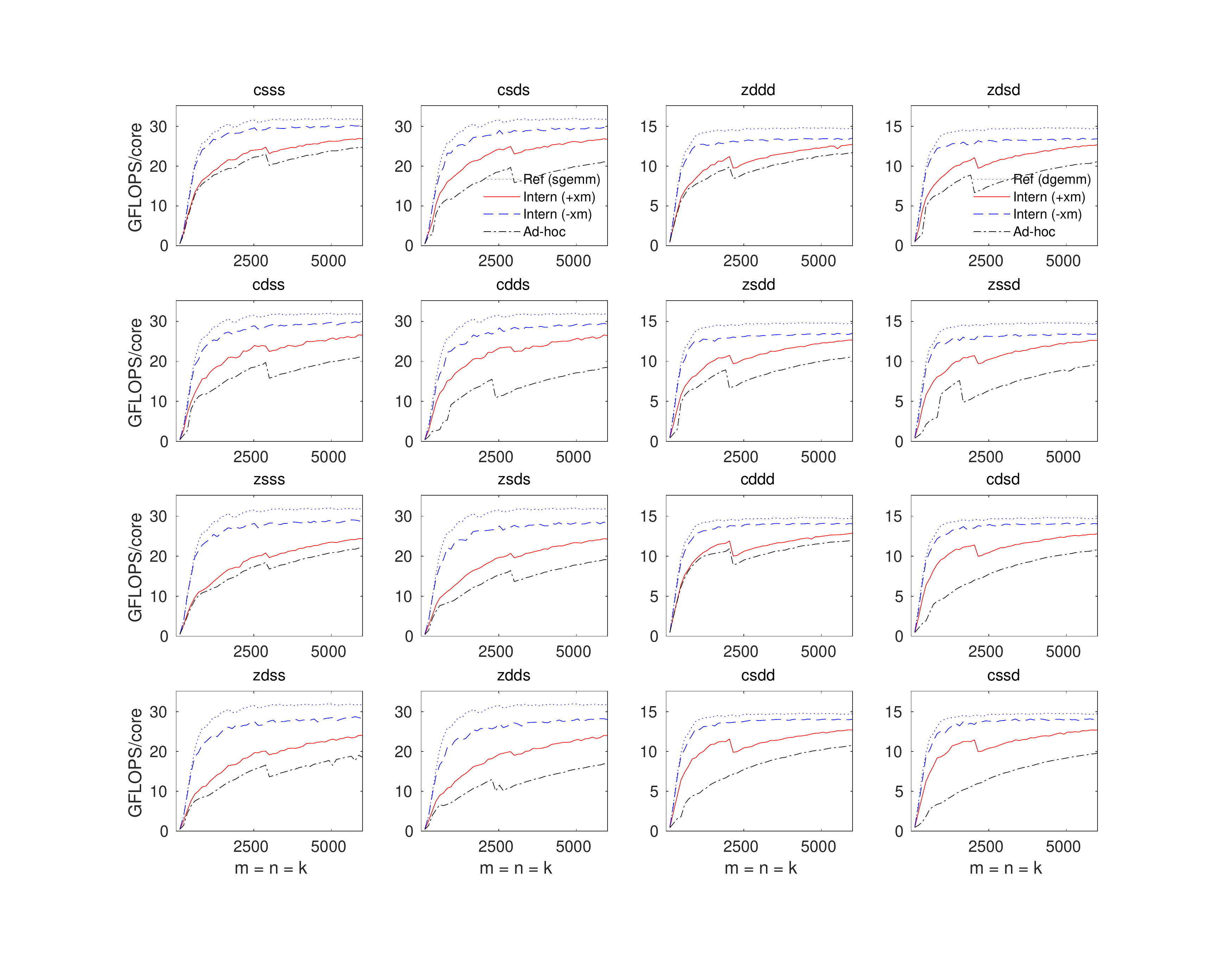}
\end{tabular}
\end{center}
\caption{
\sentencezeroa{\twoabns}{\onecns}\sentenceonea\sentencetwoa
}
\label{fig:perf_tx2_t28_2}
\end{figure}

%
%
\begin{figure}[hp!]
\begin{center}
\begin{tabular}{l}
\hspace{\graphhspace}
\includegraphics[width=\graphwidth,trim={\trimleft, \trimlower, \trimright, \trimupper},clip]{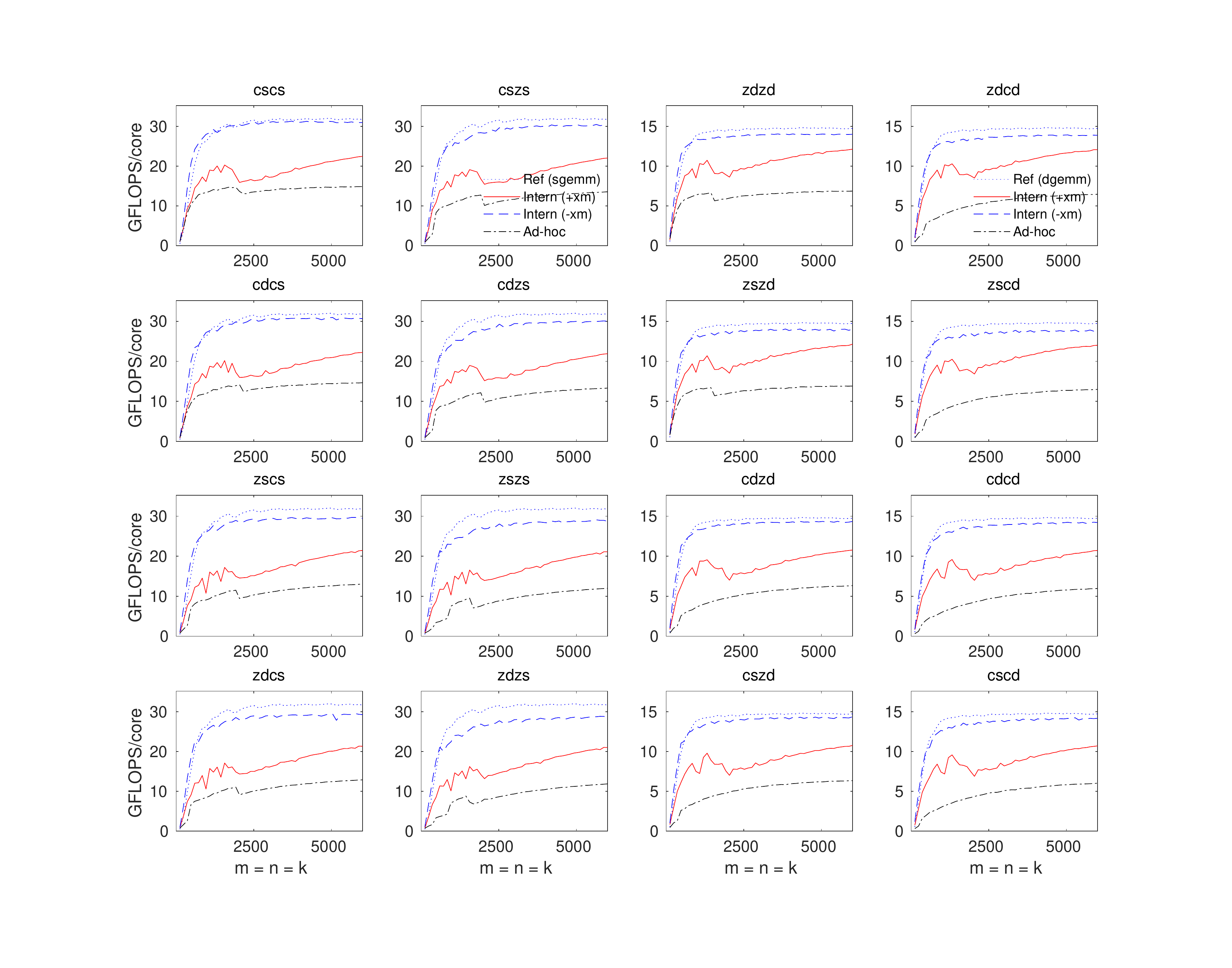} \\ \whline
\hspace{\graphhspace}
\includegraphics[width=\graphwidth,trim={\trimleft, \trimlower, \trimright, \trimupper},clip]{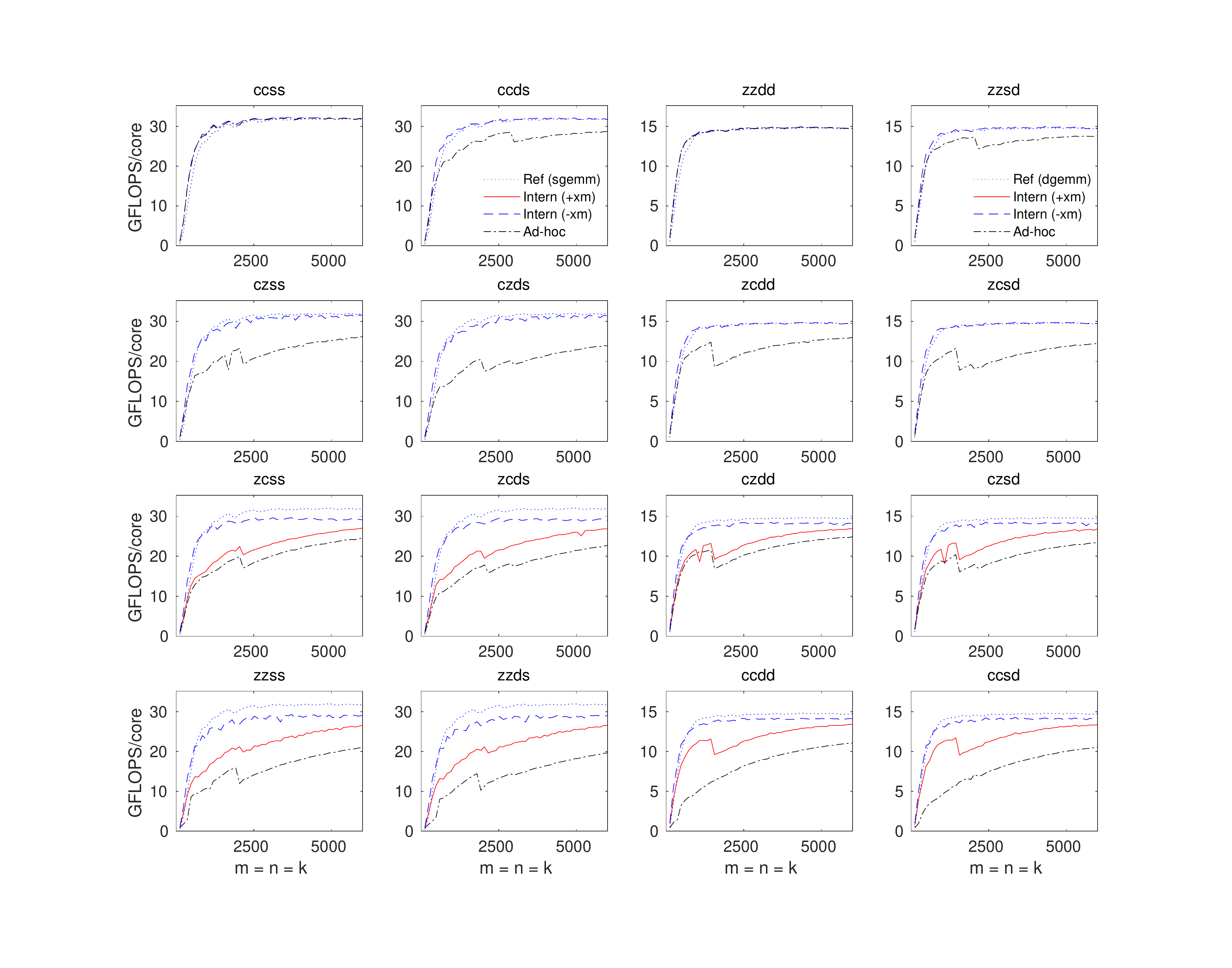}
\end{tabular}
\end{center}
\caption{
\sentencezeroa{\twobcns}{\twoacns}\sentenceonea\sentencetwoa
}
\label{fig:perf_tx2_t28_3}
\end{figure}

\renewcommand{\graphhspace}{-3.0mm}
\renewcommand{\graphwidth}{4.4in}
\renewcommand{\trimleft}{3.35cm}
\renewcommand{\trimlower}{2.7cm}
\renewcommand{\trimright}{3.7cm}
\renewcommand{\trimupper}{2.2cm}

\renewcommand{\sentencezeroa}[2]{Multithreaded (56 threads) performance of ``Internal'' and ``Ad-hoc'' implementations of \gemm for all precision combinations within mixed-domain Cases #1 (top) and #2 (bottom) on a Marvell ThunderX2 CN9975 processor. }
\renewcommand{\sentenceonea}{The 16 graphs on the left side and right sides report computation in single- and double-precision, respectively. }
\renewcommand{\sentencetwoa}{The theoretical peak performance coincides with the top of each graph. }

%
%
\begin{figure}[hp!]
\begin{center}
\begin{tabular}{l}
\hspace{\graphhspace}
\includegraphics[width=\graphwidth,trim={\trimleft, \trimlower, \trimright, \trimupper},clip]{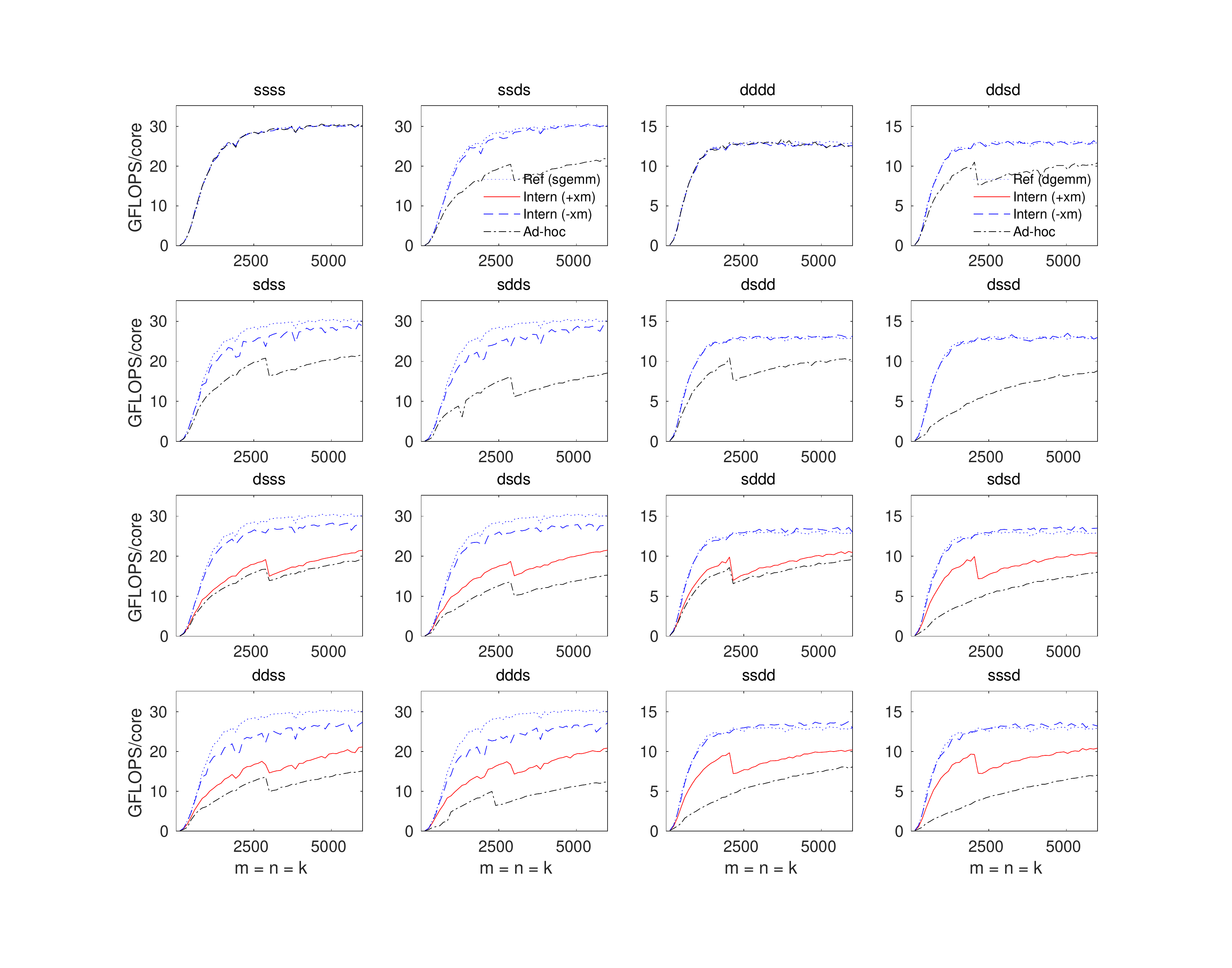} \\ \whline
\hspace{\graphhspace}
\includegraphics[width=\graphwidth,trim={\trimleft, \trimlower, \trimright, \trimupper},clip]{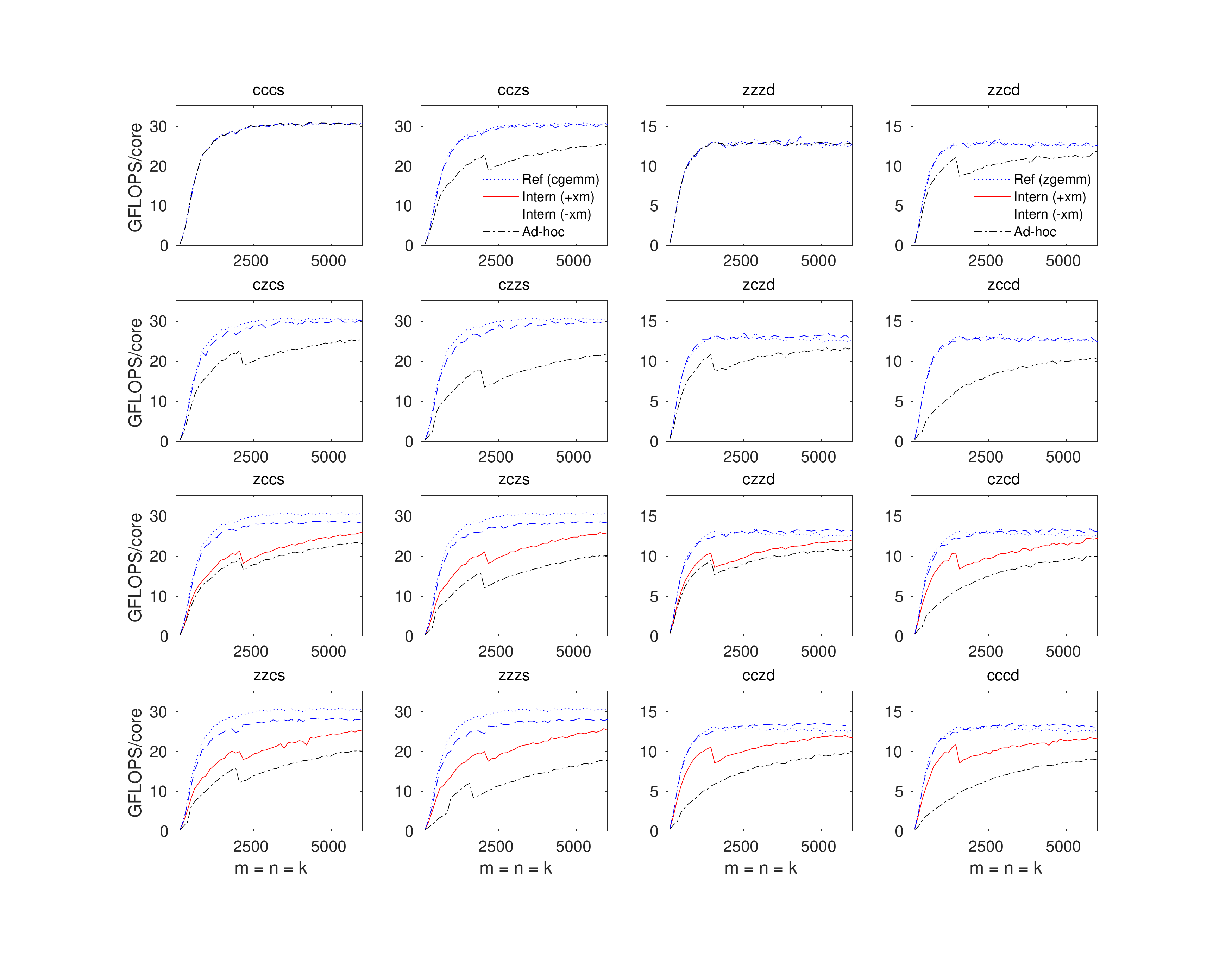}
\end{tabular}
\end{center}
\caption{
\sentencezeroa{\zerons}{\threens}\sentenceonea\sentencetwoa
}
\label{fig:perf_tx2_t56_0}
\end{figure}

%
%
\begin{figure}[hp!]
\begin{center}
\begin{tabular}{l}
\hspace{\graphhspace}
\includegraphics[width=\graphwidth,trim={\trimleft, \trimlower, \trimright, \trimupper},clip]{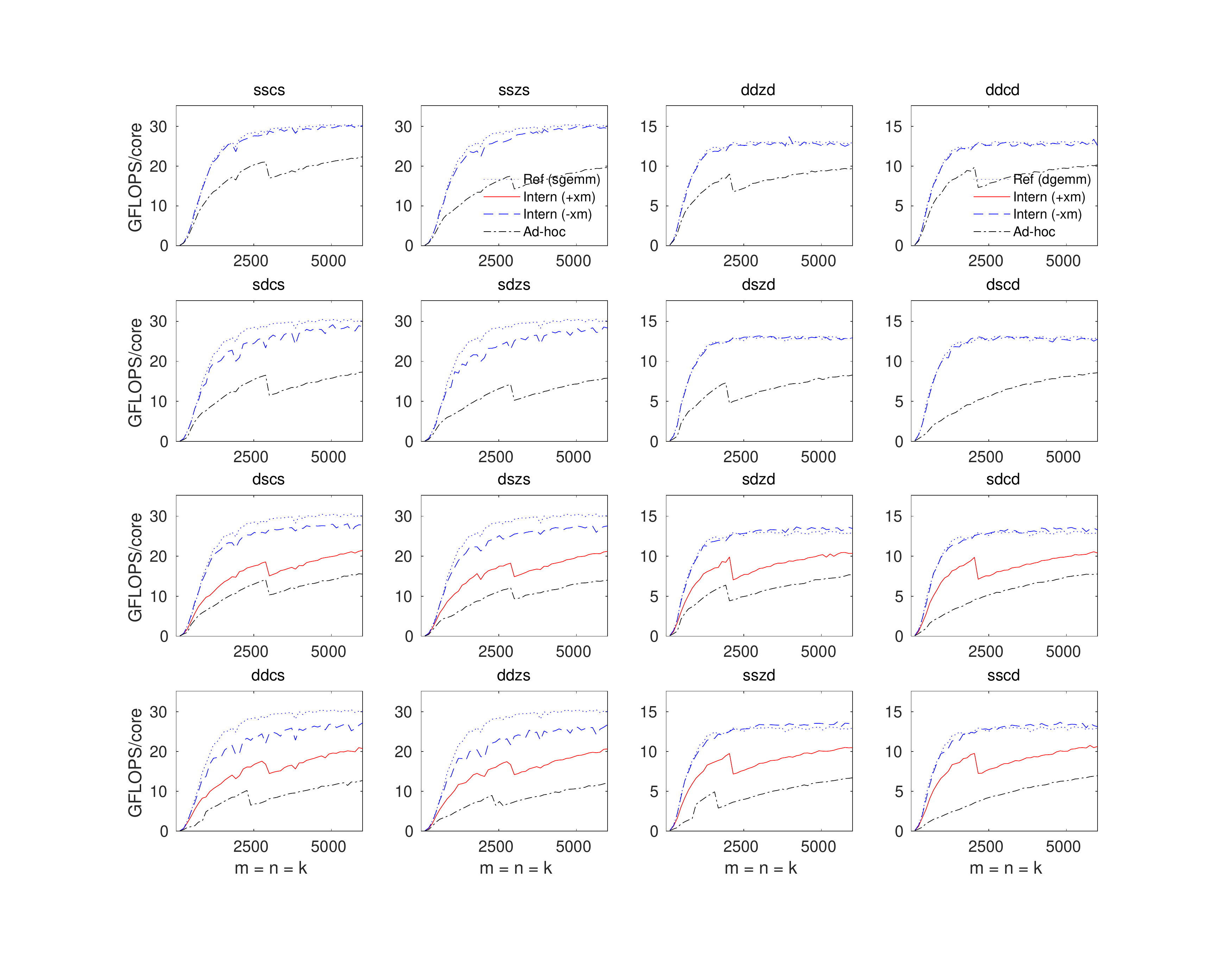} \\ \whline
\hspace{\graphhspace}
\includegraphics[width=\graphwidth,trim={\trimleft, \trimlower, \trimright, \trimupper},clip]{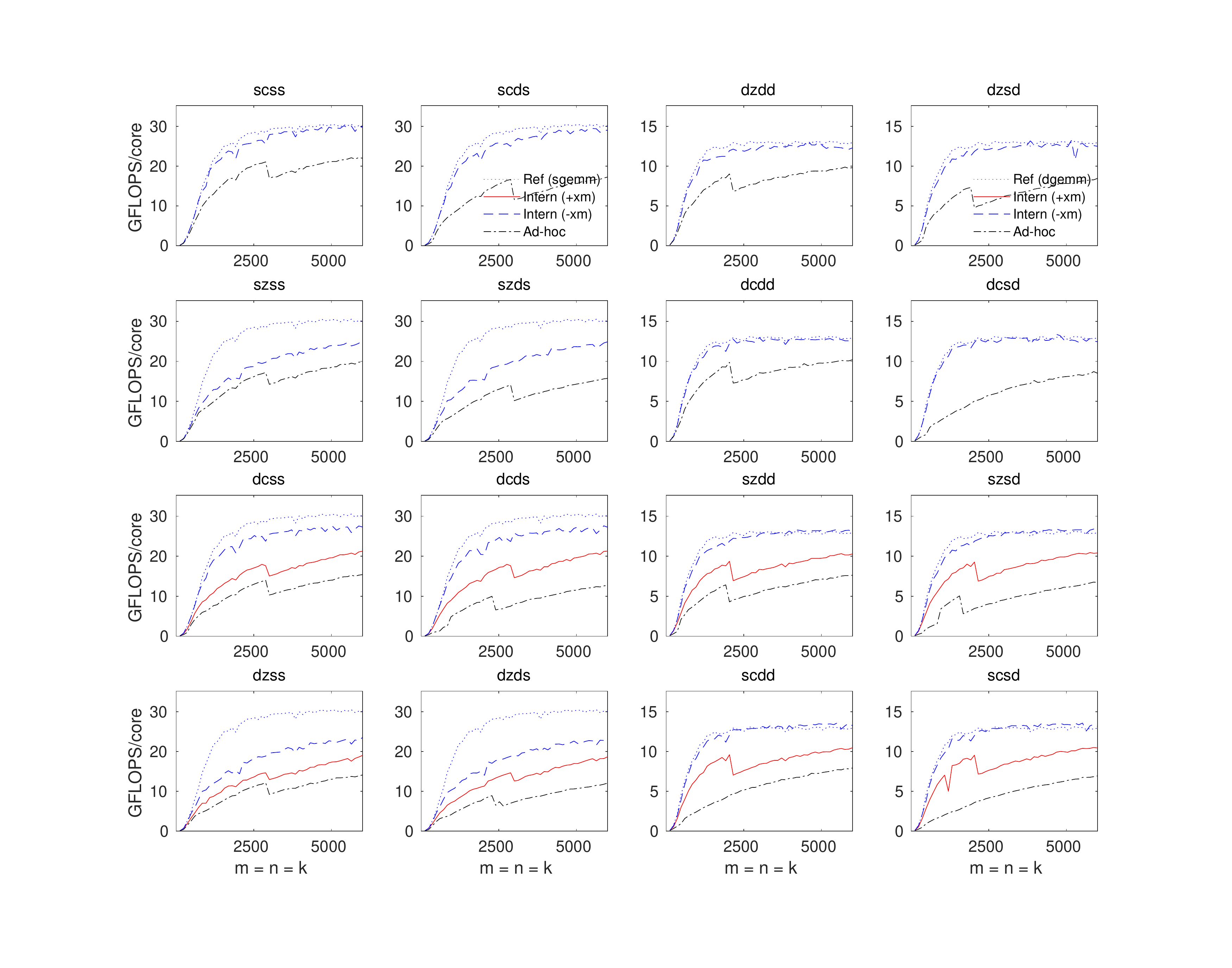}
\end{tabular}
\end{center}
\caption{
\sentencezeroa{\oneans}{\onebns}\sentenceonea\sentencetwoa
}
\label{fig:perf_tx2_t56_1}
\end{figure}

%
%
\begin{figure}[hp!]
\begin{center}
\begin{tabular}{l}
\hspace{\graphhspace}
\includegraphics[width=\graphwidth,trim={\trimleft, \trimlower, \trimright, \trimupper},clip]{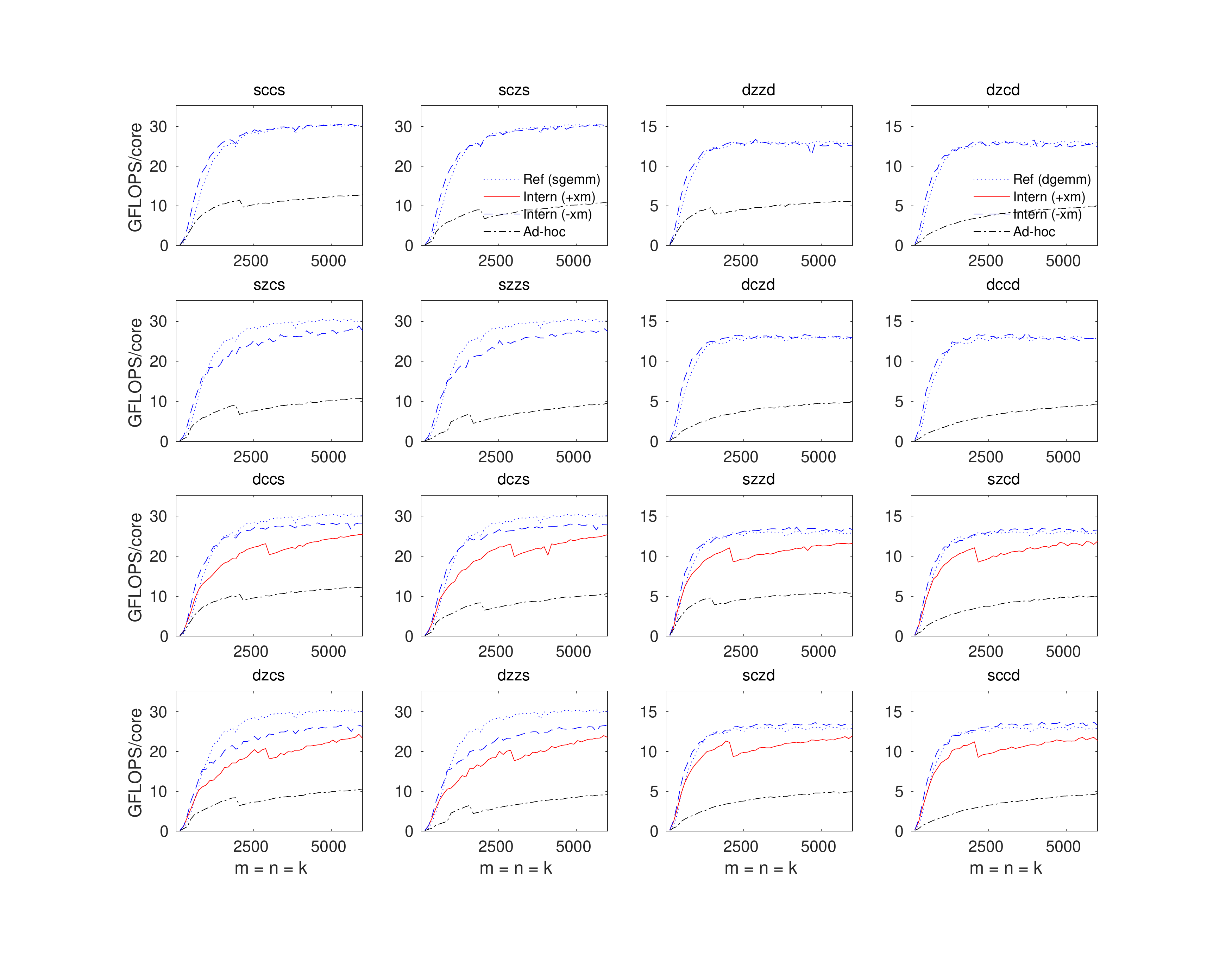} \\ \whline
\hspace{\graphhspace}
\includegraphics[width=\graphwidth,trim={\trimleft, \trimlower, \trimright, \trimupper},clip]{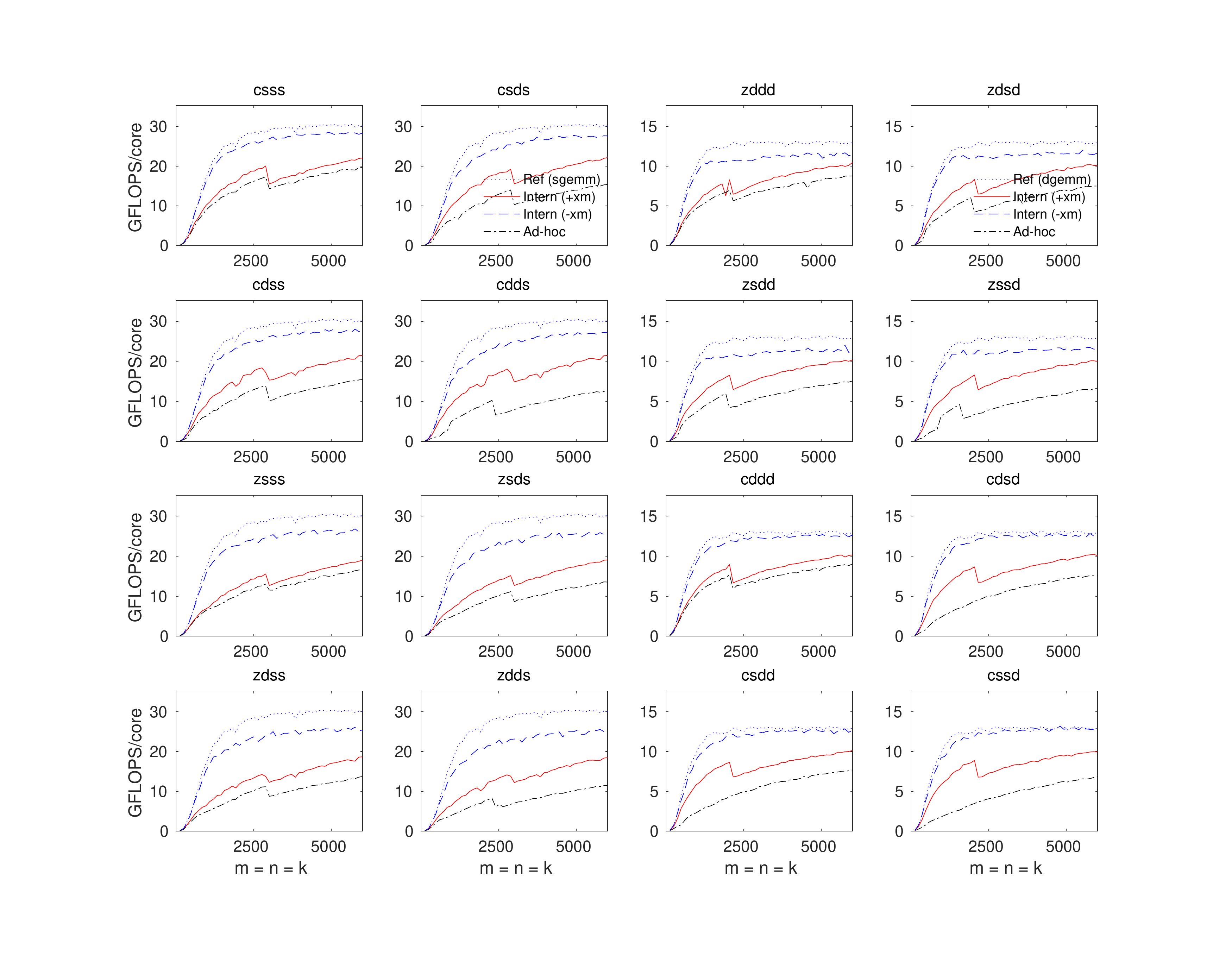}
\end{tabular}
\end{center}
\caption{
\sentencezeroa{\twoabns}{\onecns}\sentenceonea\sentencetwoa
}
\label{fig:perf_tx2_t56_2}
\end{figure}

%
%
\begin{figure}[hp!]
\begin{center}
\begin{tabular}{l}
\hspace{\graphhspace}
\includegraphics[width=\graphwidth,trim={\trimleft, \trimlower, \trimright, \trimupper},clip]{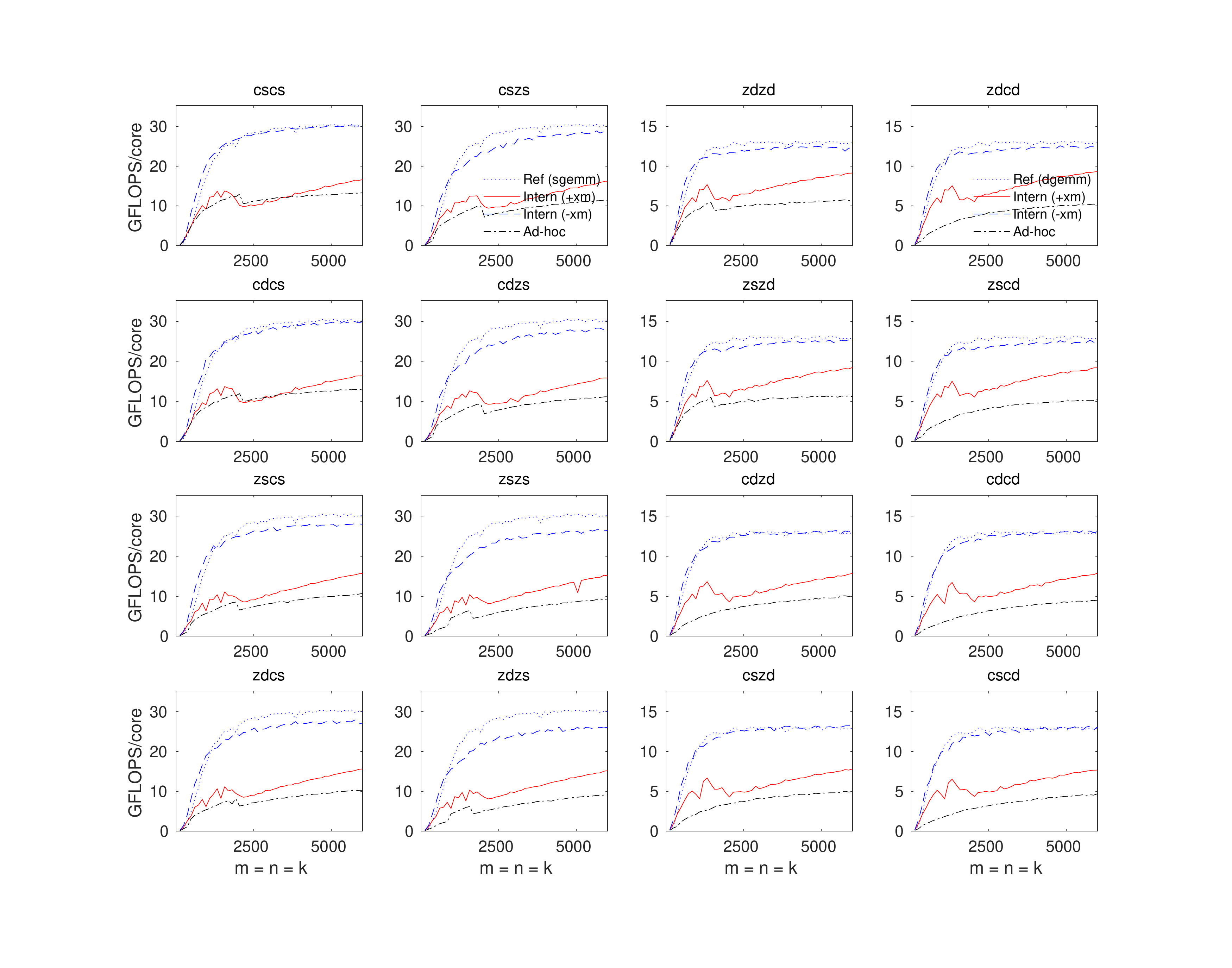} \\ \whline
\hspace{\graphhspace}
\includegraphics[width=\graphwidth,trim={\trimleft, \trimlower, \trimright, \trimupper},clip]{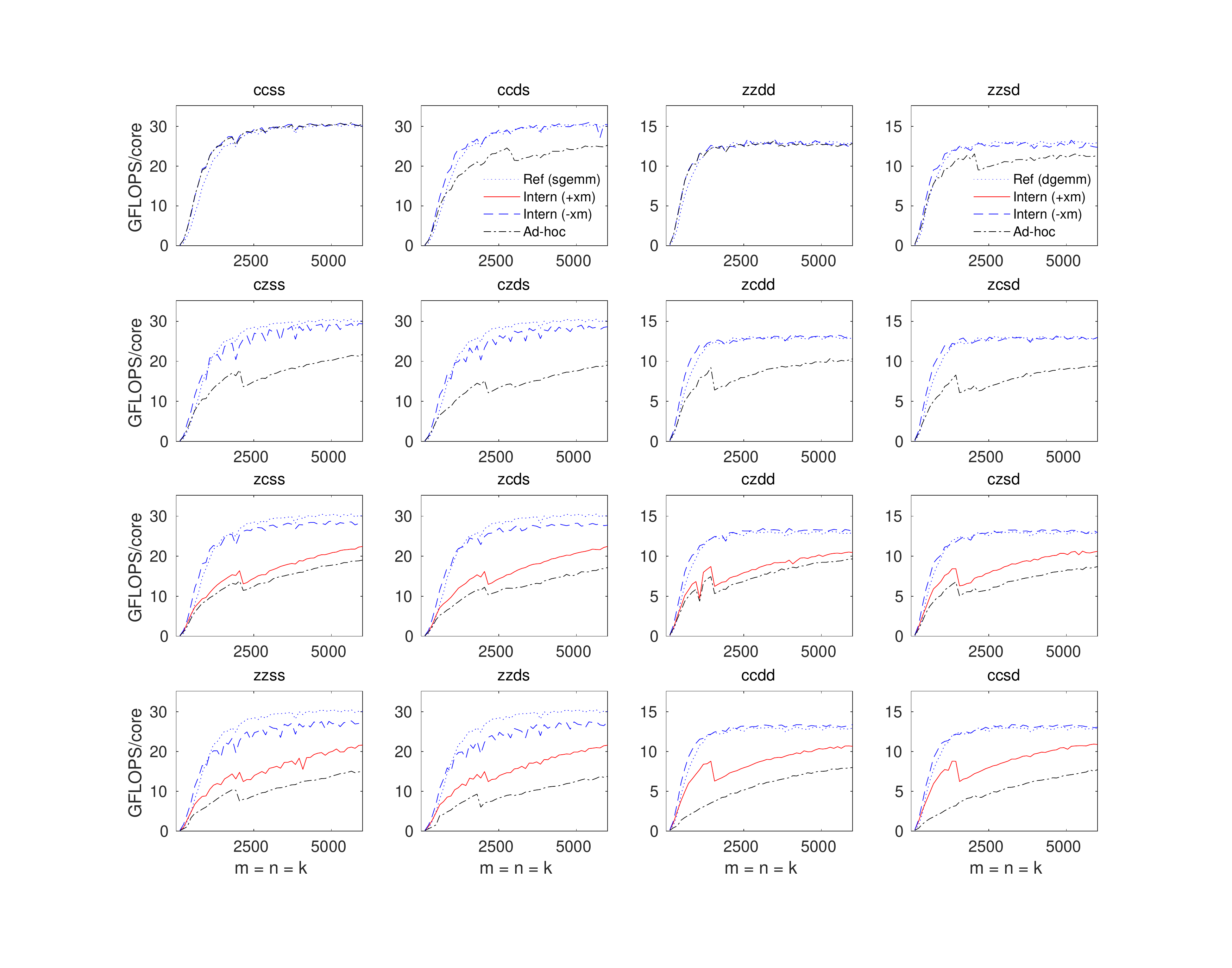}
\end{tabular}
\end{center}
\caption{
\sentencezeroa{\twobcns}{\twoacns}\sentenceonea\sentencetwoa
}
\label{fig:perf_tx2_t56_3}
\end{figure}


\renewcommand{\graphhspace}{-3.0mm}
\renewcommand{\graphwidth}{4.4in}
\renewcommand{\trimleft}{3.35cm}
\renewcommand{\trimlower}{2.7cm}
\renewcommand{\trimright}{3.7cm}
\renewcommand{\trimupper}{2.2cm}

\renewcommand{\sentencezeroa}[2]{Sequential performance of ``Internal'' and ``Ad-hoc'' implementations of \gemm for all precision combinations within mixed-domain Cases #1 (top) and #2 (bottom) on an Intel Xeon Platinum 8167M processor. }
\renewcommand{\sentenceonea}{The 16 graphs on the left side and right sides report computation in single- and double-precision, respectively. }
\renewcommand{\sentencetwoa}{The theoretical peak performance coincides with the top of each graph. }

%
%
\begin{figure}[hp!]
\begin{center}
\begin{tabular}{l}
\hspace{\graphhspace}
\includegraphics[width=\graphwidth,trim={\trimleft, \trimlower, \trimright, \trimupper},clip]{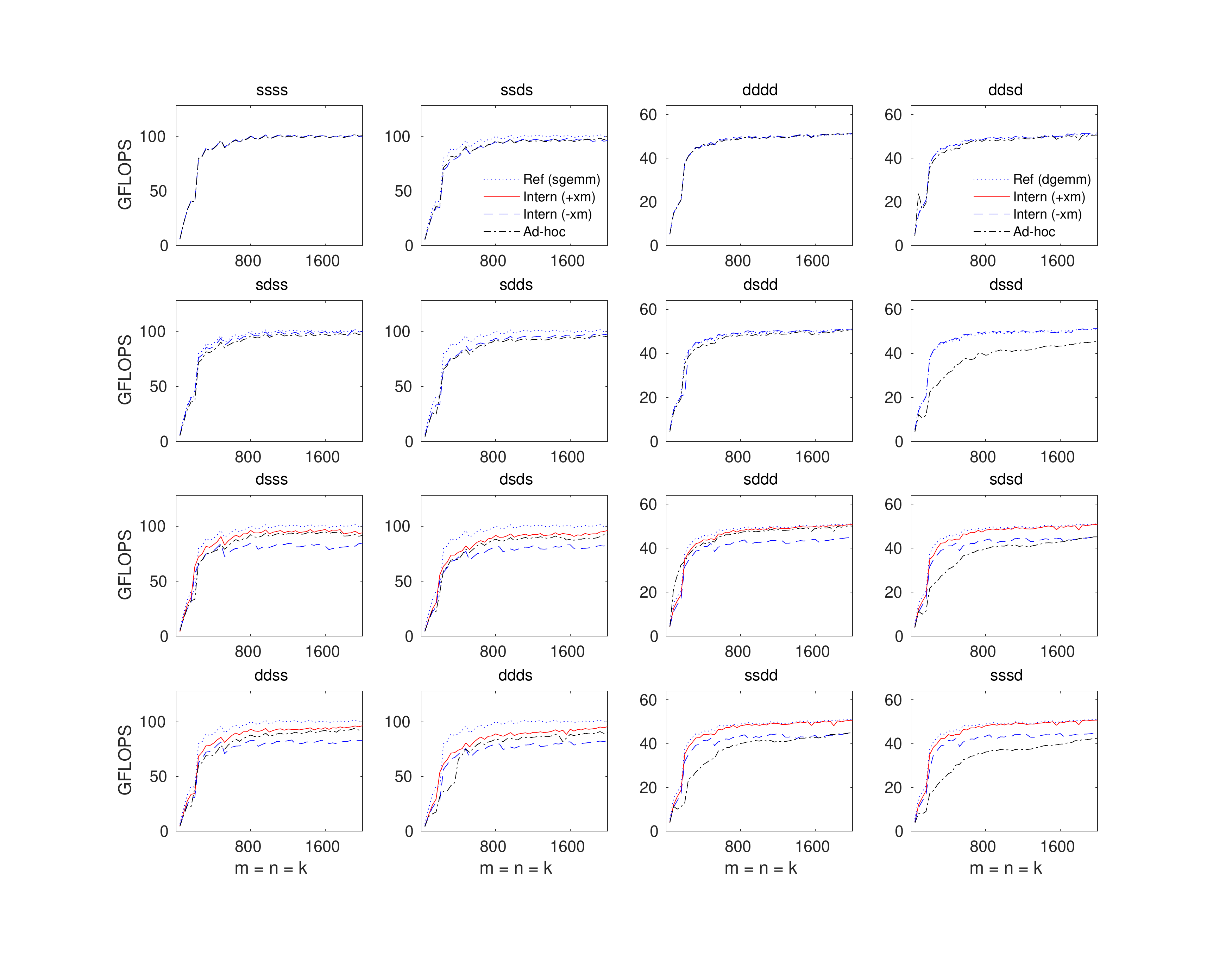} \\ \whline
\hspace{\graphhspace}
\includegraphics[width=\graphwidth,trim={\trimleft, \trimlower, \trimright, \trimupper},clip]{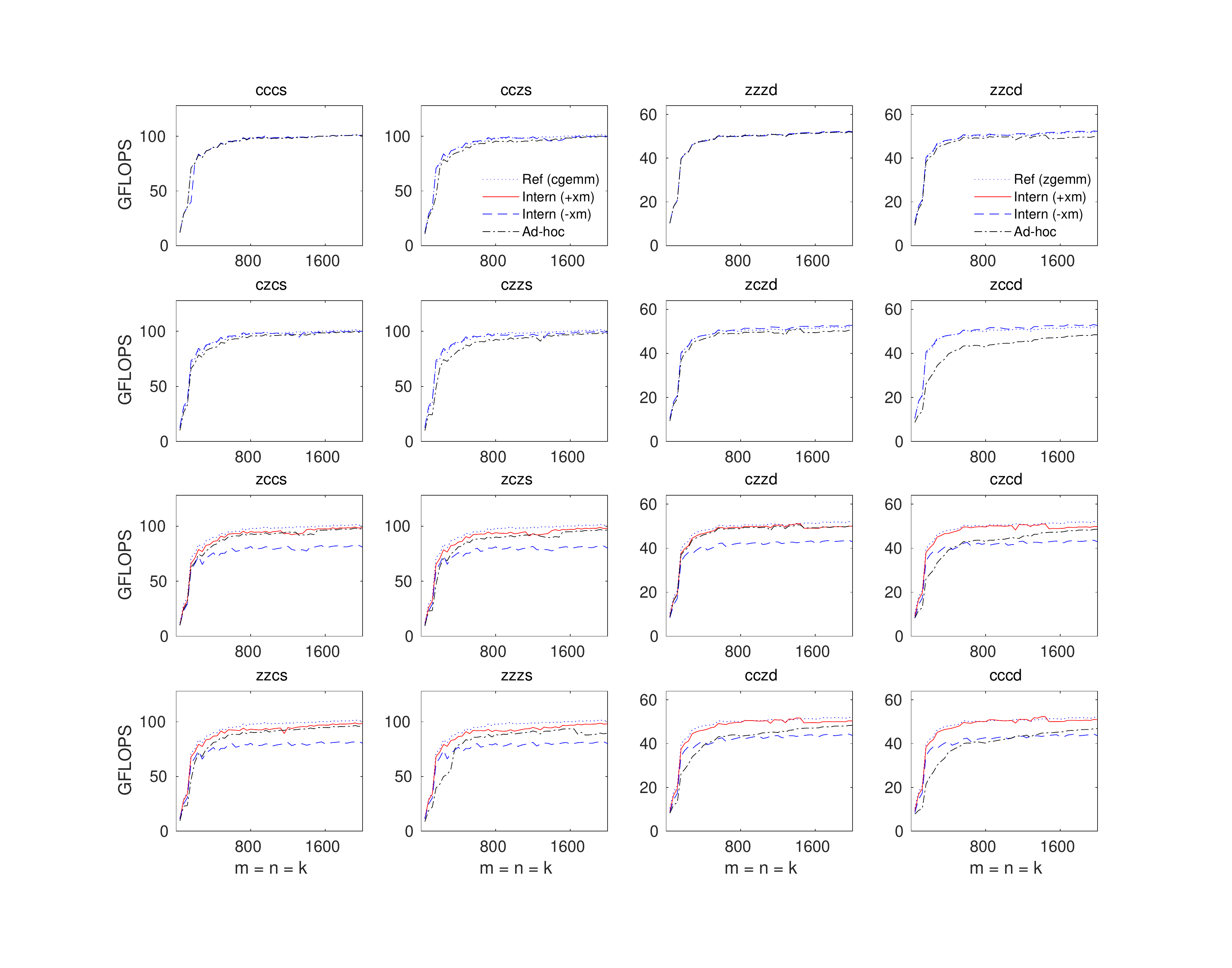}
\end{tabular}
\end{center}
\caption{
\sentencezeroa{\zerons}{\threens}\sentenceonea\sentencetwoa
}
\label{fig:perf_skx_t1_0}
\end{figure}

%
%
\begin{figure}[hp!]
\begin{center}
\begin{tabular}{l}
\hspace{\graphhspace}
\includegraphics[width=\graphwidth,trim={\trimleft, \trimlower, \trimright, \trimupper},clip]{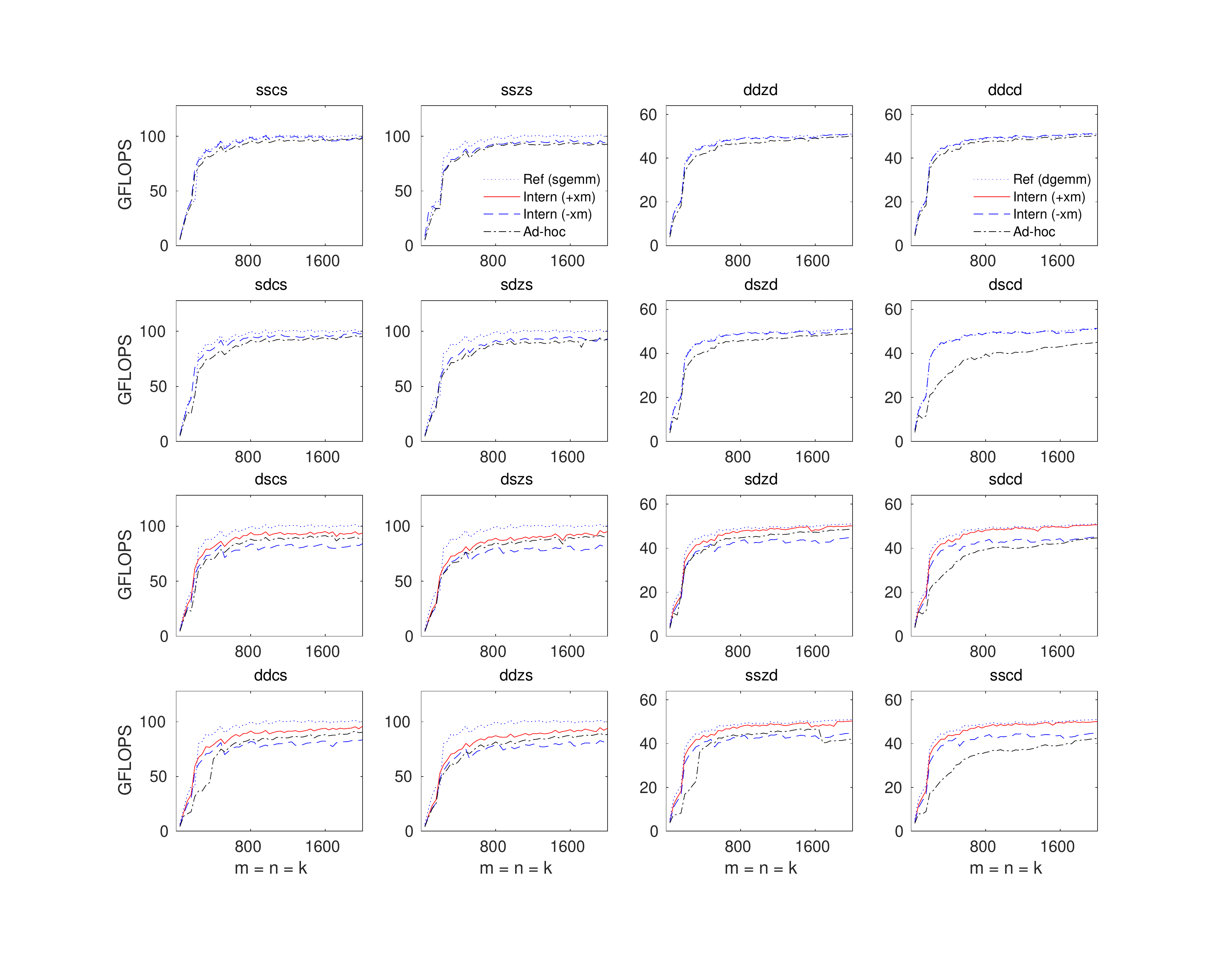} \\ \whline
\hspace{\graphhspace}
\includegraphics[width=\graphwidth,trim={\trimleft, \trimlower, \trimright, \trimupper},clip]{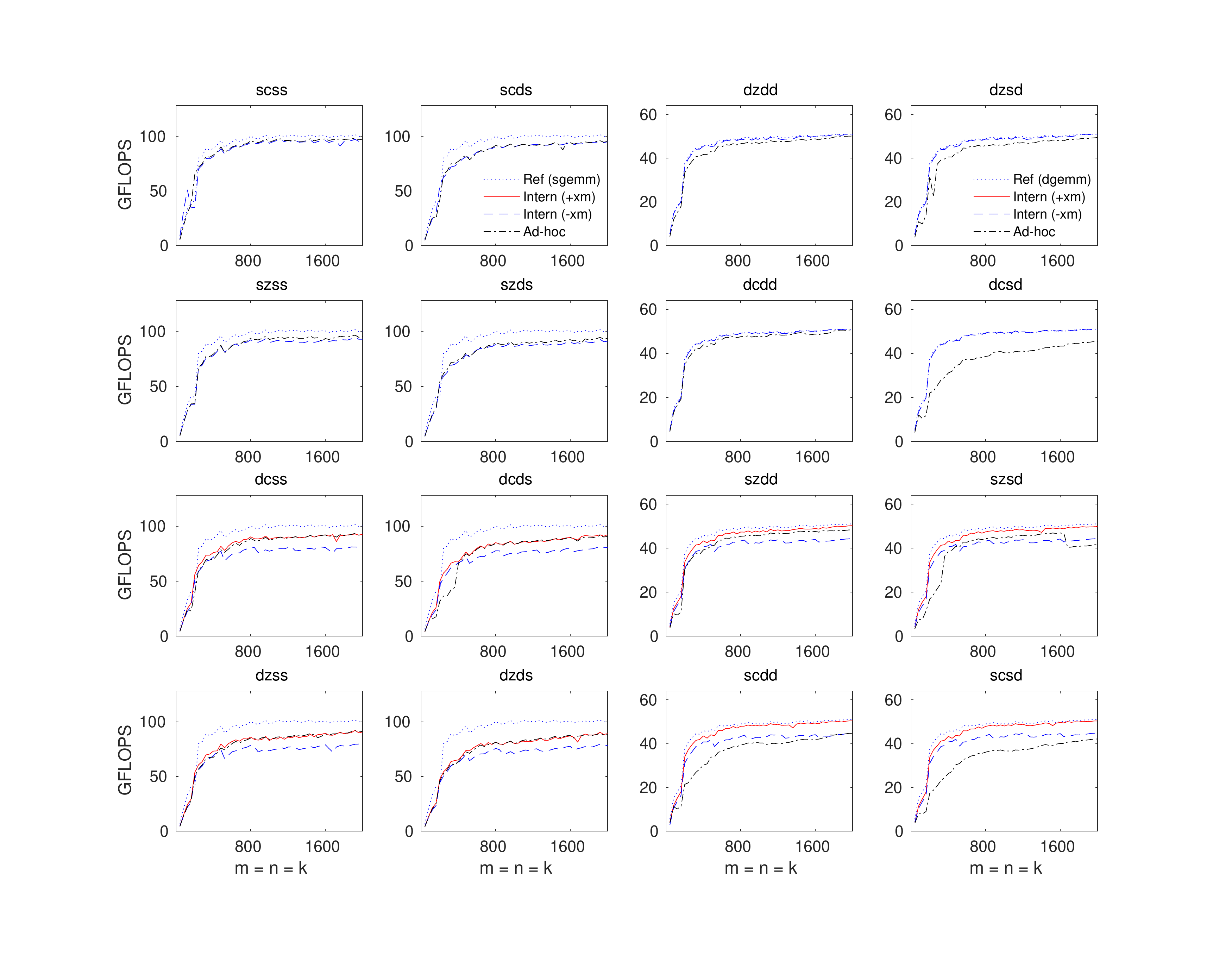}
\end{tabular}
\end{center}
\caption{
\sentencezeroa{\oneans}{\onebns}\sentenceonea\sentencetwoa
}
\label{fig:perf_skx_t1_1}
\end{figure}

%
%
\begin{figure}[hp!]
\begin{center}
\begin{tabular}{l}
\hspace{\graphhspace}
\includegraphics[width=\graphwidth,trim={\trimleft, \trimlower, \trimright, \trimupper},clip]{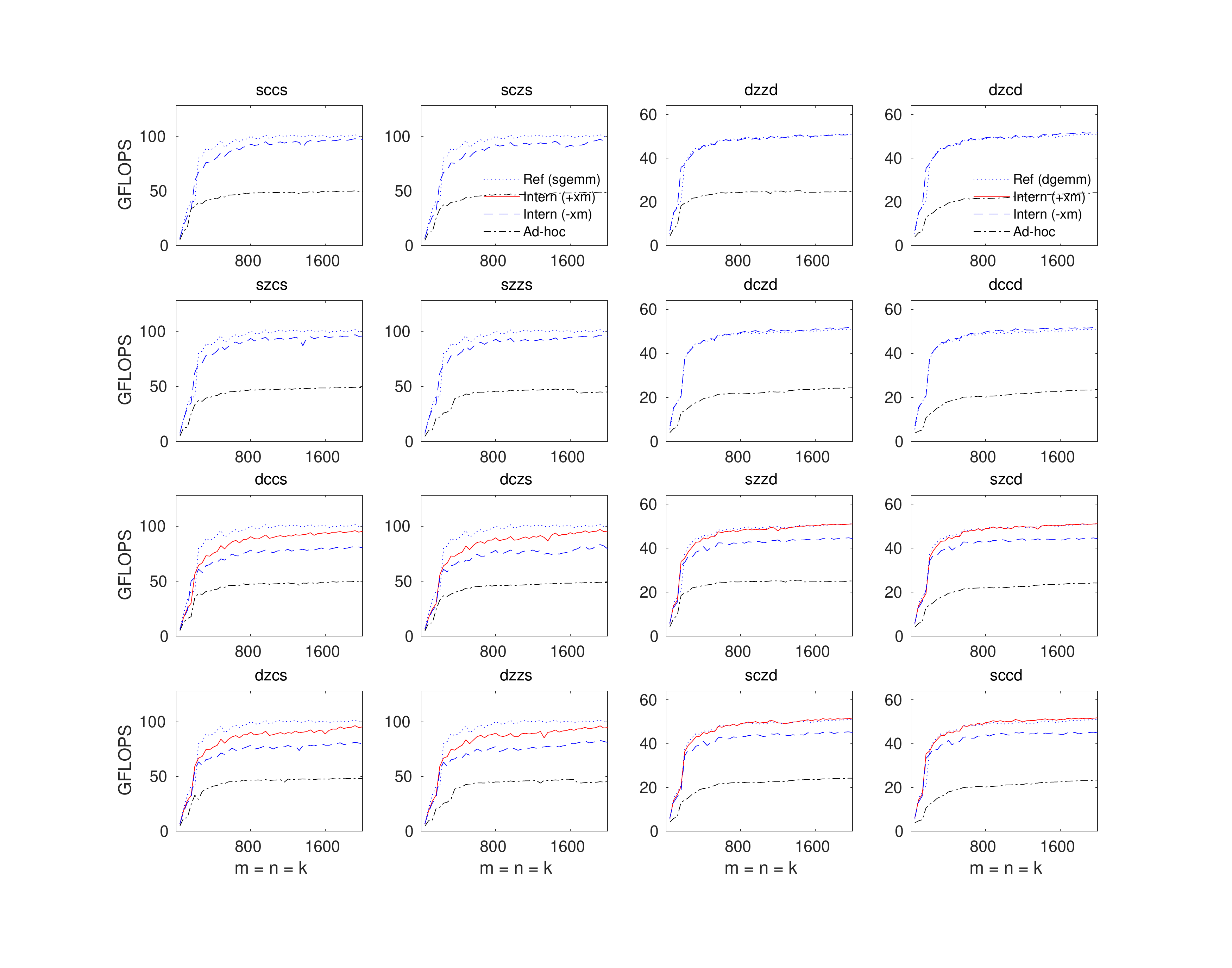} \\ \whline
\hspace{\graphhspace}
\includegraphics[width=\graphwidth,trim={\trimleft, \trimlower, \trimright, \trimupper},clip]{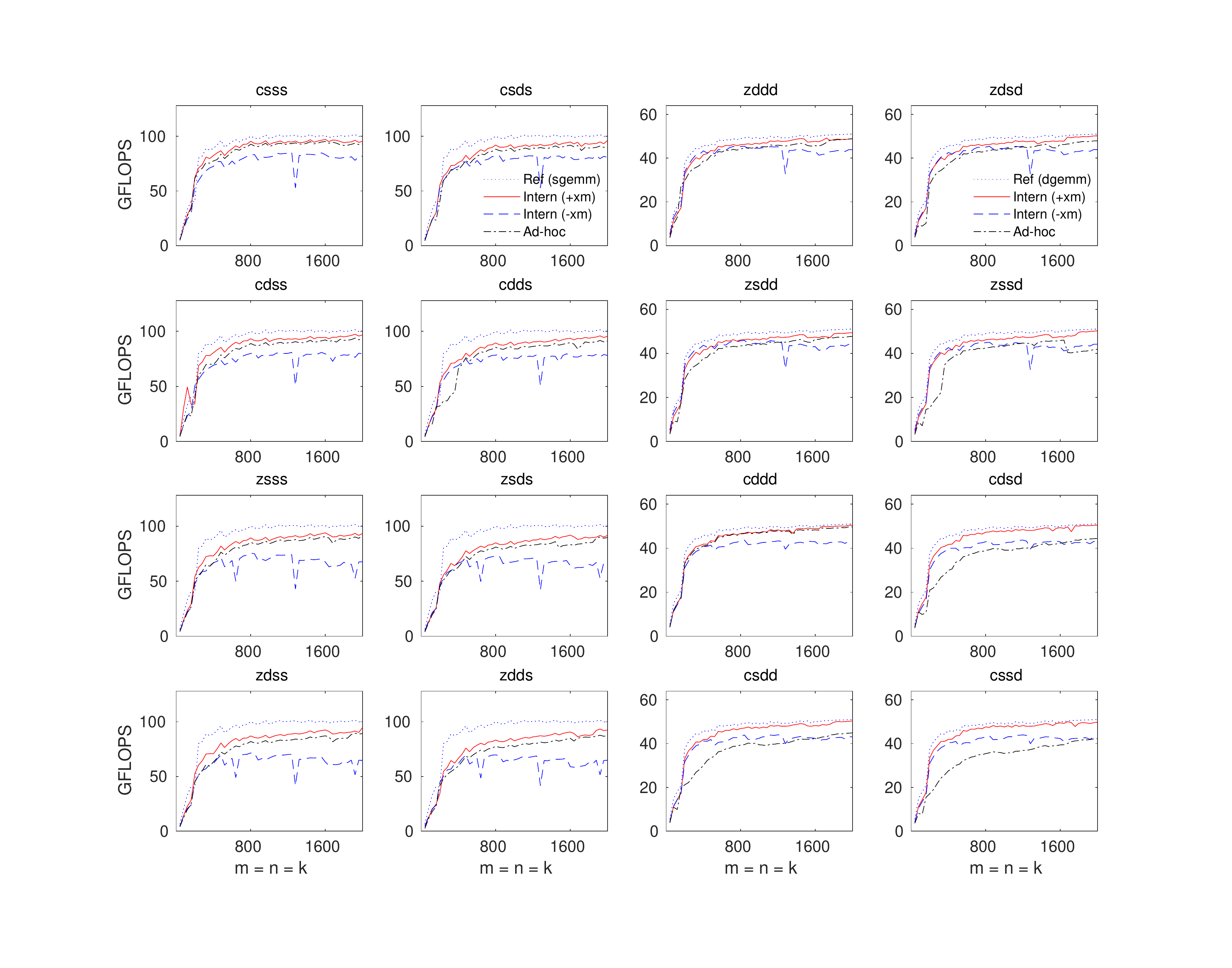}
\end{tabular}
\end{center}
\caption{
\sentencezeroa{\twoabns}{\onecns}\sentenceonea\sentencetwoa
}
\label{fig:perf_skx_t1_2}
\end{figure}

%
%
\begin{figure}[hp!]
\begin{center}
\begin{tabular}{l}
\hspace{\graphhspace}
\includegraphics[width=\graphwidth,trim={\trimleft, \trimlower, \trimright, \trimupper},clip]{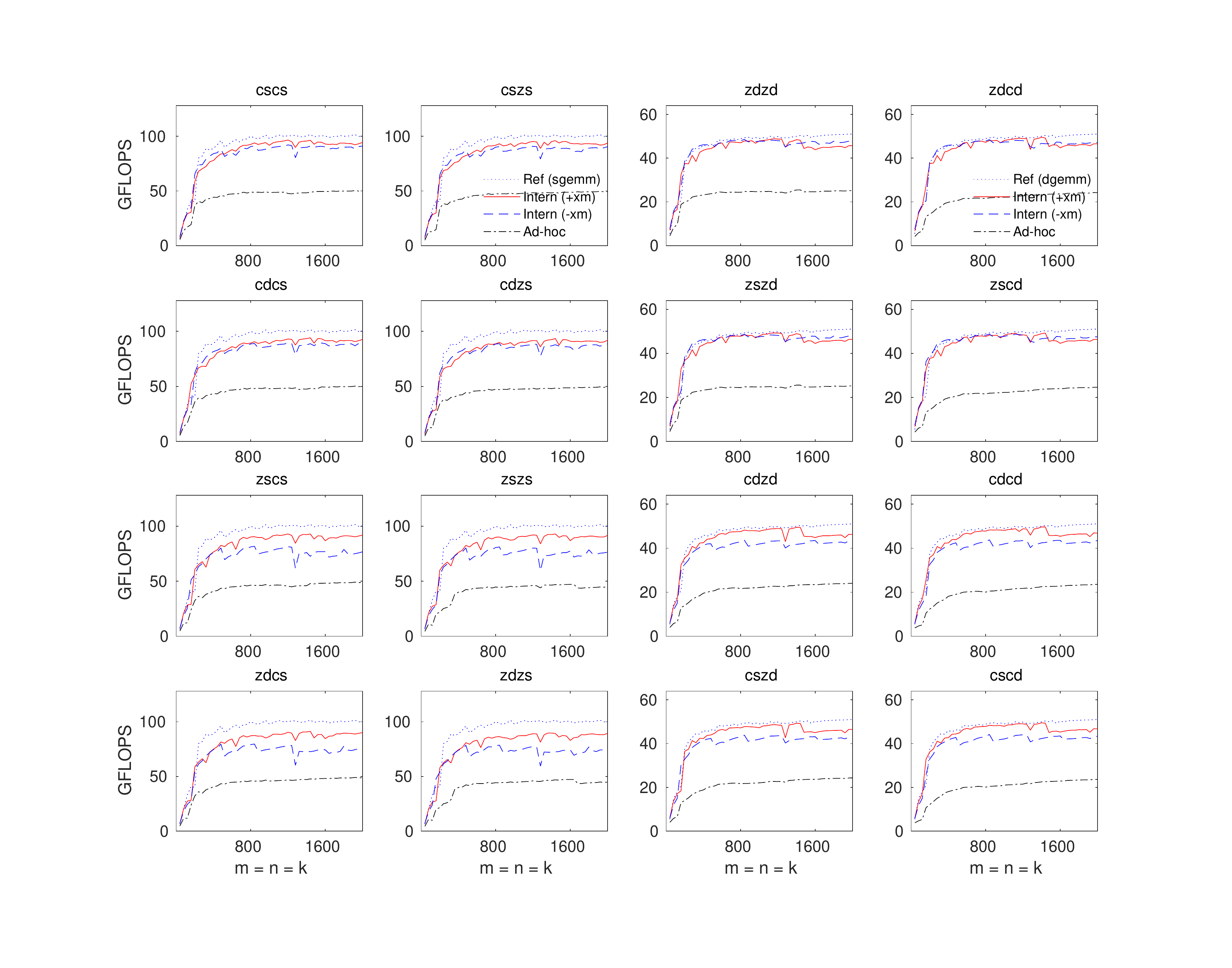} \\ \whline
\hspace{\graphhspace}
\includegraphics[width=\graphwidth,trim={\trimleft, \trimlower, \trimright, \trimupper},clip]{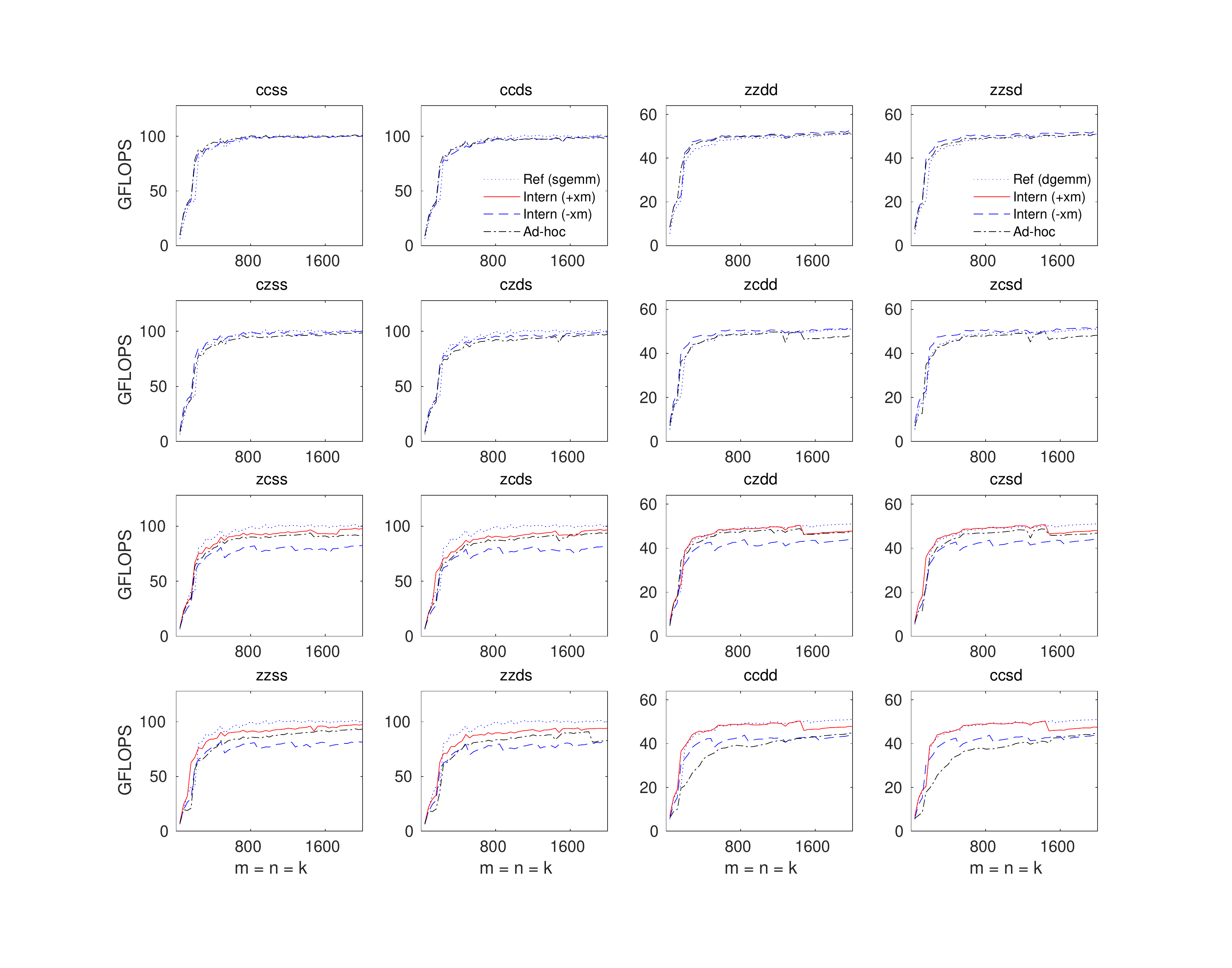}
\end{tabular}
\end{center}
\caption{
\sentencezeroa{\twobcns}{\twoacns}\sentenceonea\sentencetwoa
}
\label{fig:perf_skx_t1_3}
\end{figure}

\renewcommand{\graphhspace}{-3.0mm}
\renewcommand{\graphwidth}{4.4in}
\renewcommand{\trimleft}{3.35cm}
\renewcommand{\trimlower}{2.7cm}
\renewcommand{\trimright}{3.7cm}
\renewcommand{\trimupper}{2.2cm}

\renewcommand{\sentencezeroa}[2]{Multithreaded (26 threads) performance of ``Internal'' and ``Ad-hoc'' implementations of \gemm for all precision combinations within mixed-domain Cases #1 (top) and #2 (bottom) on an Intel Xeon Platinum 8167M processor. }
\renewcommand{\sentenceonea}{The 16 graphs on the left side and right sides report computation in single- and double-precision, respectively. }
\renewcommand{\sentencetwoa}{The theoretical peak performance coincides with the top of each graph. }

%
%
\begin{figure}[hp!]
\begin{center}
\begin{tabular}{l}
\hspace{\graphhspace}
\includegraphics[width=\graphwidth,trim={\trimleft, \trimlower, \trimright, \trimupper},clip]{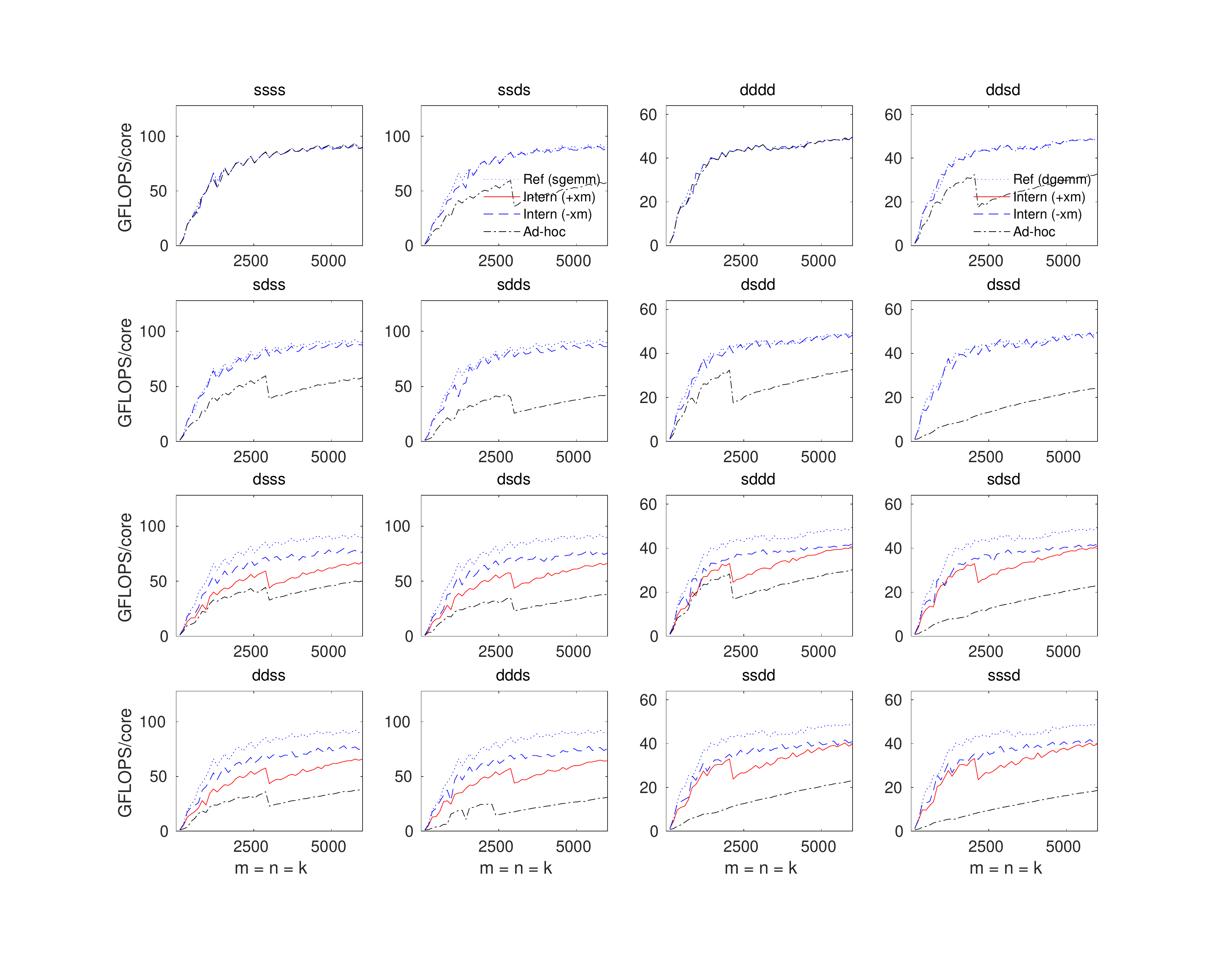} \\ \whline
\hspace{\graphhspace}
\includegraphics[width=\graphwidth,trim={\trimleft, \trimlower, \trimright, \trimupper},clip]{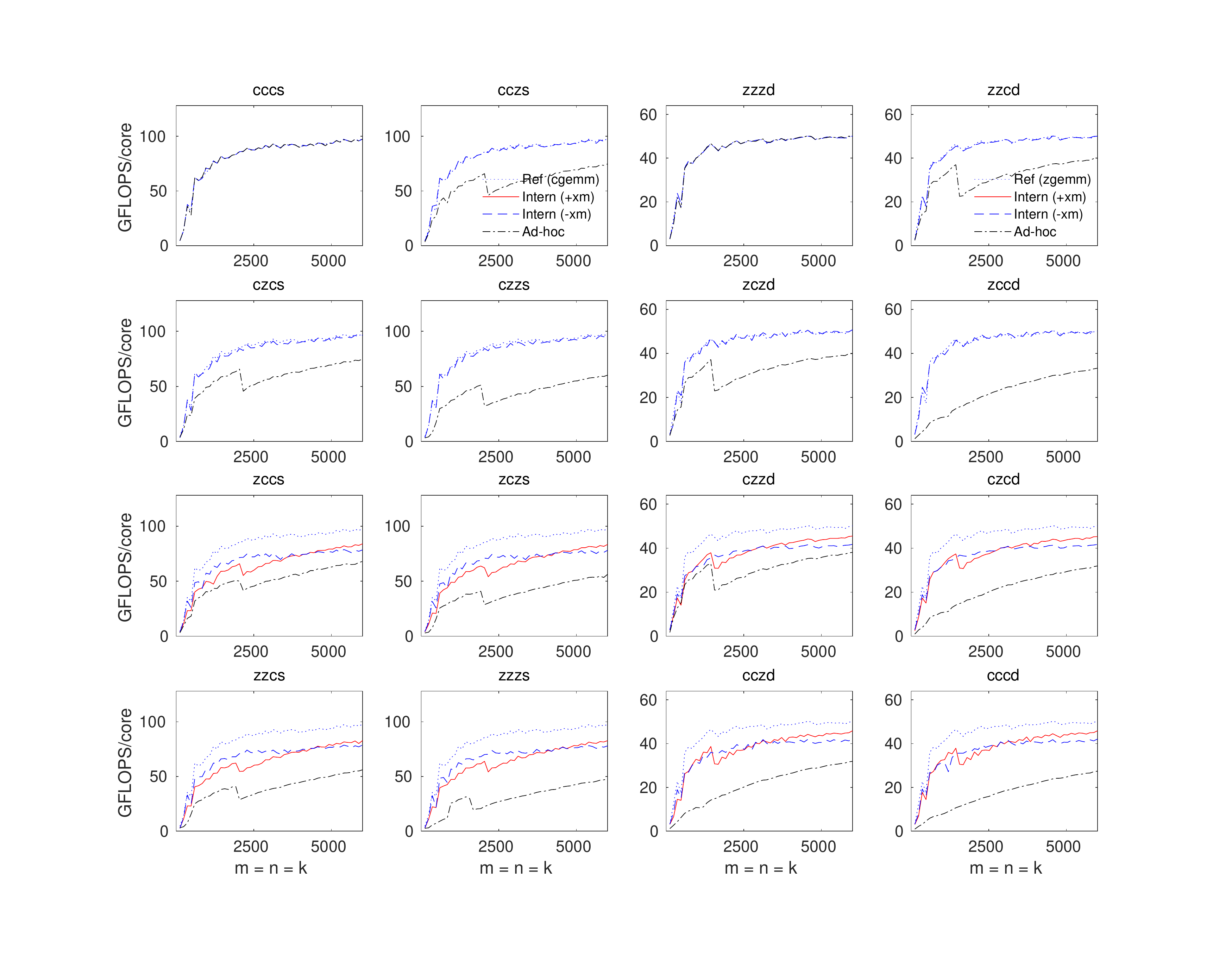}
\end{tabular}
\end{center}
\caption{
\sentencezeroa{\zerons}{\threens}\sentenceonea\sentencetwoa
}
\label{fig:perf_skx_t26_0}
\end{figure}

%
%
\begin{figure}[hp!]
\begin{center}
\begin{tabular}{l}
\hspace{\graphhspace}
\includegraphics[width=\graphwidth,trim={\trimleft, \trimlower, \trimright, \trimupper},clip]{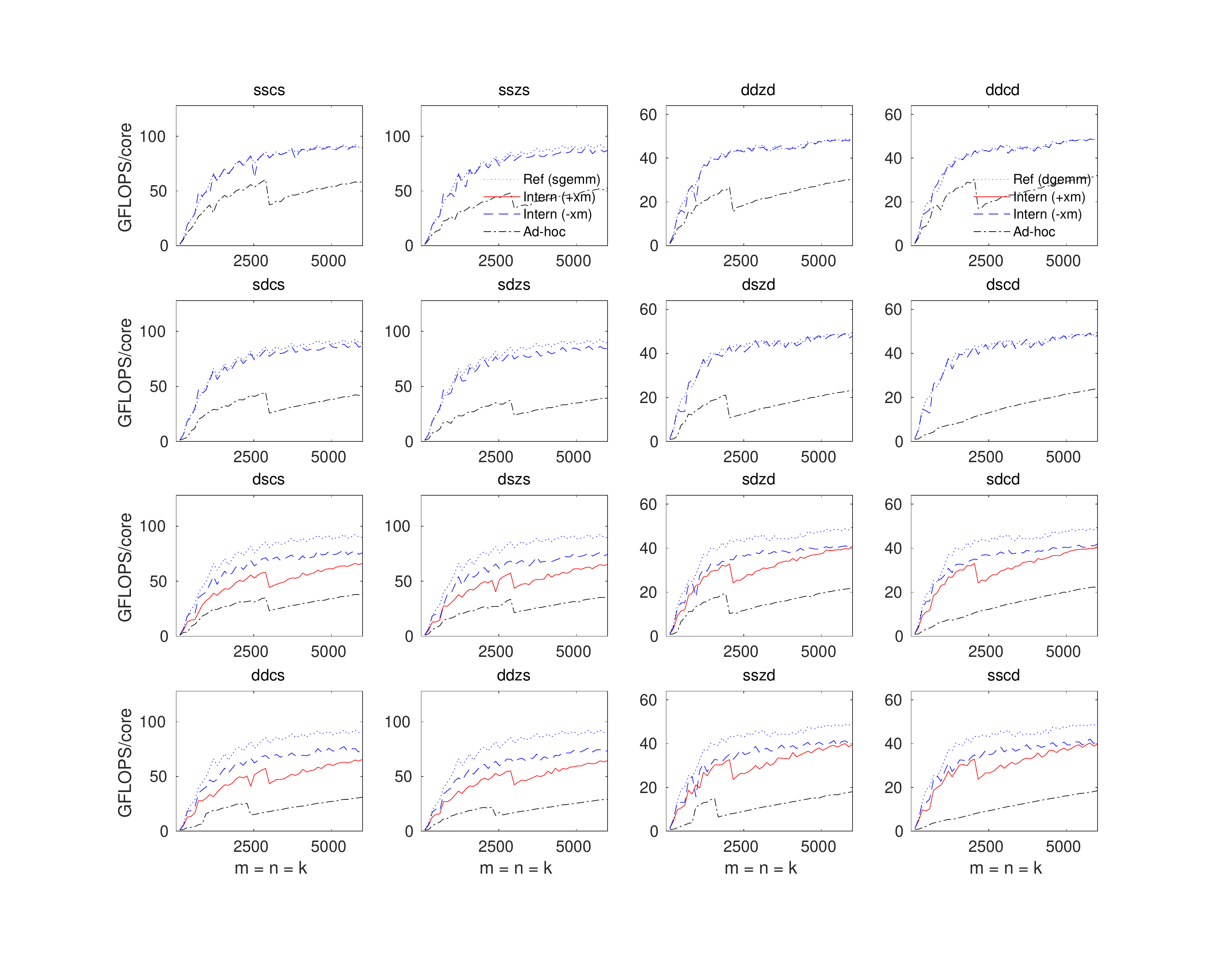} \\ \whline
\hspace{\graphhspace}
\includegraphics[width=\graphwidth,trim={\trimleft, \trimlower, \trimright, \trimupper},clip]{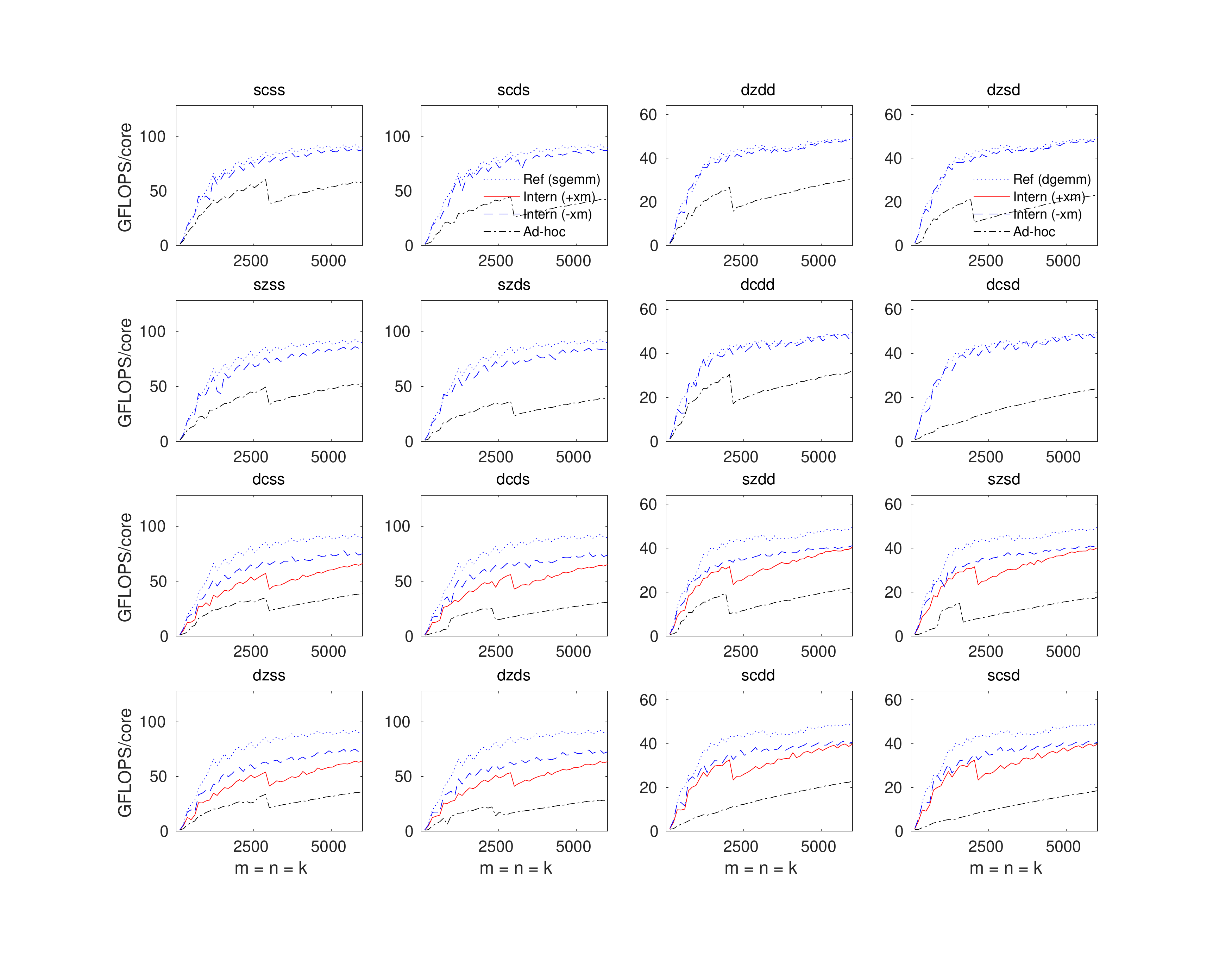}
\end{tabular}
\end{center}
\caption{
\sentencezeroa{\oneans}{\onebns}\sentenceonea\sentencetwoa
}
\label{fig:perf_skx_t26_1}
\end{figure}

%
%
\begin{figure}[hp!]
\begin{center}
\begin{tabular}{l}
\hspace{\graphhspace}
\includegraphics[width=\graphwidth,trim={\trimleft, \trimlower, \trimright, \trimupper},clip]{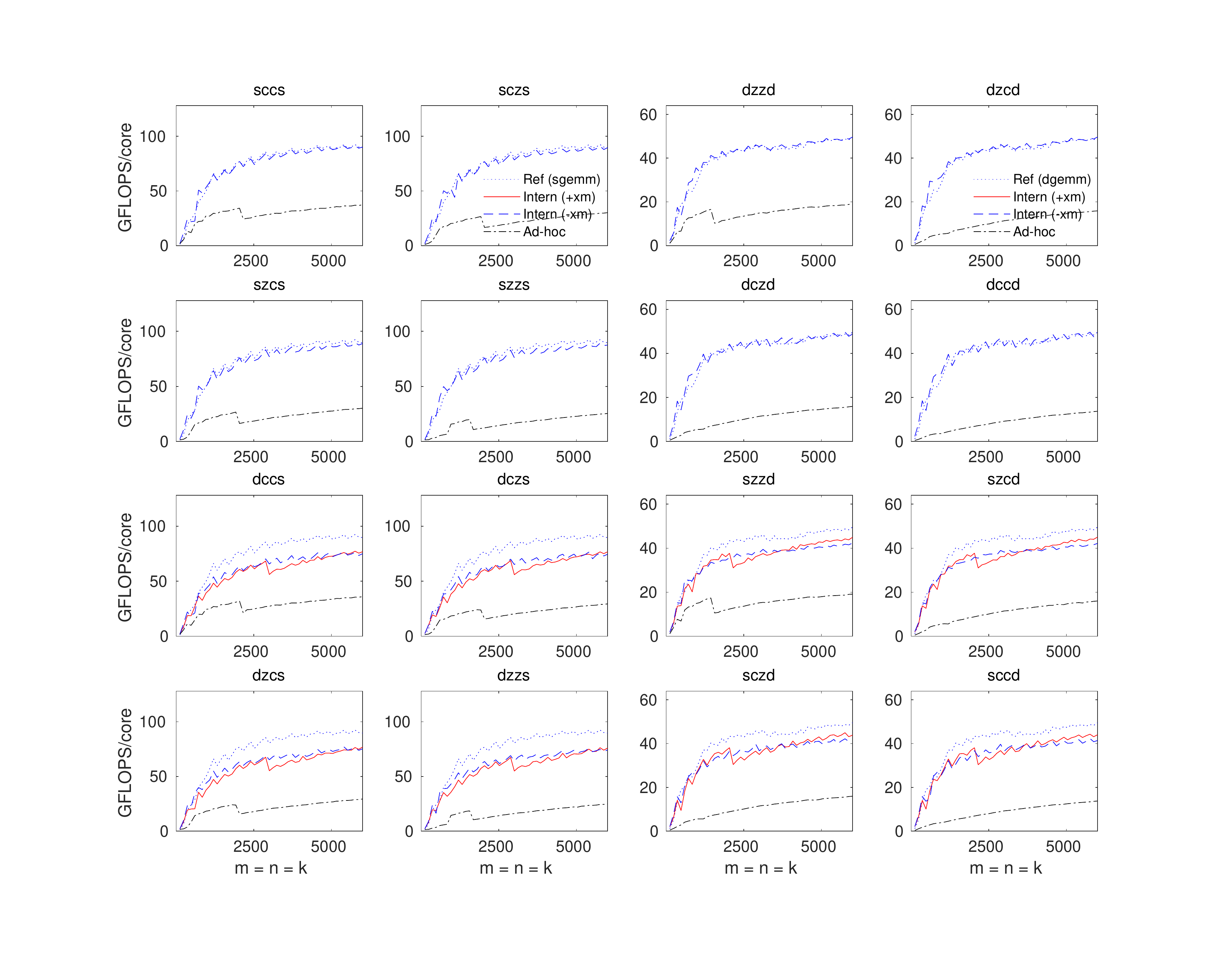} \\ \whline
\hspace{\graphhspace}
\includegraphics[width=\graphwidth,trim={\trimleft, \trimlower, \trimright, \trimupper},clip]{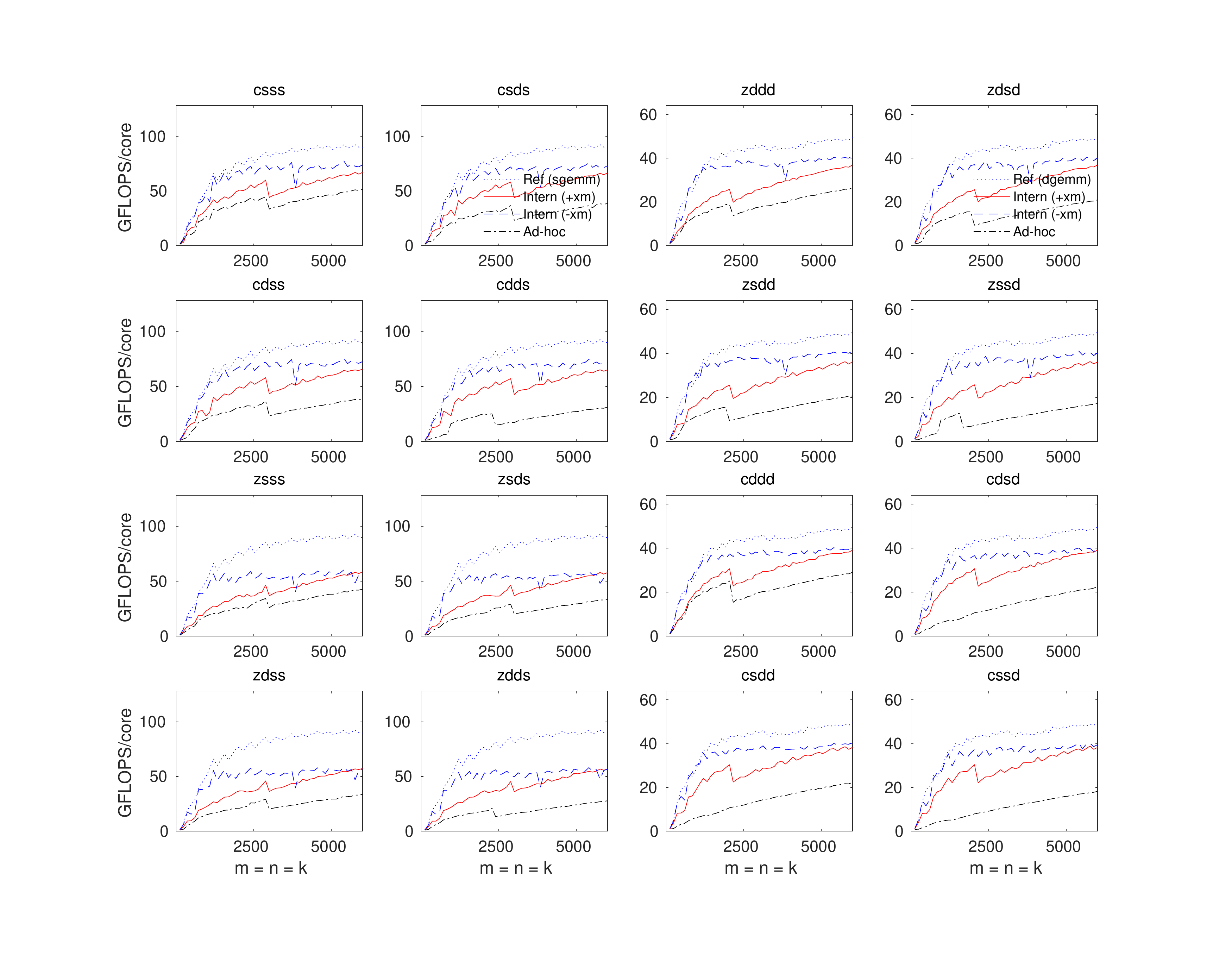}
\end{tabular}
\end{center}
\caption{
\sentencezeroa{\twoabns}{\onecns}\sentenceonea\sentencetwoa
}
\label{fig:perf_skx_t26_2}
\end{figure}

%
%
\begin{figure}[hp!]
\begin{center}
\begin{tabular}{l}
\hspace{\graphhspace}
\includegraphics[width=\graphwidth,trim={\trimleft, \trimlower, \trimright, \trimupper},clip]{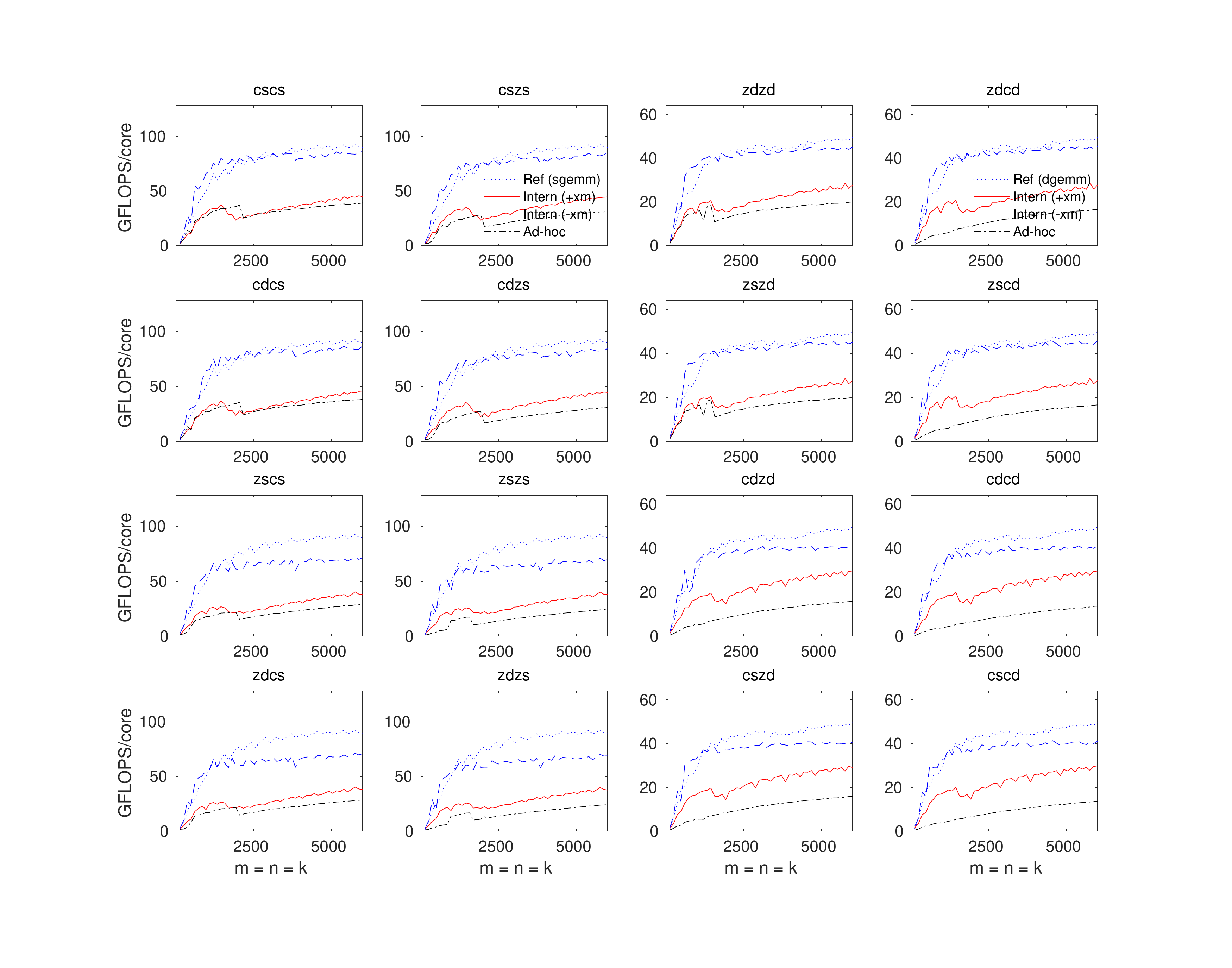} \\ \whline
\hspace{\graphhspace}
\includegraphics[width=\graphwidth,trim={\trimleft, \trimlower, \trimright, \trimupper},clip]{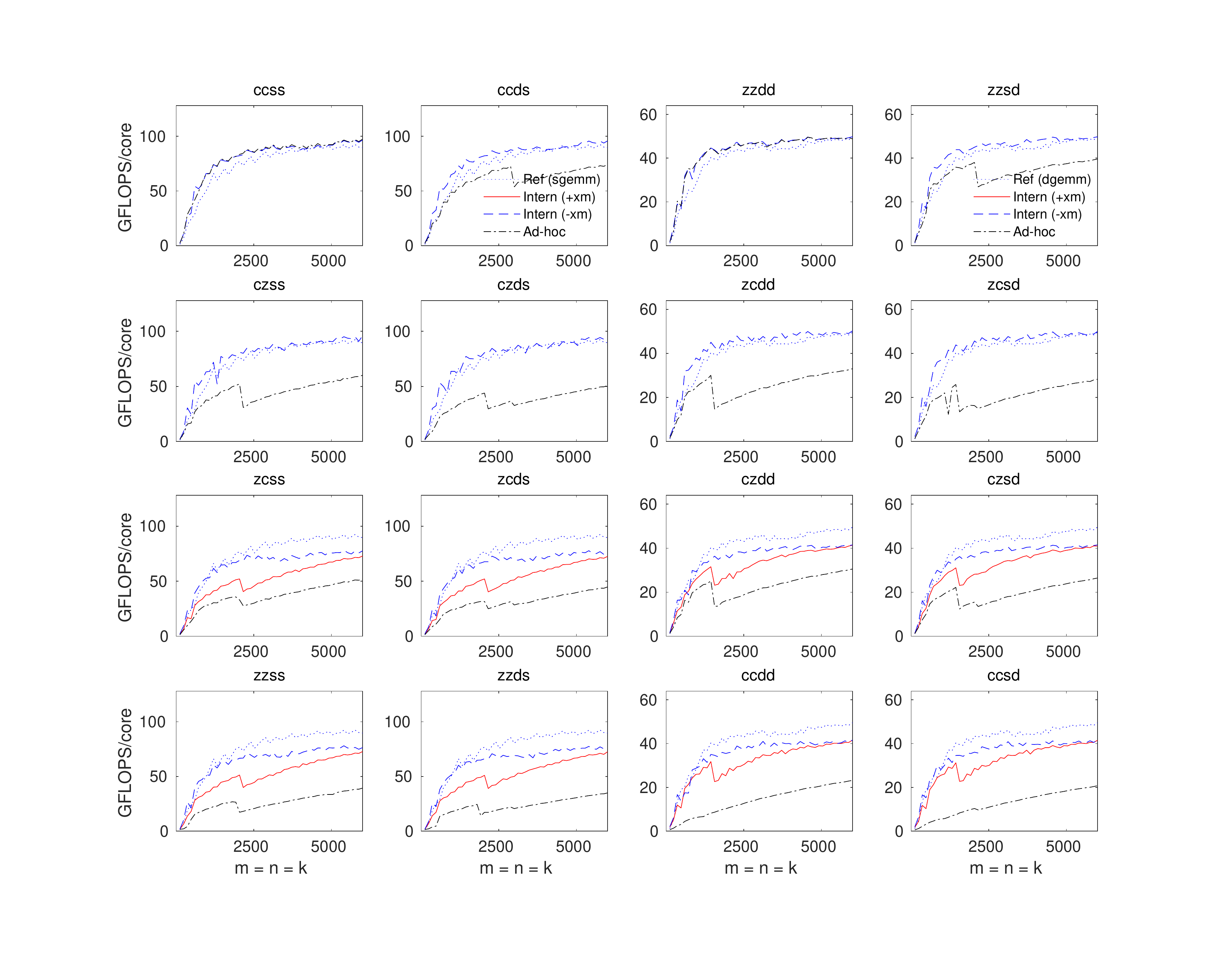}
\end{tabular}
\end{center}
\caption{
\sentencezeroa{\twobcns}{\twoacns}\sentenceonea\sentencetwoa
}
\label{fig:perf_skx_t26_3}
\end{figure}

\renewcommand{\graphhspace}{-3.0mm}
\renewcommand{\graphwidth}{4.4in}
\renewcommand{\trimleft}{3.35cm}
\renewcommand{\trimlower}{2.7cm}
\renewcommand{\trimright}{3.7cm}
\renewcommand{\trimupper}{2.2cm}

\renewcommand{\sentencezeroa}[2]{Multithreaded (52 threads) performance of ``Internal'' and ``Ad-hoc'' implementations of \gemm for all precision combinations within mixed-domain Cases #1 (top) and #2 (bottom) on an Intel Xeon Platinum 8167M processor. }
\renewcommand{\sentenceonea}{The 16 graphs on the left side and right sides report computation in single- and double-precision, respectively. }
\renewcommand{\sentencetwoa}{The theoretical peak performance coincides with the top of each graph. }

%
%
\begin{figure}[hp!]
\begin{center}
\begin{tabular}{l}
\hspace{\graphhspace}
\includegraphics[width=\graphwidth,trim={\trimleft, \trimlower, \trimright, \trimupper},clip]{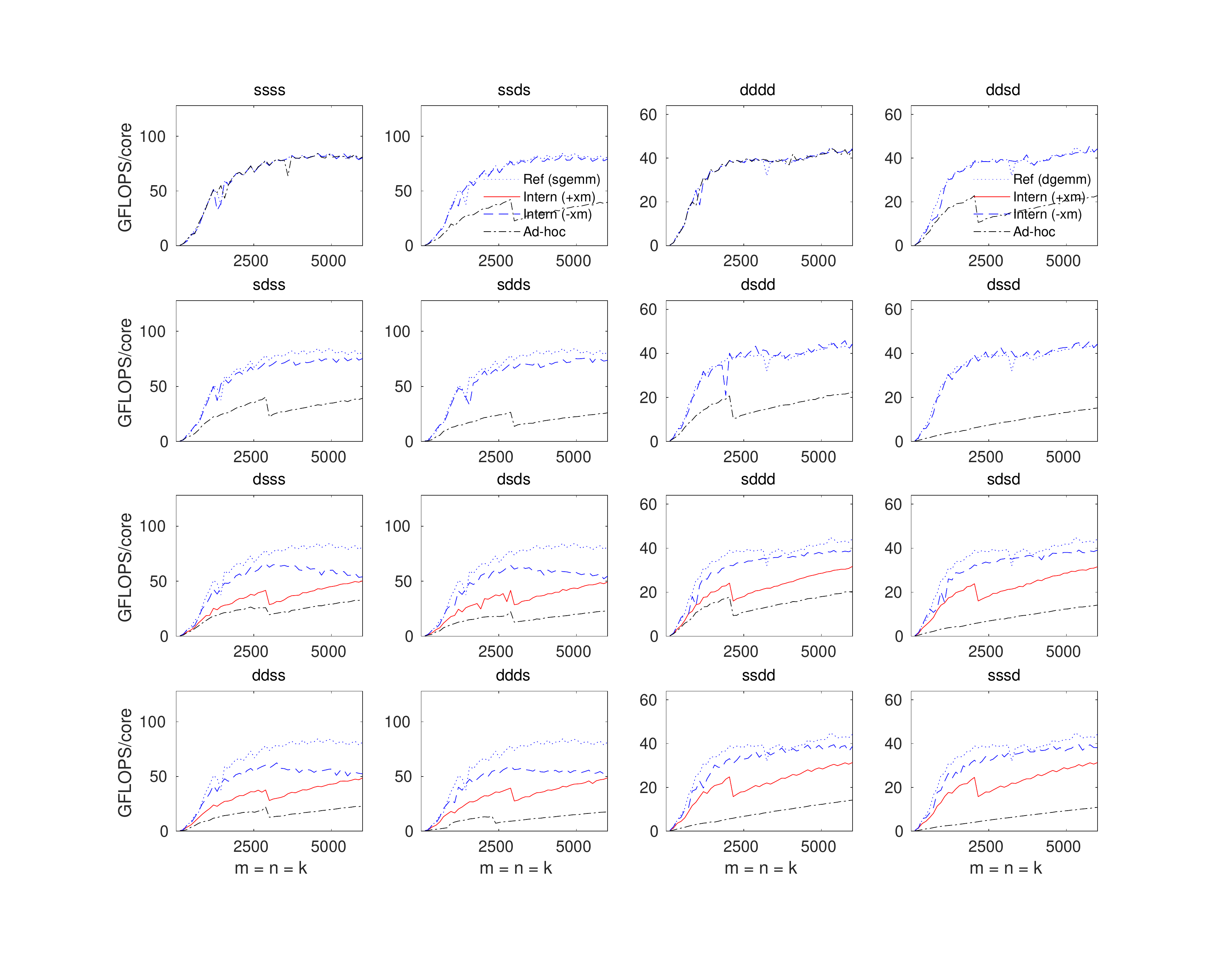} \\ \whline
\hspace{\graphhspace}
\includegraphics[width=\graphwidth,trim={\trimleft, \trimlower, \trimright, \trimupper},clip]{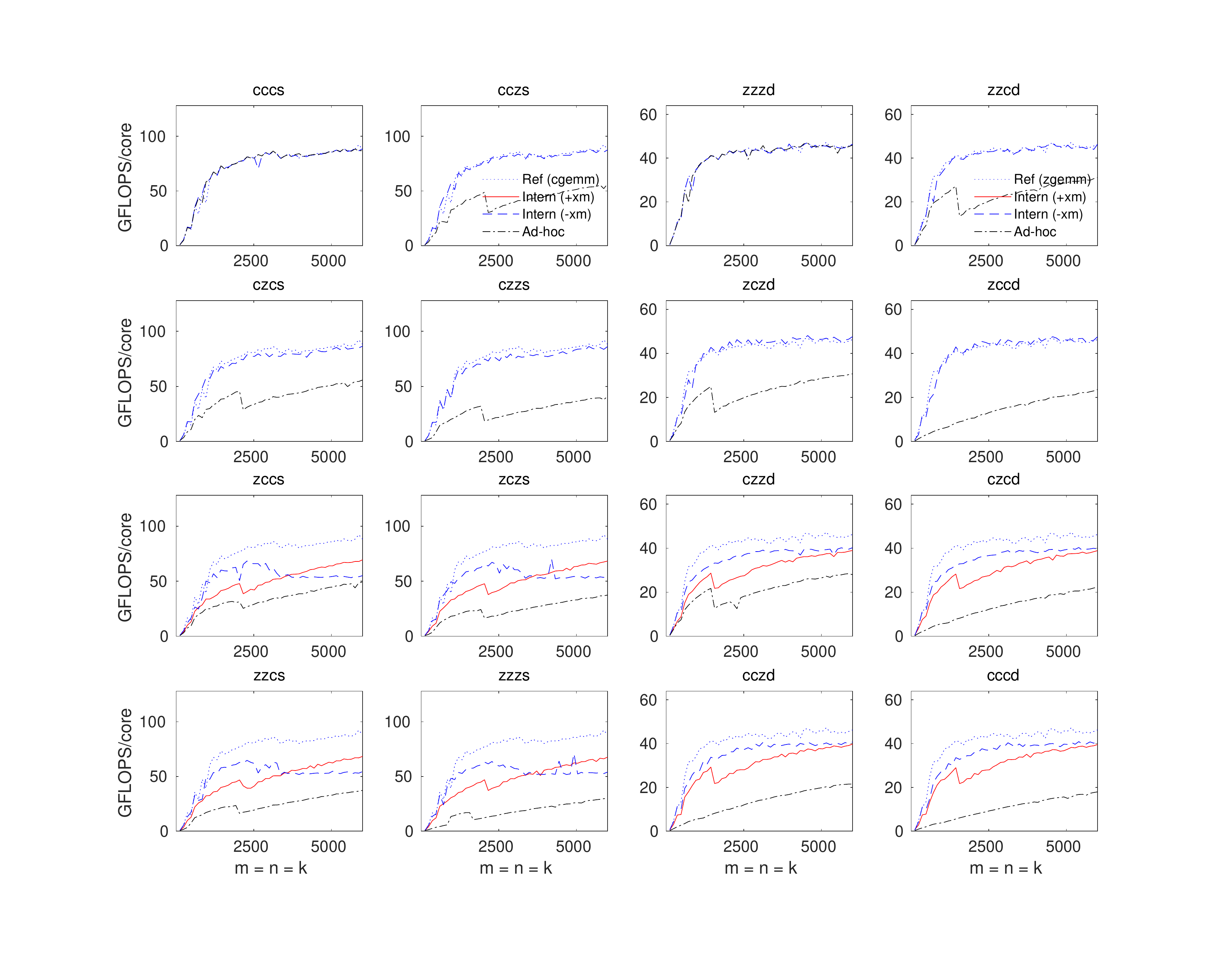}
\end{tabular}
\end{center}
\caption{
\sentencezeroa{\zerons}{\threens}\sentenceonea\sentencetwoa
}
\label{fig:perf_skx_t52_0}
\end{figure}

%
%
\begin{figure}[hp!]
\begin{center}
\begin{tabular}{l}
\hspace{\graphhspace}
\includegraphics[width=\graphwidth,trim={\trimleft, \trimlower, \trimright, \trimupper},clip]{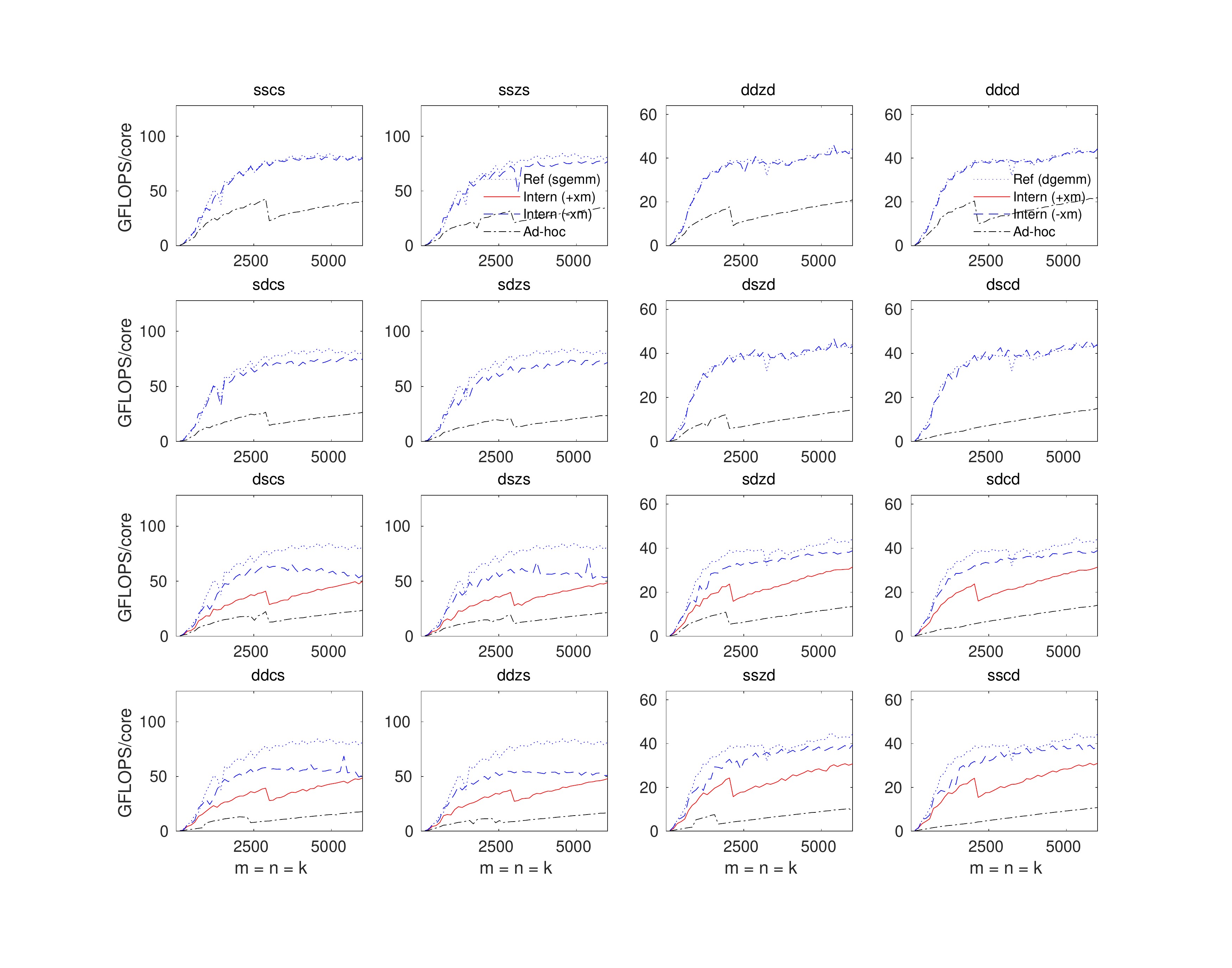} \\ \whline
\hspace{\graphhspace}
\includegraphics[width=\graphwidth,trim={\trimleft, \trimlower, \trimright, \trimupper},clip]{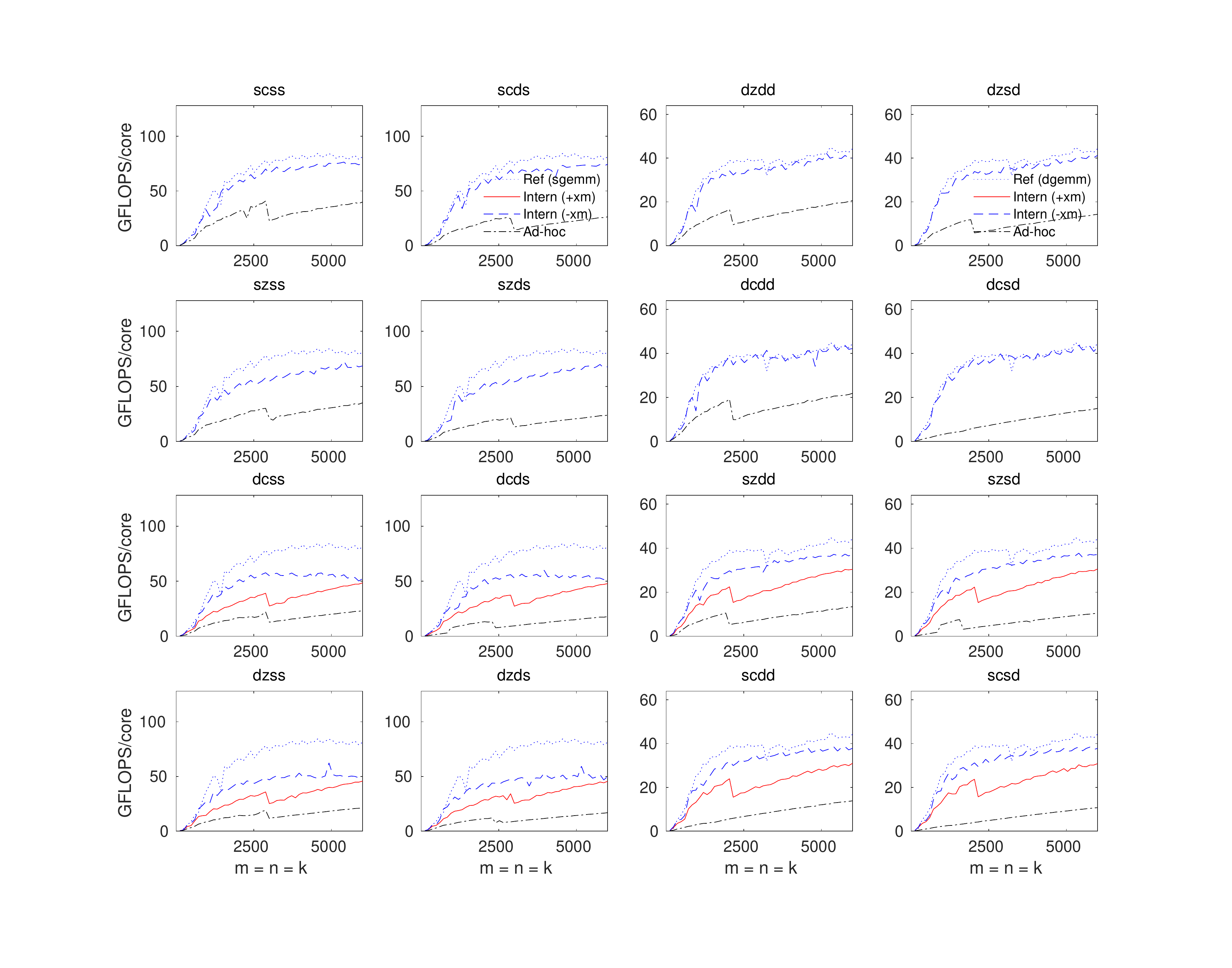}
\end{tabular}
\end{center}
\caption{
\sentencezeroa{\oneans}{\onebns}\sentenceonea\sentencetwoa
}
\label{fig:perf_skx_t52_1}
\end{figure}

%
%
\begin{figure}[hp!]
\begin{center}
\begin{tabular}{l}
\hspace{\graphhspace}
\includegraphics[width=\graphwidth,trim={\trimleft, \trimlower, \trimright, \trimupper},clip]{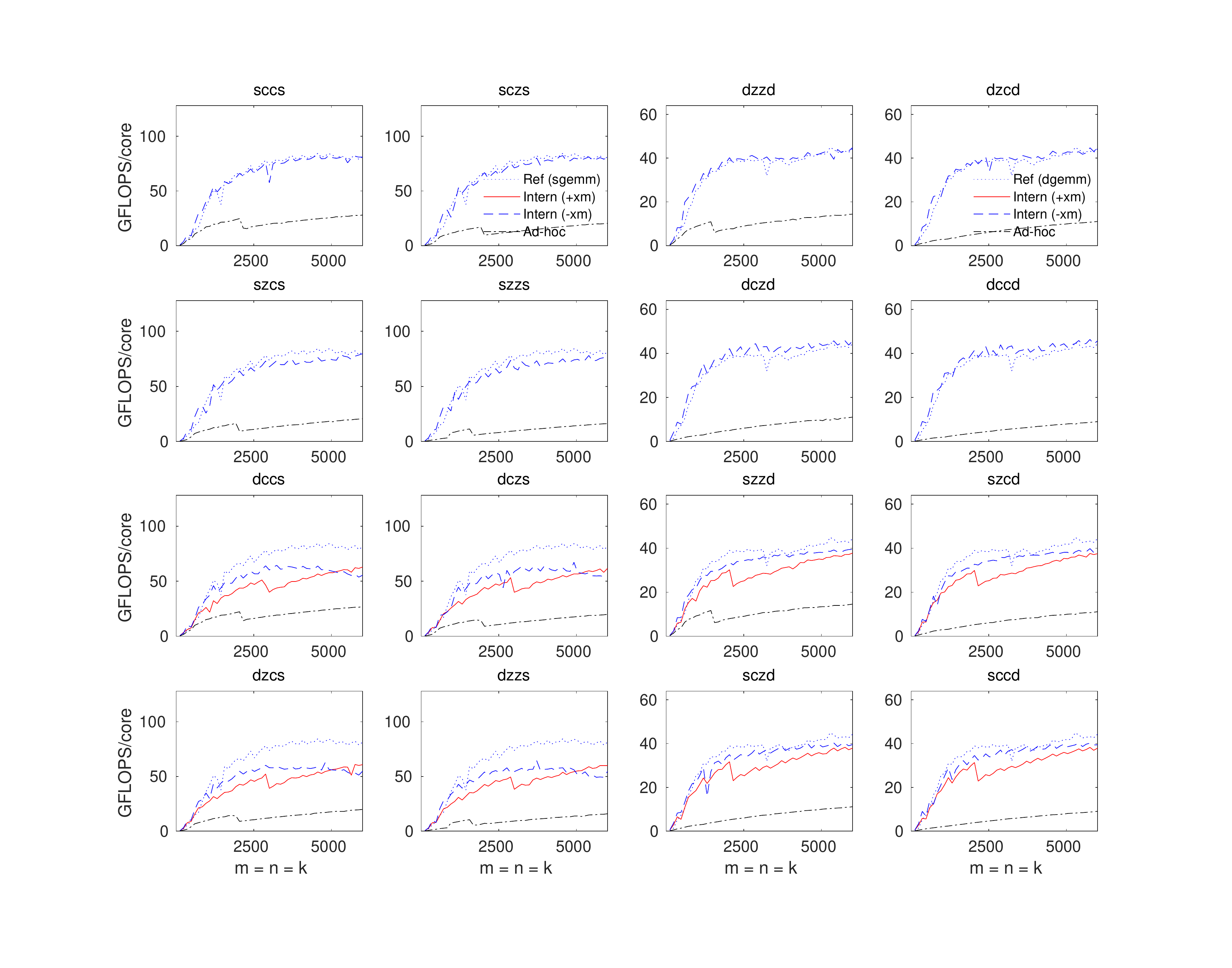} \\ \whline
\hspace{\graphhspace}
\includegraphics[width=\graphwidth,trim={\trimleft, \trimlower, \trimright, \trimupper},clip]{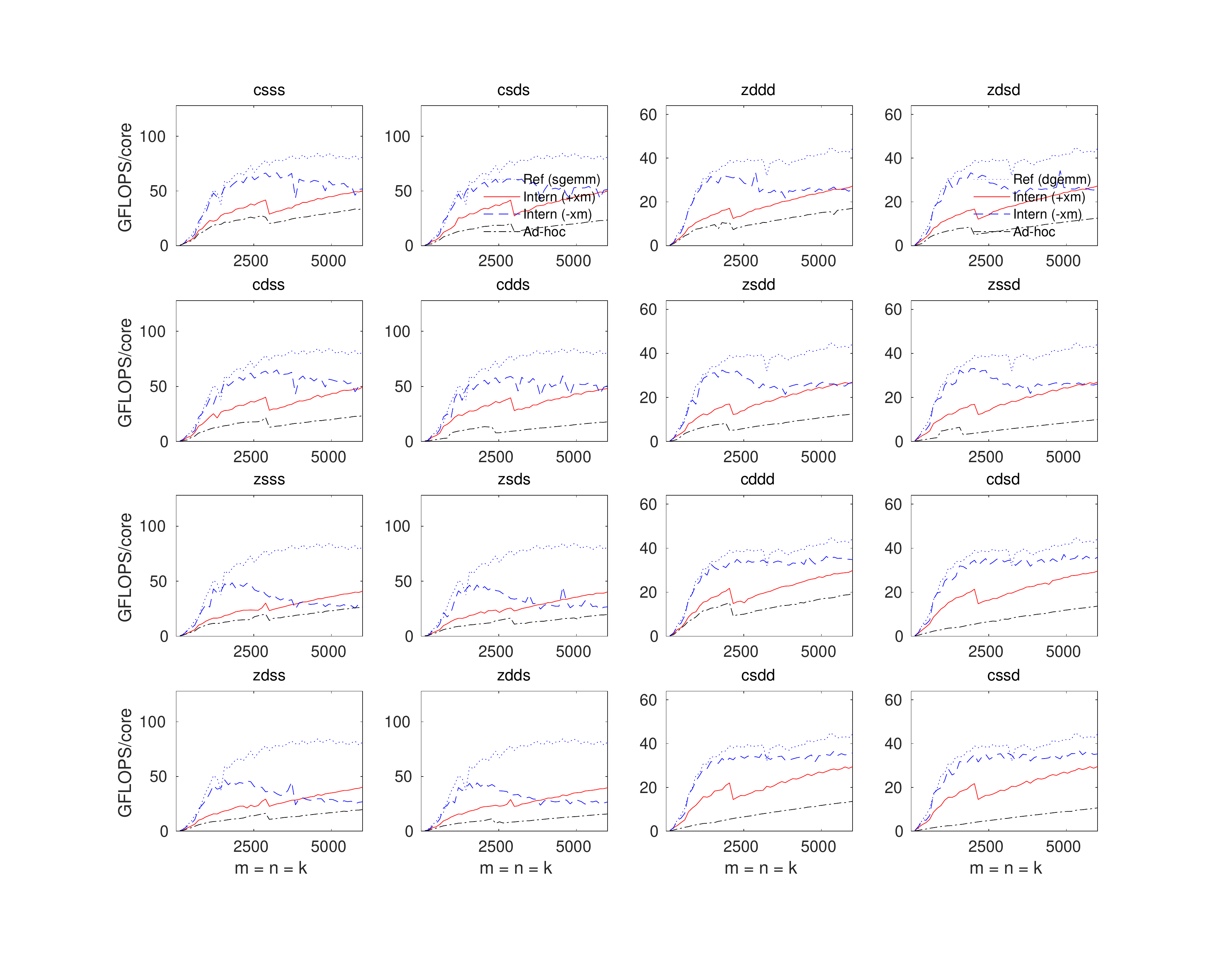}
\end{tabular}
\end{center}
\caption{
\sentencezeroa{\twoabns}{\onecns}\sentenceonea\sentencetwoa
}
\label{fig:perf_skx_t52_2}
\end{figure}

%
%
\begin{figure}[hp!]
\begin{center}
\begin{tabular}{l}
\hspace{\graphhspace}
\includegraphics[width=\graphwidth,trim={\trimleft, \trimlower, \trimright, \trimupper},clip]{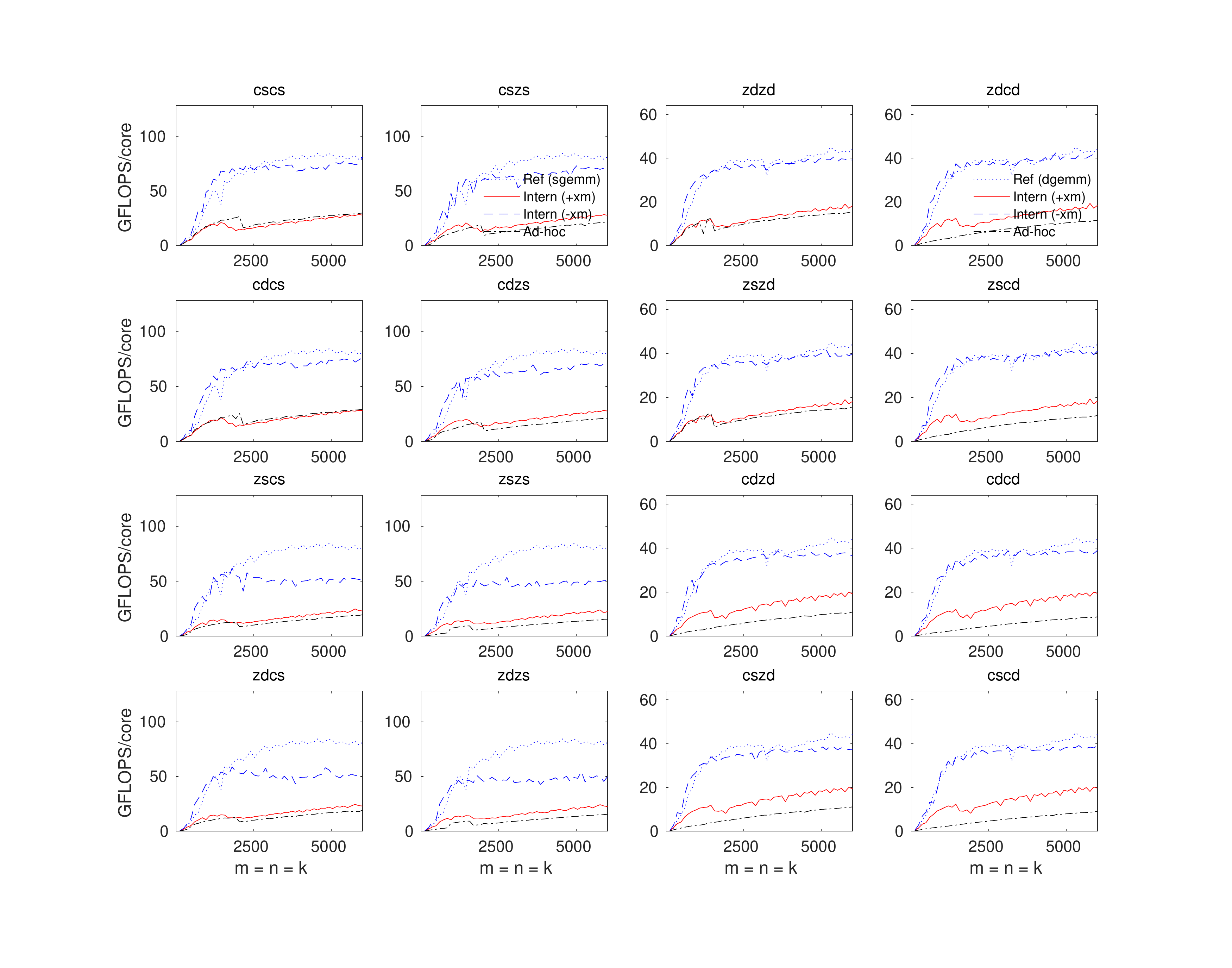} \\ \whline
\hspace{\graphhspace}
\includegraphics[width=\graphwidth,trim={\trimleft, \trimlower, \trimright, \trimupper},clip]{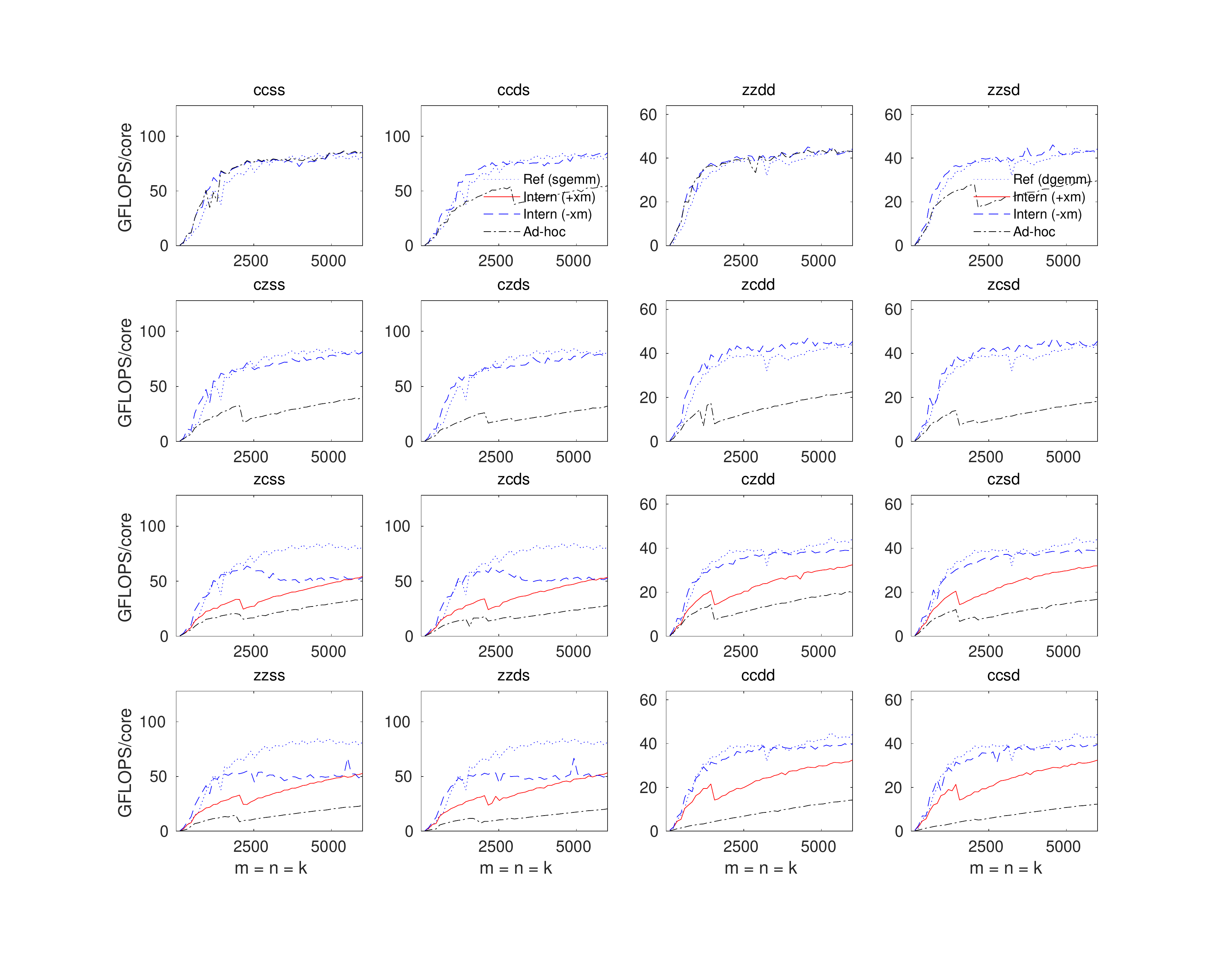}
\end{tabular}
\end{center}
\caption{
\sentencezeroa{\twobcns}{\twoacns}\sentenceonea\sentencetwoa
}
\label{fig:perf_skx_t52_3}
\end{figure}

\end{appendices}

\end{document}